\pdfoutput=1
\documentclass[useAMS,usenatbib]{mn2e}
\usepackage{graphicx}
\title[Mergers vs collisions: the multi-messenger picture]
{The multi-messenger picture of compact object encounters:
binary mergers versus dynamical collisions}
\author[S. Rosswog, et al.]{S. Rosswog$^{1,2,3}$\thanks{E-mail:
s.rosswog@jacobs-university.de}, T. Piran$^{4}$\thanks{E-mail:tsvi.piran@mail.huji.ac.il} and 
E. Nakar$^{5}$\thanks{E-mail:udini@wise.tau.ac.il}\\
$^{1}$School of Engineering and Science, Jacobs University Bremen, Germany\\
$^2$TASC, Department of Astronomy and Astrophysics, University of California, Santa Cruz, CA 95064\\
$^3$The Oskar Klein Centre, Department of Astronomy, AlbaNova, Stockholm University, SE-106 91 Stockholm, Sweden\\
$^4$Racah Institute of Physics, The Hebrew University, Jerusalem 91904, Israel\\
$^5$Raymond and Beverly Sackler School of Physics \& Astronomy, Tel Aviv University, 
Tel Aviv 69978, Israel}

\begin{document}

\date{Accepted 2012. Received 2012; in original form 2012}

\pagerange{\pageref{firstpage}--\pageref{lastpage}} \pubyear{2012}

\def\msun{M$_{\odot}$}
\def\Msun{M$_{\odot}$ }
\def\be{\begin{equation}}
\def\ee{\end{equation}}
\def\bi{\begin{itemize}}
\def\i{\item}
\def\ei{\end{itemize}}
\def\ben{\begin{enumerate}}
\def\een{\end{enumerate}}
\def\bea{\begin{eqnarray}}
\def\eea{\end{eqnarray}}
\def\gcc{gcm$^{-3}$}
\def\ye{$Y_e$}
\def\Ye{$Y_e$ }

\maketitle

\label{firstpage}

\begin{abstract}
We explore the multi-messenger signatures of encounters between two neutron stars (ns$^2$)
and between a neutron star and a stellar-mass black hole (nsbh). We focus on the differences 
between gravitational wave driven binary mergers and dynamical collisions that occur, 
for example, in globular clusters. Our discussion is based on Newtonian hydrodynamics
simulations that incorporate a nuclear equation of state and a multi-flavour neutrino
treatment.
For both types of encounters we compare the gravitational wave and neutrino emission 
properties. We also calculate the rates at which nearly unbound mass is delivered back to the 
central remnant in a ballistic-fallback-plus-viscous-disk model and we analyze the properties
of the dynamically ejected matter. 
Last but not least we address the electromagnetic transients that accompany each type of encounter.\\ 
We find that dynamical collisions are at least as promising as binary mergers for producing (short) 
gamma-ray bursts, but they also share the same possible caveats in terms of baryonic pollution. 
All encounter remnants produce peak neutrino luminosities of at least $\sim 10^{53}$ erg/s,
some of the collision cases exceed this value by more than an order of magnitude.
The canonical ns$^2$ merger case ejects more than 1\% of a solar mass of extremely 
neutron-rich ($Y_e\sim 0.03$) material, an amount that is consistent with double neutron 
star mergers being a major source of r-process in the galaxy. nsbh collisions eject very 
large amounts of matter ($\sim 0.15$ \msun) which seriously constrains their admissible
occurrence rates. The compact object {\em collision} rate (sum of ns$^2$ and nsbh) must therefore 
be less, likely much less, than 10\% of the ns$^2$ {\em merger} rate. 
The radioactively decaying ejecta produce  optical-UV ``macronova''  which, 
for the canonical merger case, peak after $\sim 0.4$ days with a luminosity of $\sim 5 \times 10^{41}$ 
erg/s.  ns$^2$ (nsbh) collisions reach up to 2 (4) times larger peak luminosities.
The dynamic ejecta deposit a kinetic energy comparable to a supernova in the ambient medium. 
The canonical merger case releases approximately $2 \times 10^{50}$ erg, the most extreme 
(but likely rare) cases deposit kinetic energies of up to $10^{52}$ erg. The deceleration of 
this  mildly relativistic material by the ambient medium produces long lasting radio flares. 
A canonical ns$^2$ merger at the detection horizon of advanced LIGO/Virgo produces a radio 
flare that peaks on a time scale of one year with a flux of $\sim$0.1 mJy at 1.4 GHz. 
Collisions eject more material at higher velocities and therefore produce brighter and 
longer lasting flares. 
\end{abstract}


\section{Introduction}

The encounter of a neutron star (ns) with another neutron star or with a stellar mass black hole
(bh) is a fascinating events that involve a large variety of different physical processes, 
see \cite{faber09}, \cite{duez10a}, \cite{shibata11}, \cite{rosswog11b} and \cite{faber12} for recent reviews 
on various aspects of this topic and for a guide to the current literature.\\
The detection of the gravitational waves (GWs) emitted during the inspiral of ns$^2$ and 
nsbh binary mergers are prime targets of ground-based gravitational wave detectors 
such as LIGO, Virgo and GEO600 \citep{willke06,acernese08,grote08,smith09}. In their 
advanced states, the interferometers target to detect the signals 
of ns$^2$ coalescences out to  hundreds of megaparsecs, corresponding 
to redshifts of $z\approx 0.1$. 
Once detected, the gravitational waves offer the exciting possibility of 
filling gaps in our understanding of the neutron star equation of state 
in the high density, low temperature regime that is experimentally hardly accessible.\\
Compact binaries have for a long time been the prime candidates for the
central engines of (short) gamma-ray bursts (GRBs; \cite{eichler89,paczynski91,narayan92})
and this hypothesis has survived being confronted with a wealth of observational 
results in recent years. Several challenges remain, however, and the case is far from being closed
\citep{piran05,nakar07,lee07,gehrels09}.\\
\begin{figure*}
   \centerline{
   \includegraphics[width=10cm,angle=0]{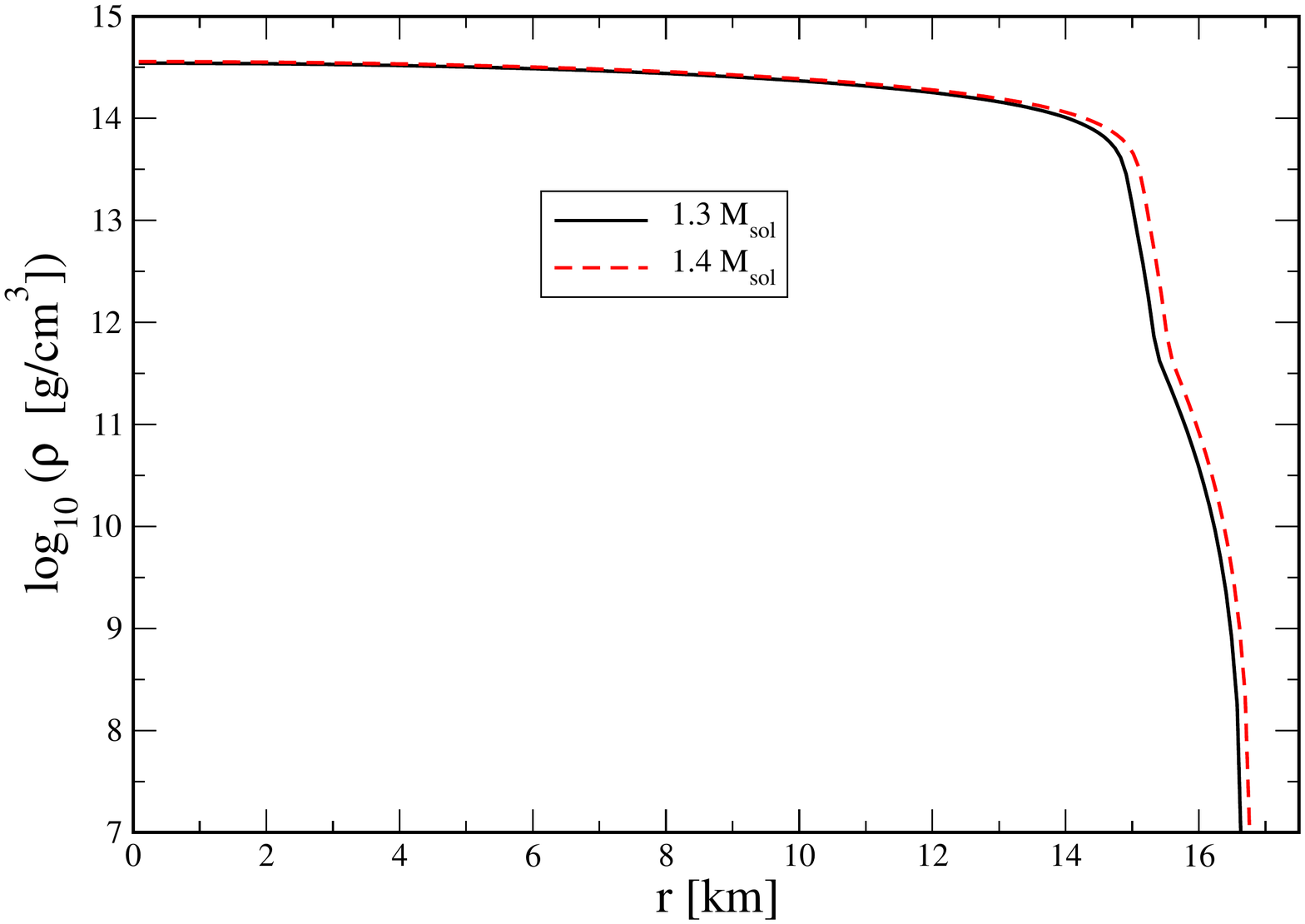} \hspace*{-0.5cm}
   \includegraphics[width=10cm,angle=0]{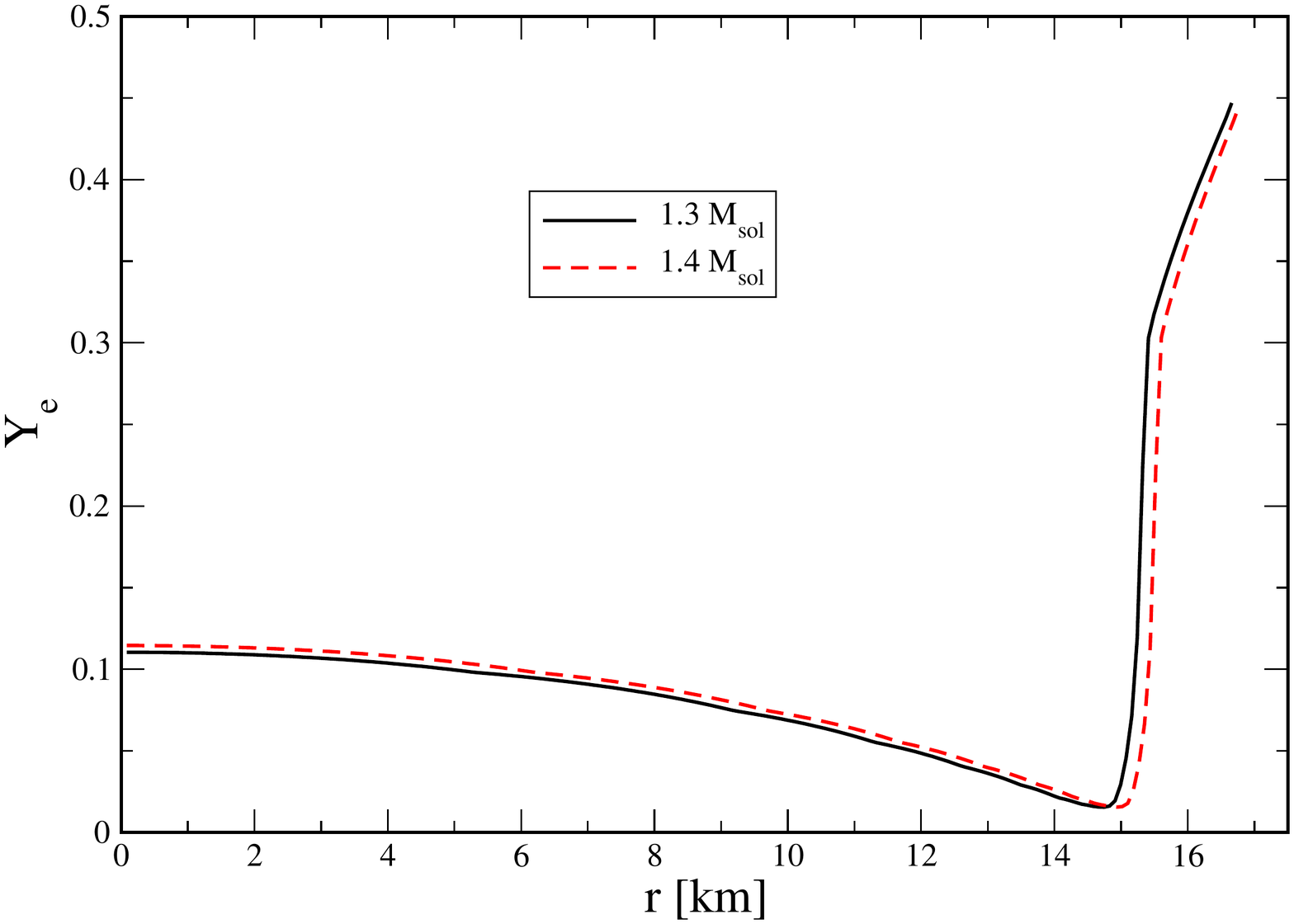}}  
   \caption{Initial density and $Y_e$ profiles (in hydrostatic and $\beta$-equilibrium) 
            of the  neutron stars used in this study (1.3 and 1.4 \msun, nuclear equation 
            of state of Shen et al. \citep{shen98a}).}
   \label{fig:initial_profiles}
\end{figure*}
They are also promising sources for the heaviest elements in the 
Universe that are formed via rapid neutron capture
\citep{lattimer74,lattimer76,eichler89,freiburghaus99b,rosswog99,roberts11,goriely11a}.
The textbook r-process source, core-collapse supernovae, have been found to be 
seriously challenged in providing the physical conditions (high entropy, low electron 
fraction together with rapid expansion) that are required to produce the heavy ($A>90$) r-process
elements \citep{roberts10,fischer10,arcones11a}\footnote{A possible exception may
be magnetorotationally driven supernova jets where interestingly low electron fraction
values seem to be reachable \citep{winteler12b}. It remains to be explored, however, how robust
this scenario is with respect to the stellar parameters and with respect to its nucleosynthetic 
yields.}. The main r-process nucleosynthesis
contenders are compact binary mergers which release neutron-rich matter in at least
three ways. Apart from the matter that is ejected dynamically 
via gravitational torques, there is an additional contribution due to neutrino-driven 
winds \citep{dessart09} and from the late-time dissolution of accretion disks
\citep{chen07,beloborodov08,metzger08}. While the initial starting point is the same, 
cold neutron star matter in $\beta$-equilibrium, the three channels differ in the 
amounts of released matter, in their entropies, expansion time scales and electron 
fractions. Therefore they might possibly produce different nucleosynthetic signatures.\\
The ejecta are responsible for two types of electromagnetic transients: the decompressed
neutron star matter is subject to radioactive decays \citep{li98,rosswog05a,metzger10b,roberts11} 
and at some stage the ejecta dissipate their kinetic energy in the ambient medium. The former 
is expected to produce an optical display not too 
different from a supernova, but much shorter, sometimes
referred to as ``macronova'', the latter has been shown to produce detectable, isotropic 
radio emission that peaks near one gigahertz and persists on a detectable (sub-milliJansky) 
level for weeks out to a distance of $z\approx0.1$ \citep{nakar11a}. Especially if the true 
GW detection rates should be near the lower end of the predictions, additional 
electromagnetic signatures would be crucial to confirm marginal gravitational wave
detections and they would therefore enhance the effective detector sensitivities 
\citep{kochanek93,hughes03,dalal06,arun09}.\\ 
Recently, dynamical collisions between compact objects
as they may occur in the core of globular clusters have been studied
by several authors \citep{kocsis06a,oleary09,lee10a,kocsis12}.
\cite{lee10a} concluded that collisions could produce GRBs at a detectable rate. 
To demonstrate the viability as GRB engines they performed hydrodynamic simulations 
of compact object collisions, though without use of detailed microphysics such as a 
nuclear equation of state or neutrino emission.\\
The main questions that we want to address in this paper are: 
\bi
\i [a)] In which ways do the remnants and signatures of dynamical 
        collisions differ from those of binary mergers? 
\i [b)] To which extent can their rate be constrained by nucleosynthetic yields?
\i [c)] How different are electromagnetic transients following mergers and collisions?
\ei 
To address these issues we performed a sizeable set of Newtonian hydrodynamics
simulations which include a nuclear equation of state and an opacity-dependent neutrino 
cooling scheme. These simulations are subsequently explored to predict the 
electromagnetic signatures of compact binary encounters.
Clearly, the investigated systems are relativistic and ultimately
General Relativity should be applied. It is, however, not
gravity alone that shapes the dynamics and observable signatures of these encounters.
Instead, also the remaining fundamental interactions contribute their share: 
a) the strong interaction via the nuclear equation of state, b) the weak 
interaction since it determines the neutrino emission rates and thus 
the evolution of the electron fraction $Y_e$ and c) the 
electromagnetic interaction which is, for example, responsible for producing radio-flares once 
ejected material dissipates its kinetic energy in the ambient medium. Given this complexity,
we consider Newtonian gravity as a tolerable approximation for the time being.
The presented simulations are meant to serve as benchmarks for future
simulations that may include general relativity and the relevant microphysics.\\ 
We discuss the viability of dynamical collisions as GRB central engines, in particular
the properties of remnant disks, their neutrino emission and
the prospects for magnetic field amplification. We further calculate the mass and the return time 
scales of fallback and we present in detail the properties of dynamical ejecta 
as a basis for subsequent nucleosynthesis calculations. Based on these findings we discuss the
question ``What is the electromagnetic signature of a ns$^2$ and a nsbh encounter?'' \\
The paper is structured as follows. In Sec. \ref{sec:sim} we briefly summarize the ingredients
of our simulations both in terms of physics and numerical methods. In Sec. \ref{sec:results}
we describe the main findings. Sec. \ref{subsec:dynamics_GW} discusses
the dynamics and its imprint on gravitational waves, Sec. \ref{subsec:nu}
explores the neutrino signal and Sec. \ref{subsec:fallback} the differences
between merger and collision fallback. In Sec. \ref{subsec:ejecta} we discuss the
properties of the dynamically ejected material such as mass, electron 
fraction and velocity structure. They all  shape the electromagnetic display,
which is addressed in Sec. \ref{subsec:elmag}. Our results are summarized and 
discussed in Sec.~\ref{sec:summary}.

\section[]{Simulations}
\label{sec:sim}
The simulations of this paper make use of the Smooth Particle Hydrodynamics (SPH) 
method, see \cite{monaghan05} and \cite{rosswog09b} for recent reviews. Our code 
is an updated version of the one that was used in earlier studies
\citep{rosswog02a,rosswog03a,rosswog03c,rosswog05a}.
We solve the following evolution equations for each particle $a$
\bea
\frac{{d\vec{v}}_{a}}{dt}&=& \hspace*{-0.3cm}- \sum_b m_b \left(
\frac{P_a}{\rho_a^2} + \frac{P_b}{\rho_b^2}  + \Pi_{ab}
 \right)\nabla_a W_{ab} + \vec{f}_{a,\rm
   g} +  \vec{f}_{a,\rm GW} \label{eq:momentum_equation}\\ 
\frac{d u_{a}}{dt}&=&  \sum_b m_b \left( \frac{P_a}{\rho_a^2} + \frac{1}{2}  \Pi_{ab} \right) \vec{v}_{ab} \cdot
\nabla_a W_{ab} - \frac{du_a^\nu}{dt} \\
\frac{d Y_{e,b}}{dt}  &=& \lambda_{\rm PC} \; Y_{\rm n} - \lambda_{\rm EC} \; Y_{\rm n},
\label{eq:energy_equation}
\eea
the mass density is calculated by summing up contributions from neighboring particles
\be
\rho_a= \sum_b m_b W_{ab}\label{eq:basic:summary_sum_rho}.\\
\ee
Here  $m_b$ is the (constant) mass of particle $b$ and $W_{ab}= W(|\vec{r}_a-\vec{r}_b|, h_{ab})$
denotes the cubic spline kernel \citep{monaghan85b} evaluated with the average smoothing length
$h_{ab}=(h_a+h_b)/2$. $\vec{f}_{a,\rm g}$ is the additional acceleration due to self-gravity that 
we evaluate using a binary tree \citep{benz90b} and $\vec{f}_{a,\rm GW}$ results from gravitational 
wave back-reaction. The relative particle velocity is denoted as $\vec{v}_{ab}= \vec{v}_a-\vec{v}_b$. 
To produce entropy in shocks artifical dissipation is included via the tensor $\Pi_{ab}$.
It has the standard form \cite{monaghan92} but particular care has been taken to avoid
possible artifacts due to artificial viscosity, this has been outlined in detail
in \cite{rosswog08b}. The quantity $u_a$ denotes the specific internal energy of particle 
$a$ the evolution of which is determined by $PdV$-work and viscous heating (summation term) 
and the energy loss to neutrinos, $\frac{du_a^\nu}{dt}$. The quantities 
\be
\lambda_{\rm PC}= \frac{R^{\rm eff}_{\rm PC}}{\eta_{\rm np}} \quad {\rm and } \quad
\lambda_{\rm EC}= \frac{R^{\rm eff}_{\rm EC}}{\eta_{\rm pn}}
\ee
are the electron and positron capture rates per neutron/proton. $R^{\rm eff}_{\rm EC/PC}$ are the effective
neutrino number emission rates and the quantities $\eta_{\rm np}$/$\eta_{\rm pn}$
reduce in the non-degenerate limit to the number densities $n_n$ and $n_p$ \citep{bruenn85},
for a more detailed account on the neutrino treatment we refer to the original paper \citep{rosswog03a}.
The pressure at a particle $b$, $P_b(\rho_b,T_b,Y_{e,b})$, is calculated using the 
Shen et al. equation of state (EOS) \citep{shen98a,shen98b} extended to lower densities as 
described in \cite{rosswog02a}. \\
Compact binary systems are driven towards coalescence via the emission of gravitational waves.
The corresponding radiation reaction forces for a slow-motion, weak field source can
be calculated as the gradient of a radiation reaction potential which contains the
fifth time derivatives of the reduced quadrupole moments \citep{burke71}.
Simple backreaction prescriptions usually rely on reducing the order of the time derivatives  
by averaging over several orbital periods. This procedure is well justified during the 
secular inspiral stages of a compact binary merger, but it has no well-defined meaning 
in the case of a parabolic encounter. For this reason we ignore the exerted 
backreaction for the {\em collision} cases of this first study, for the merger cases we use a
simple point mass description \citep{davies94} for $\vec{f}_{a,\rm GW}$. The possible
impact of this technical shortcoming will be discussed below.\\
As a diagnostics of the dynamical evolution we will use below the gravitational wave 
amplitudes as seen by an observer located along the binary rotation axis. Consistent with the
Newtonian treatment of gravity we extract gravitational waves in quadrupole approximation
\bea
h^{\rm TT}_+ &=& \frac{1}{d} \frac{G}{c^4} \left( \ddot{I}^{xx} - \ddot{I}^{yy} \right)\\
h^{\rm TT}_\times &=&  \frac{2}{d} \frac{G}{c^4}  \ddot{I}^{xy},
\eea
where $I^{ij}$ is the reduced quadrupole moment tensor evaluated at retarded times,
\be
I^{ij}= \sum_b m_b (x^i_b x^j_b - \frac{1}{3}\delta^{ij} r_b^2).\label{eq:quad_mom}
\ee
The needed time derivatives can be obtained by straight forward
differentiation of Eq.(\ref{eq:quad_mom}) so that the amplitudes can be calculated as sums
involving particle masses, positions, velocities and forces.\\
In the presented simulations we restrict our collision study to parabolic orbits. The strength
of such an encounter is parametrized by the parameter
\be
\beta \equiv \frac{R_1+R_2}{r_P},\label{eq:beta}
\ee
where the $R_i$ are the neutron star/Schwarzschild radii of the involved objects and 
$r_P$ is the pericenter  distance. Thus $\beta=1$ corresponds to a grazing impact, 
stronger (weaker) impacts have larger (smaller) values. Since the collision rates are 
proportional to the pericenter distance $r_P$, we consider run A with $\beta=1$ as the 
most likely ns$^2$ collision case.
Collisions with pericenter distances $\beta<1$ can still form tidal capture 
binaries \citep{fabian75,lee10a} which can lead to final collisions after a sequence of pericenter passages.
Collisions with $\beta<1$, however, become computationally increasingly cumbersome
due to the discrepancy between orbital and internal dynamical time scales. Therefore
we only consider  collisions with $\beta \ge 1$. To keep the explored parameter space
under controle, all investigated nsbh collisions possess a fixed impact strength of
 $\beta=1$ and we only vary the black hole mass.\\
It has long been known that the ns mass distribution possesses a narrow peak
near 1.35 \Msun \citep{thorsett99}. Recent studies find an additional
broader peak around 1.5-1.7 \Msun \citep{kiziltan10,valentim11} for neutron stars with
white dwarf companions. There may be an additional low-mass peak near 1.25 \Msun 
produced by electron capture supernovae \citep{podsiadlowski04,vandenheuvel04,schwab10}.
The mass distributions for neutron stars of different evolutionary paths have
recently been discussed in \cite{oezel12}. In our simulations we restrict ourselves 
conservatively to masses near the 1.33 \Msun peak that the latter authors find for double
neutron star systems. \\
In ns$^2$ collision cases we use masses of 1.3 and 1.4 \Msun and 
explore the dependence on the impact strength parameter $\beta$. For nsbh collisions we use 
a neutron star of 1.3 \Msun and we investigate the dependence on the black hole 
mass ($m_{\rm bh}=$ 3, 5 and 10 \msun) while keeping the impact parameter at $\beta=1$.\\
We consider run H with $m_1= 1.3$ \msun, $m_2= 1.4$ \msun, i.e. $q\approx 0.923$
and negligible spins \citep{bildsten92,kochanek92} as the generic merger case and we 
will use it frequently as a reference point to compare the other cases against.  
As a somewhat academic case, we explore an initially tidally locked binary neutron 
star system in run G.  In runs I and J we briefly touch upon neutron star black
hole mergers where the black hole is treated as a Newtonian point mass with an absorbing 
boundary at the Schwarzschild radius.\\
The initial neutron stars are constructed from spherically symmetric, zero-temperature, 
$\beta$-equilibrium profiles, see Fig.~\ref{fig:initial_profiles}. The SPH particles 
are placed in a close packed, hexagonal lattice configuration and they are subsequently
relaxed so that they can find their true numerical equilibrium state, see 
Sect. 3.1 of \cite{rosswog07c}. Some of our simulations have been run up to 0.5 s, 
more than an order of magnitude longer than existing simulations on this topic.\\
All performed simulations are summarized in Tab.~\ref{tab:runs}.

\begin{table*}
 \centering
 \begin{minipage}{140mm}
  \caption{Overview over the performed simulations. The impact strength parameter $\beta$ is defined in Eq.~(\ref{eq:beta}).}
  \begin{tabular}{@{}ccrccrlcccclr@{}}
  \hline
   Run   &  $m_1$ [\msun] & $m_2$ [\msun] & $\beta$ & $N_{\rm SPH}\; [10^6]$ & $t_{\rm end}$ [ms]& objects/comment \\
\hline 
\\
       \underline{Collisions} \\
   A     &   1.3          & 1.4           & 1       & 2.7      &  21.2    & ns-ns\\
   B     &   1.3          & 1.4           & 2       & 8.0      &   9.0    & ns-ns\\
   C     &   1.3          & 1.4           & 5       & 2.7      &  13.2    & ns-ns\\
   D     &   1.3          & 3.0           & 1       & 1.3      & 127.5    & ns-bh\\
   E     &   1.3          & 5.0           & 1       & 1.3      & 143.6    & ns-bh\\
   F     &   1.3          & 10.0          & 1       & 1.3      & 540.3    & ns-bh\\
\\
 \underline{Mergers} \\
   G     &   1.3          & 1.4           & n.a.    & 2.7      &  20.3    & ns-ns, corot.\\
   H     &   1.3          & 1.4           & n.a.    & 2.7      &  19.1    & ns-ns, no spins\\
   I     &   1.4          & 5.0           & n.a.    & 0.2      & 138.7    & ns-bh, no spins\\
   J     &   1.4          & 10.0          & n.a.    & 0.2      & 139.3    & ns-bh, no spins\\
\hline
\label{tab:runs}
\end{tabular}
\end{minipage}
\end{table*}

\section{Results}
\label{sec:results}             
\subsection[]{Encounter dynamics and gravitational wave emission}
\label{subsec:dynamics_GW}
\begin{figure*}
   \centerline{
   \includegraphics[width=10cm,angle=0]{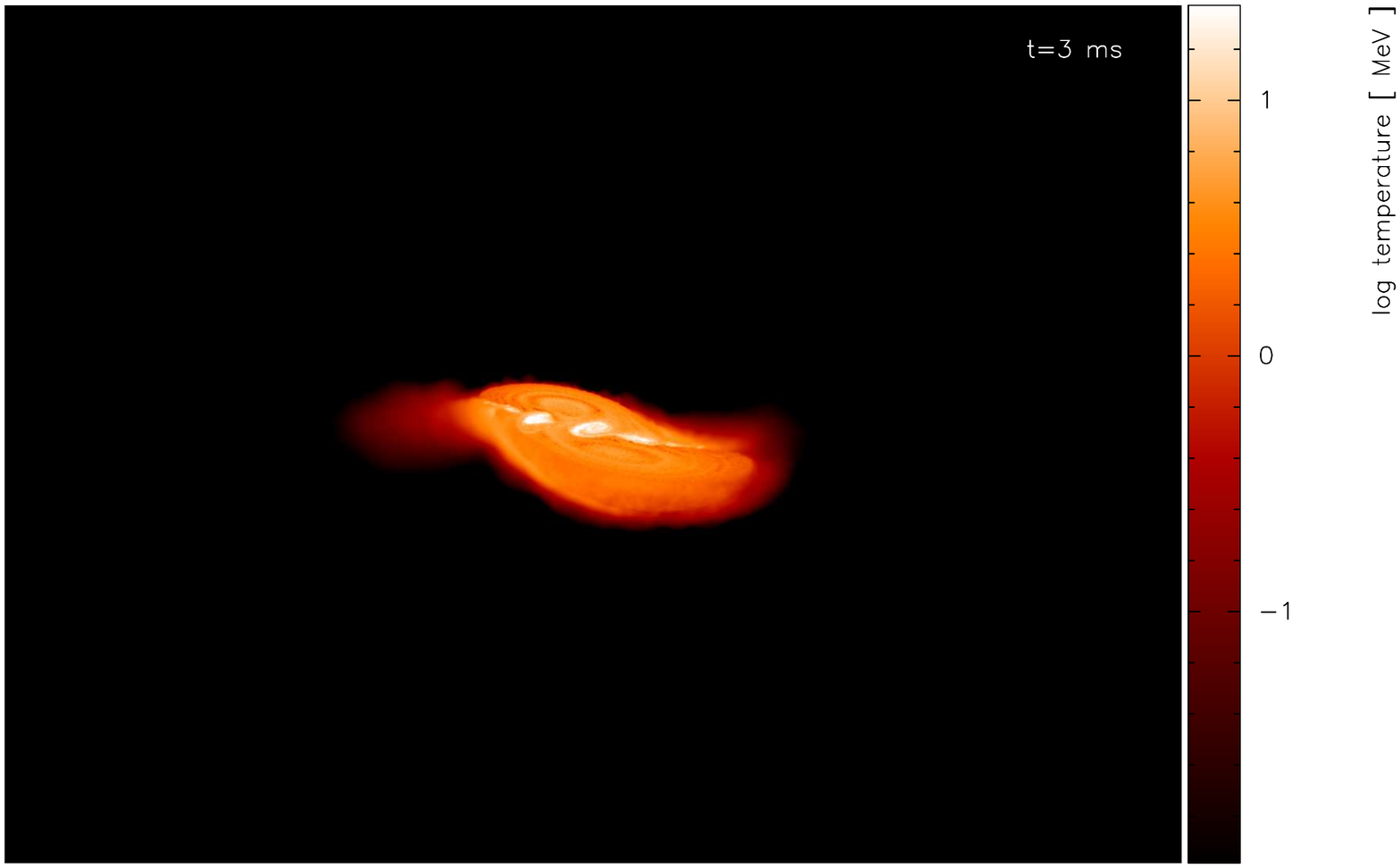} \hspace*{-0.5cm}
   \includegraphics[width=10cm,angle=0]{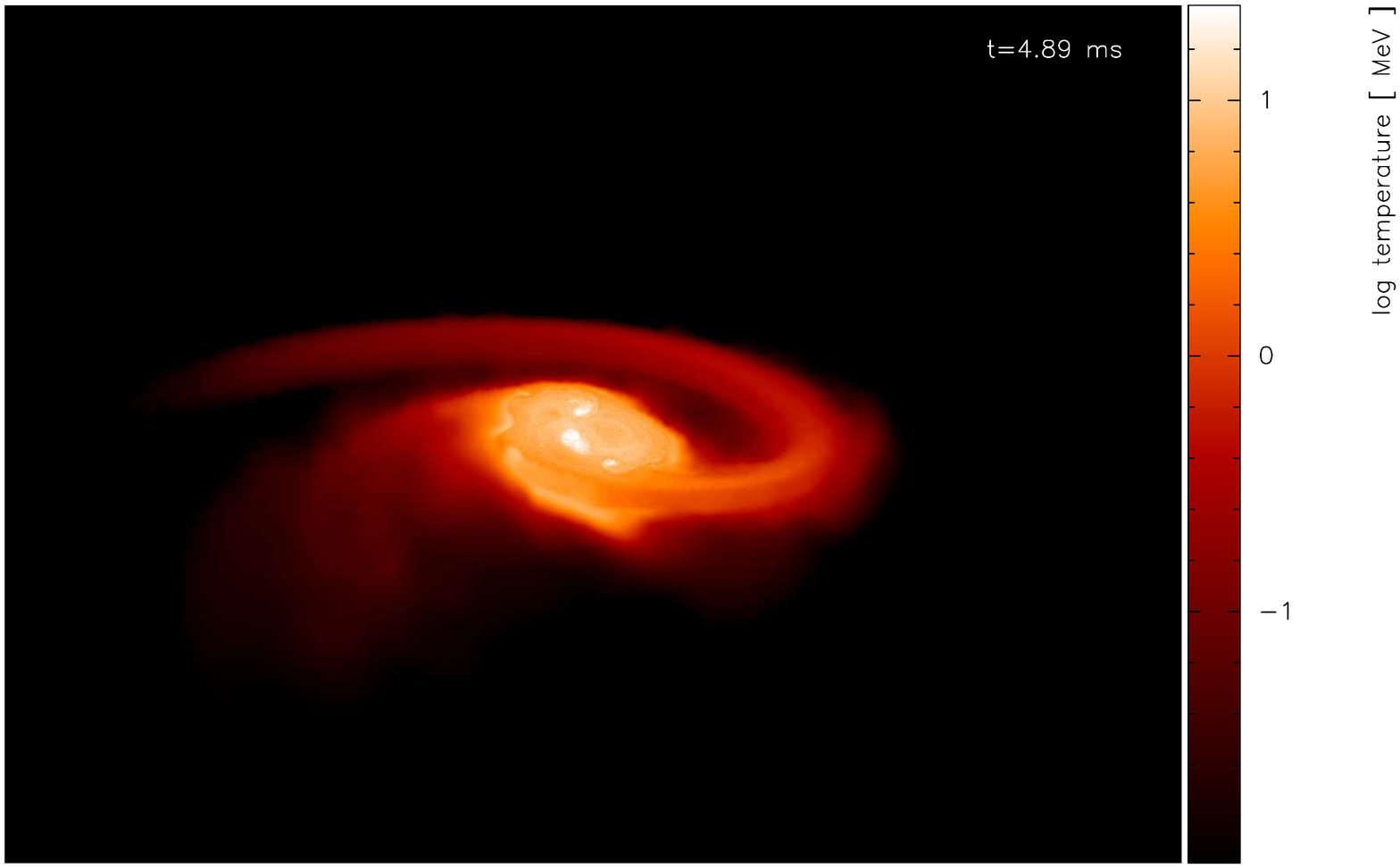}}
   \vspace*{-0.5cm}
   \centerline{
   \includegraphics[width=10cm,angle=0]{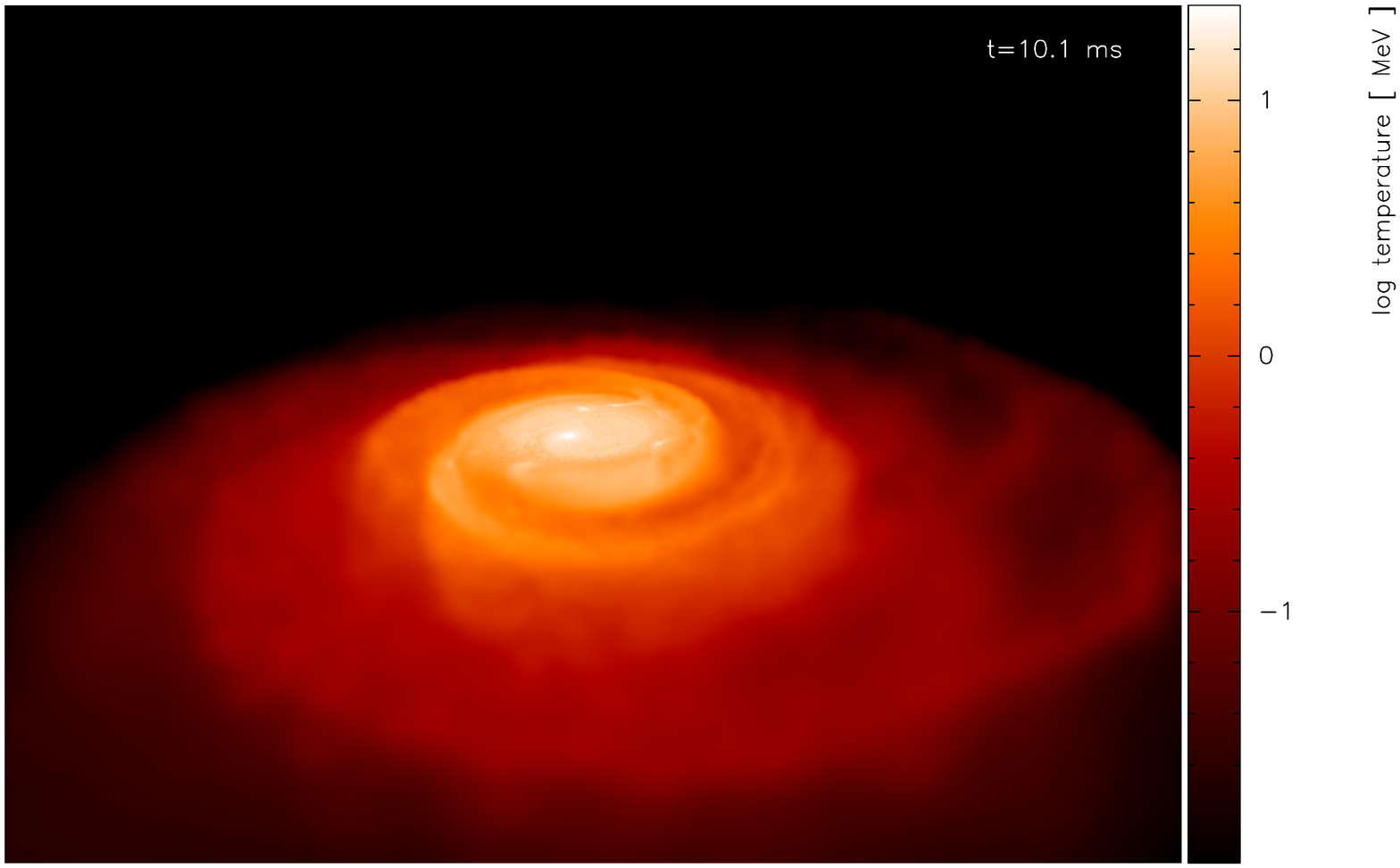} \hspace*{-0.5cm}
   \includegraphics[width=10cm,angle=0]{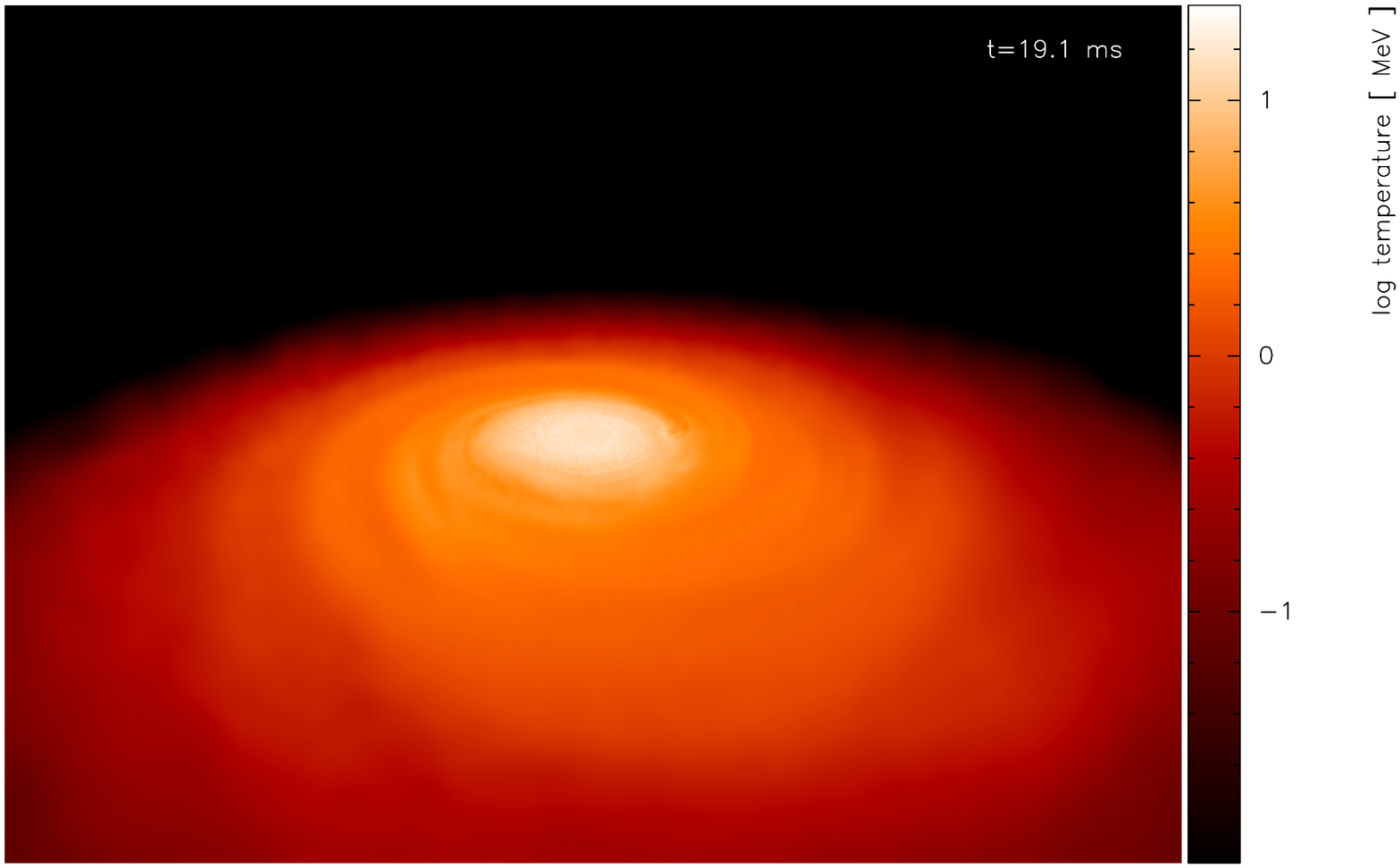}}    
   \caption{3D rendering of the temperature distribution in the standard neutron star merger
            case (1.3 and 1.4 \msun, no spin; run H). The upper half of the matter distribution has been "chopped off" 
            to allow for a view into the stars. 
            To enhance the contrast, the upper limit of the colourbar has been fixed to 20 MeV.
            In the various vortices that emerge due to Kelvin-Helmholtz instabilities peak
            temperatures in excess of 60 MeV are temporarily reached.}
   \label{fig:temp_rendering_ns13_ns14_irrot}
\end{figure*}

\begin{figure*}
   \centerline{
   \includegraphics[width=10cm,angle=0]{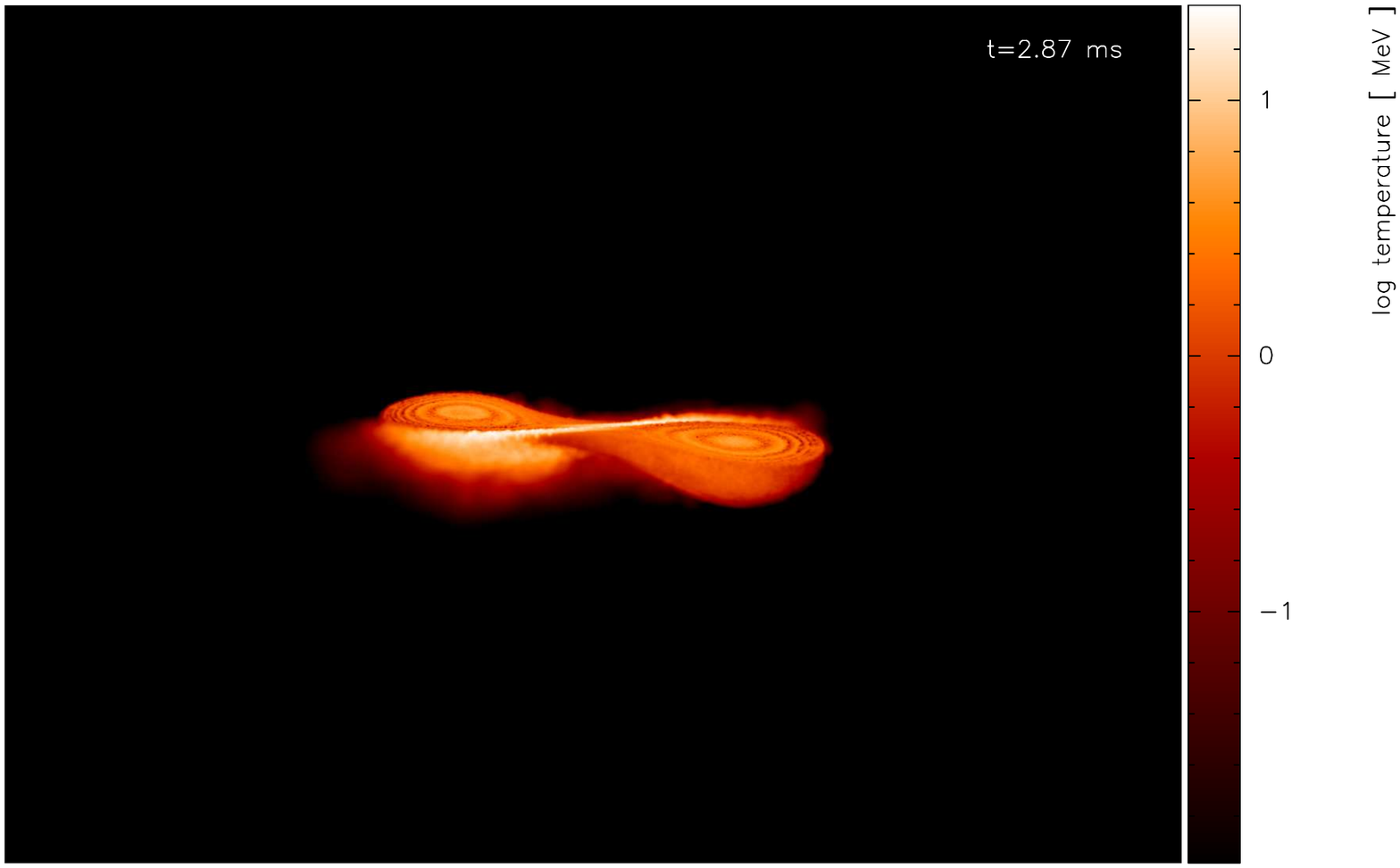}  \hspace*{-0.5cm}
    \includegraphics[width=10cm,angle=0]{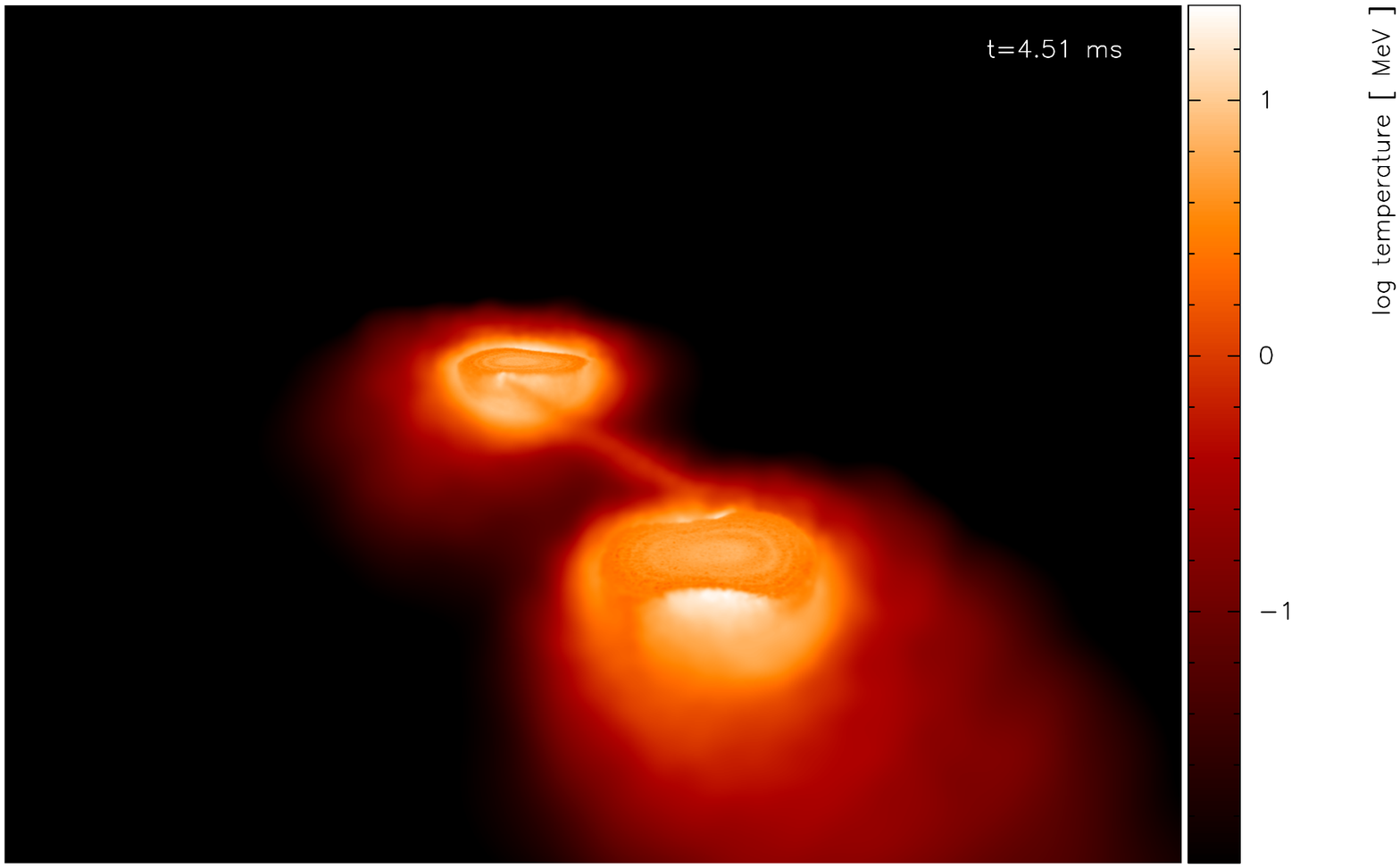}} 
   \vspace*{-0.5cm}  
   \centerline{
   \includegraphics[width=10cm,angle=0]{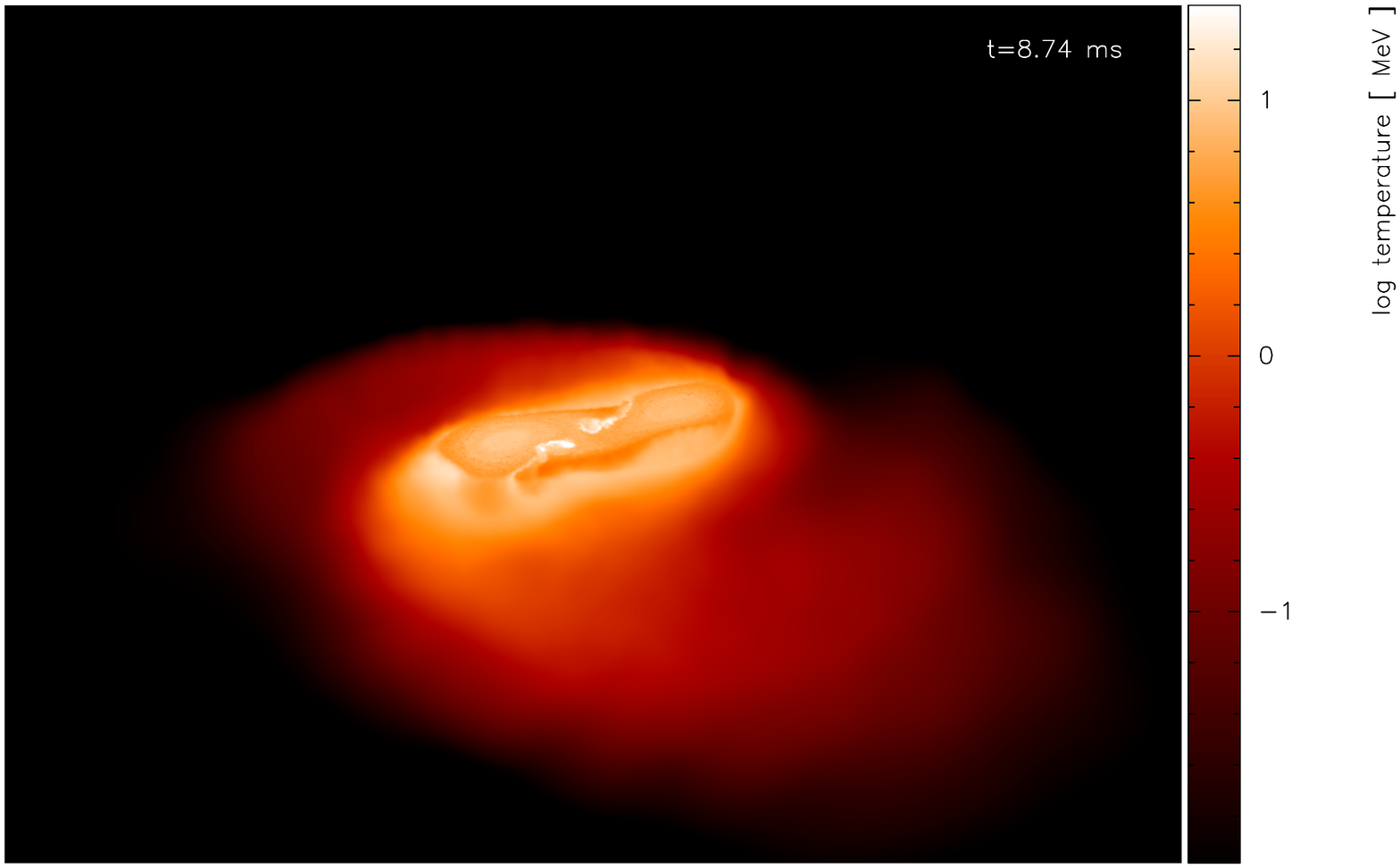}  \hspace*{-0.5cm}
    \includegraphics[width=10cm,angle=0]{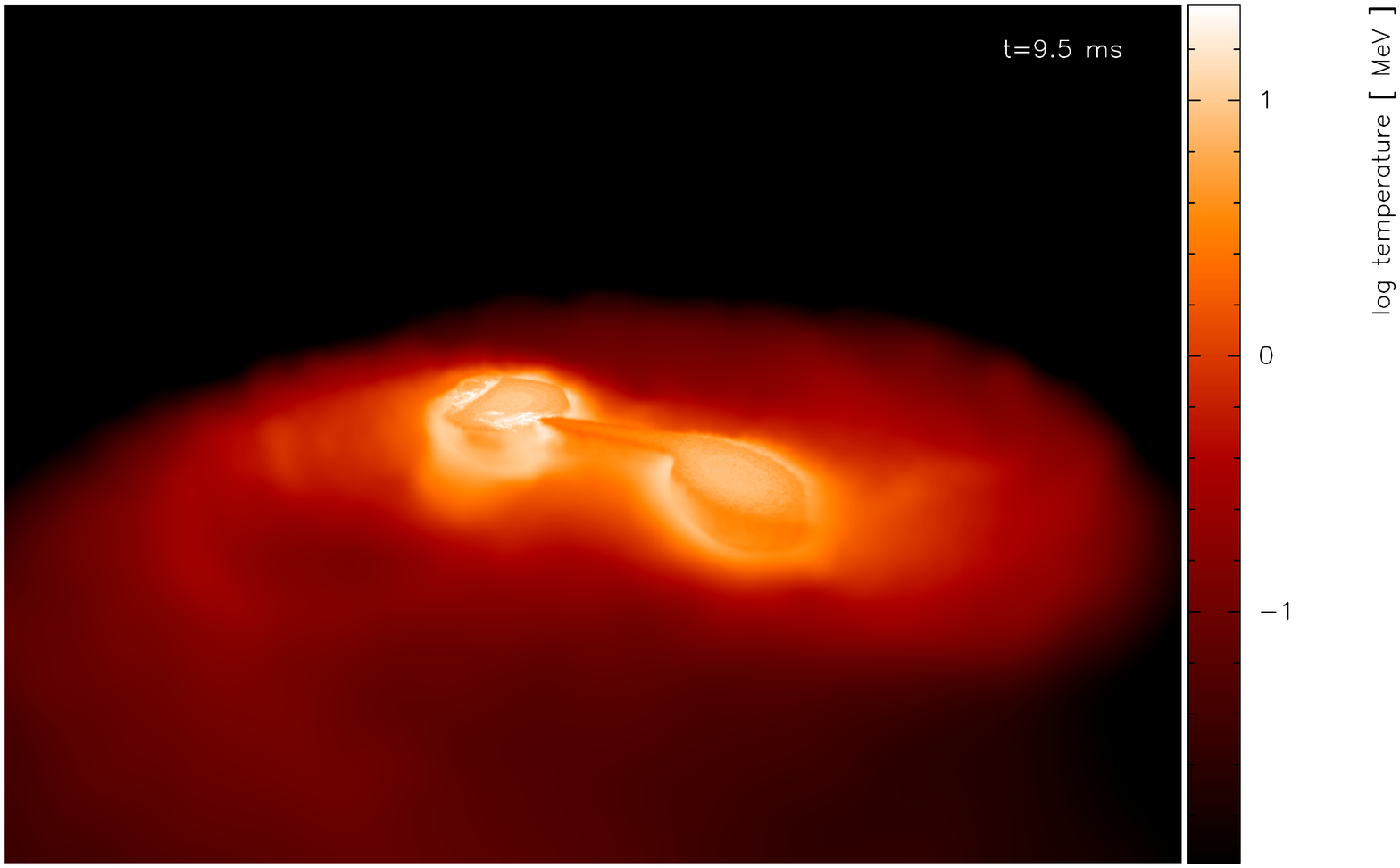}}   
    \vspace*{-0.5cm}
    \centerline{
   \includegraphics[width=10cm,angle=0]{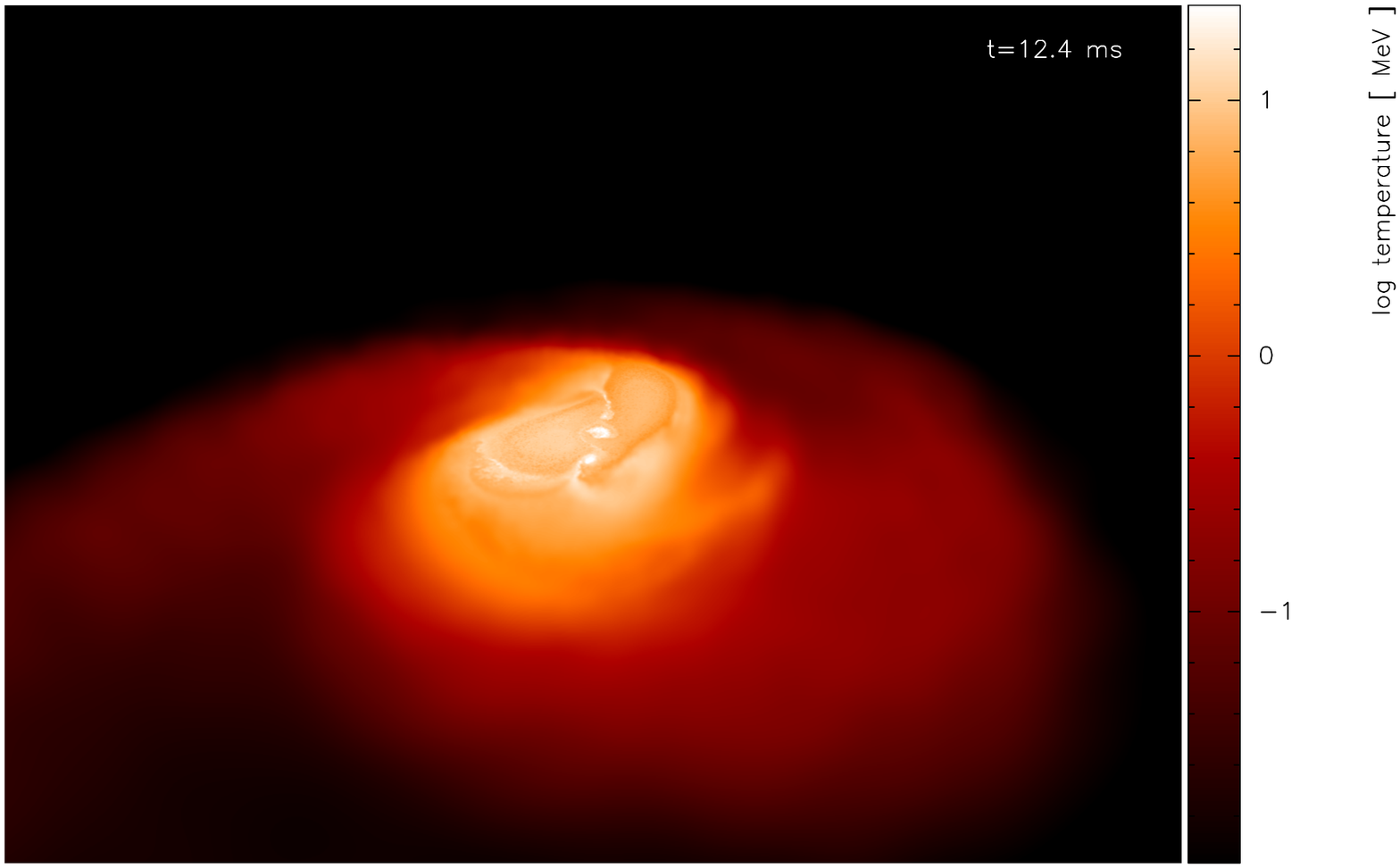}  \hspace*{-0.5cm}
    \includegraphics[width=10cm,angle=0]{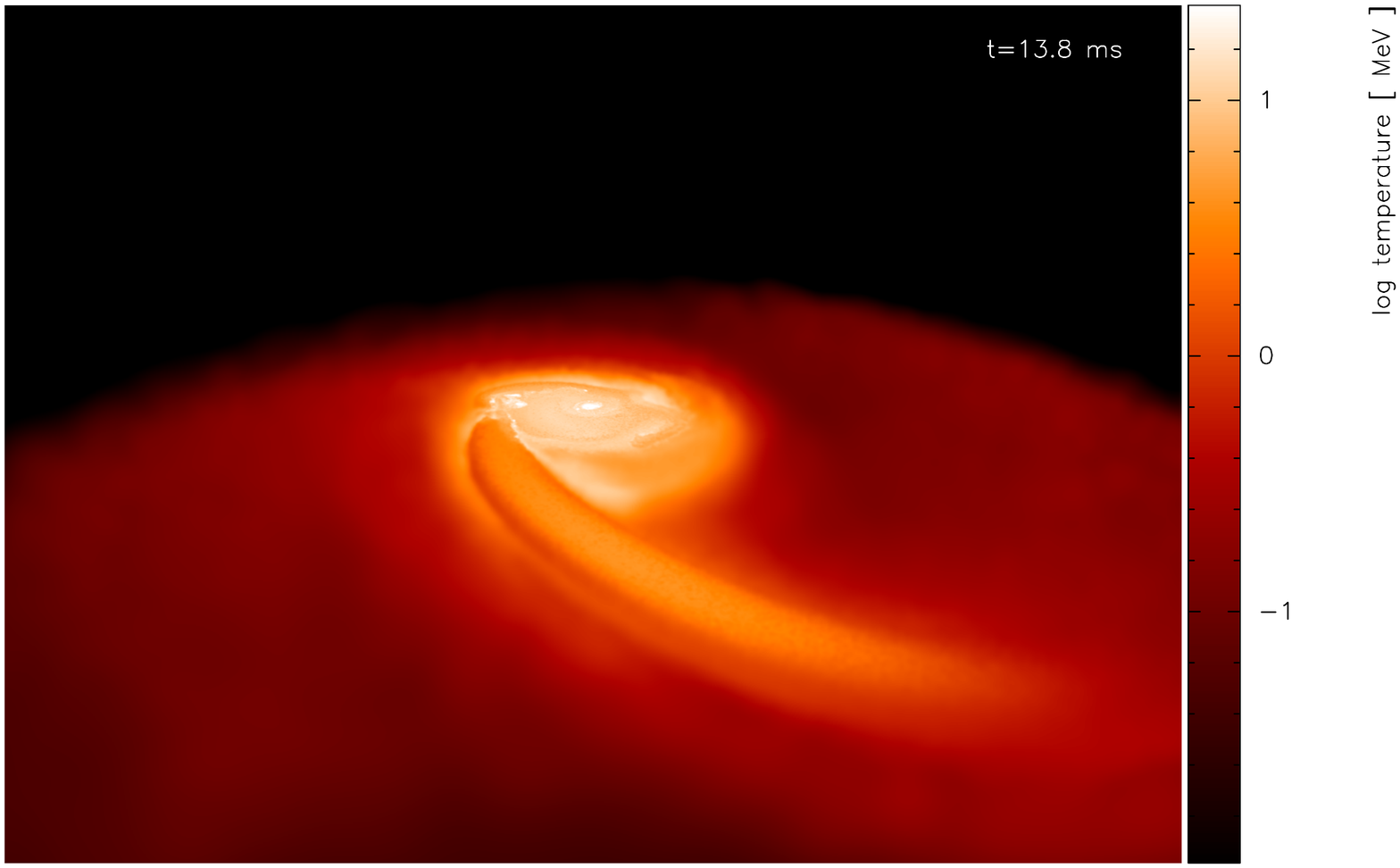}}   
   \caption{3D rendering of the temperature distribution during the grazing impact 
            of two neutron stars (1.3 and 1.4 \msun, $\beta= 1$; run A). It is only in the
            third close encounter (panel 5) that finally a single object forms. In each 
            close encounter a slew of Kelvin-Helmholtz vortices forms at the interface 
            between the stars. For display reasons only matter below the orbital plane is 
            shown and the colour bar has been restricted to values below 20 MeV.}
   \label{fig:nsns_coll_beta1}
\end{figure*}

To set the stage for later comparisons, we start with a brief description of the ``standard'' binary merger
case, run H. In Fig.~\ref{fig:temp_rendering_ns13_ns14_irrot} we show a 3D rendering of its temperature
distribution. We only display matter below the orbital plane so that the temperatures and flow structures inside the 
central remnant can be easily grasped. About one orbital period after contact two asymmetric spiral arms have 
formed (panel one and two), which evolve during the next $\sim 15$ milliseconds into a nearly axisymmetric torus (panel 4).
When the stars come into contact a shear interface forms between them. Such Kelvin-Helmholtz unstable
interfaces have long been known to emerge in neutron star mergers, see, for example, \cite{ruffert96,rosswog99,rasio99}. 
The resulting vortices have also been found to locally amplify pre-existing magnetic 
fields  \citep{price06,anderson08b,obergaulinger10} and inside of them the (SPH particle) temperatures
can temporarily reach values in excess of 60 MeV. The somewhat 
academic case of an initially tidally locked binary shows more pronounced tidal tails (due to larger angular momentum), 
but similar temperatures. Both double neutron star merger cases produce reasonably well-defined massive 
tori of 0.25 \Msun in the irrotational ``standard'' case and 0.30 \Msun for tidal locking, see Tab.~\ref{tab:mass}.\\
The collision cases in contrast  can suffer several close encounters before finally
merging into a single object and during these passages the neutron stars are efficiently 
tidally spun up. In the $\beta=1$ case a single object only forms after the third close encounter, 
see Fig.~\ref{fig:nsns_coll_beta1}. In the first, grazing impact the stars' obital energy is used
to spin up the stars to close to their breakup period, e.g. panel 2, and now they
form an eccentric tidal capture binary. The next encounter near $t= 8$ ms is more central and 
again produces strong Kelvin-Helmholtz vortices at the interface in which (SPH particle) temperatures 
locally exceed 80 MeV. The stars separate once more, with the 1.3 \Msun star now transferring mass 
in a direct impact phase into the primary, see panel 4. The final encounter occurs around 
$t\approx 12$ ms, again forming a string of Kelvin-Helmholtz vortices (panel 5) and finally shedding
mass from the secondary neutron star (panel 6). During the encounter the density never exceeds the initial 
value ($\approx 3.6 \times 10^{14}$ \gcc). \\ 
The more central encounters form a single object after two ($\beta=2$) 
and just one encounter ($\beta=5$). In both cases strong shocks form in which the (SPH particle) temperatures
reach values in excess of 80 MeV. In such shocks the neutron stars are substantially
compressed, to values of $\approx 4.52 \times 10^{14}$ \gcc ($\beta=2$) and $\approx 5.55 \times 10^{14}$ 
\gcc ($\beta=5$). In a superposition of rapid rotation and violent, stellar-radius amplitude oscillations, 
the central objects produce a multitude of interacting shocks in a string of mass shedding episodes, 
see Fig.~\ref{fig:nsns_coll_beta2_denscut}. The oscillations 
are also imprinted on the neutrino signal, see below.\\
\begin{figure*}
   \centerline{
   \includegraphics[width=15cm,angle=0]{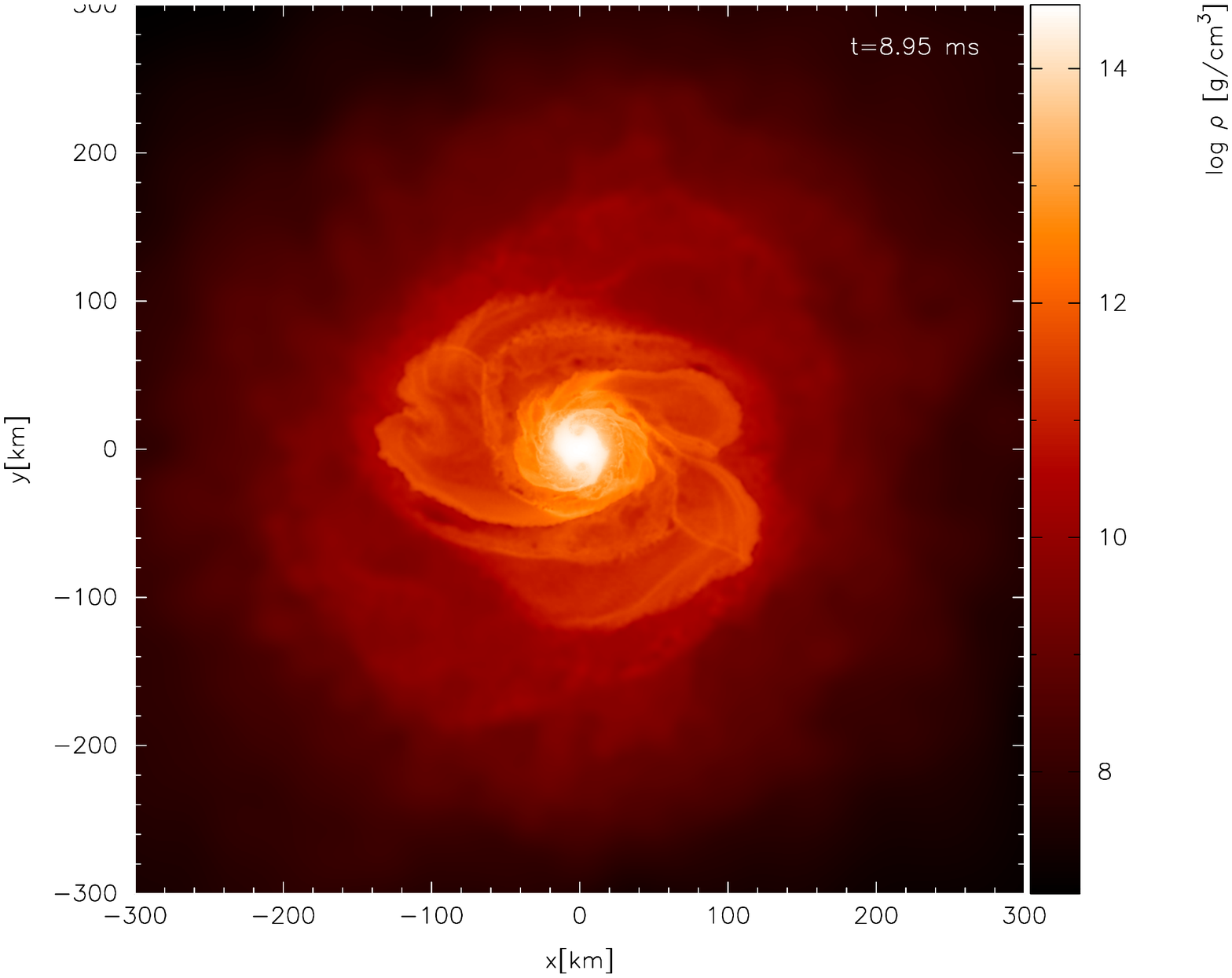}}
   \caption{Density cut through the orbital plane at the end of simulation run B (neutron stars
            with 1.3 and 1.4 \msun, $\beta=2$). The rotating and pulsating central object
            undergoes a sequence of mass shedding episodes, thereby producing various interacting
            shocks around the central remnant.}
   \label{fig:nsns_coll_beta2_denscut}
\end{figure*}
For the neutron star black hole cases we also begin with binary mergers as a reference point. Since 
they have been explored in detail before \citep{rosswog04b,rosswog05a,rosswog07b}, we restrict ourselves to a 
brief summary. In the case of a 1.4 \Msun ns and a 5 \Msun bh (run I) the neutron star starts 
transferring mass into the hole after 1.5 orbital periods. Consistent with our earlier studies 
this does not lead to the disruption of the neutron star on a dynamical time 
scale. Instead, self-gravity overcomes tidal forces again and the neutron star enters a long-lived
phase of episodic mass transfer during which it transfers mass periodically towards the hole while 
shedding mass through its outer Lagrange point\footnote{Phases of stable mass transfer are not  
restricted to the case of Newtonian gravity. A stiff equation of state \citep{rosswog04b}, small
mass ratios and large bh spin parameters make systems particularly prone to stable mass transfer, 
see \cite{shibata11} for a further discussion.}. 
This phase continues for as many as 25 orbital
revolutions before the neutron star is finally completely disrupted. The remnant at the end of the 
simulation ($t= 138.7$ ms) consists of a ``disk inside a disk'' with a mass of 0.16 \Msun for the inner,
high density disk ($\rho > 10^{11}$ \gcc, $r < 120$ km) and 0.22 \Msun if also the outer disk 
($\rho > 10^8$ \gcc, $r< 700$ km) is counted. 
The dynamics of the 1.4 \msun(ns) - 10 \msun(bh) system proceeds in a similar manner, here after 15 
orbital revolutions the neutron star is finally disrupted and leaves a $0.20$ \Msun disk together 
with a rapidly expanding one armed  spiral structure. All the numerically determined mass trasnfer
durations must be considered as robust lower limits on the true values \citep{dan11}.\\
For the {\em neutron star black hole collision} cases we only explore  the dependence on the black 
hole mass and keep the impact strength ($\beta=1$) and neutron star mass ($m_{\rm ns}= 1.3$ \msun) 
constant. During the first pericenter passage of the $m_{\rm bh}=3$ \Msun case, run D, the neutron 
star survives as a tidally spun up (close to break up, $P\approx 0.95$ ms) self-gravitating object, 
but sheds some of its mass in  a tidal tail. When the neutron star passes the black hole after 
about 5 ms for a second time another tidal tail is produced. Once more, the core of the neutron 
star survives  as a gravitationally bound object. It is only completely disrupted during the 
third and final pericenter passage at $t\approx 11$ ms. After 127 ms the remnant consists of the 
bh with 3.98 \msun, surrounded by a massive disk ($\approx 0.15$ \msun, see Tab.~\ref{tab:mass}) 
which is externally fed by three spiral arms. Qualitatively, the $m_{\rm bh}=5$ \Msun case evolves 
in a similar manner, see Fig.~\ref{fig:density_ns13_bh5}, but now the neutron star core survives 
even the third passage. At the time when we have to stop the simulation ($t= 144$ ms), the core is, 
according to its radial velocity, unbound from the black hole.  The neutron star core, however, 
is embedded into the debris gas and might therefore be further braked during its subsequent 
evolution so that it will possibly fall back towards the bh. In the $m_{\rm bh}=10$ \Msun 
case the neutron star is already completely disrupted during the second pericenter passage.
Consistent with the findings of \cite{lee10a}, all nsbh encounters have in common that they 
all leave behind a bh with a massive remnant disk (see Tab.~\ref{tab:mass}) and one tidal tail 
per close encounter.\\
The orbital dynamics is imprinted on the gravitational wave (GW) signal, for its calculation
see Sec.~\ref{sec:sim}. For the neutron star 
mergers (run G and H) the gravitational wave amplitudes $h_+$ (times the distance to the source 
$d$ as measured in code units of 1.5 km)  are shown in Fig.~\ref{fig:GW_amplitudes}, upper left 
panel. Both cases show the characteristic ``chirp'', up to $\approx 3$ ms for the non-spinning 
and up to $\approx 6$ ms for the tidally locked case, and the subsequent ``ringdown'' phase of 
the non-axisymmetric central object. The amplitudes of the ns$^2$ collisions are displayed in 
panel two of  Fig.~\ref{fig:GW_amplitudes}. Each close encounter produces a pronounced GW spike, 
for example in the $\beta=1$ case, run A, the encounters produce spikes at $t= 2.3, 8.4$ and 12.4 ms. 
The cases involving black holes show substantially longer activity after the first  encounter.
In the merger cases, run I and J, the episodic mass transfer is visible for tens of orbits until 
the neutron star is finally disrupted. The nsbh collision cases are essentially ``GW-quiescent''
(only a small contribution from the close-to-breakup rotation of the neutron star core) when 
the neutron star is receding from the bh, but produce another GW burst at the next close 
encounter. For the case with the 5 \Msun bh (run E) the longest encountered quiescent phase 
lasts as long as $\sim$ 60 ms.\\
The GW amplitude spikes produced by the close encounters coincide with peaks in the neutrino luminosities, 
see below. Although the GWs from collisions are comparable in amplitude to those from mergers, the large 
diversity and the lack of a ``standard waveform'' will make the detection of collision signals by 
current and future ground-based gravitational wave detectors extremely challenging.

\begin{figure*}
   \centerline{
   \includegraphics[width=10cm,angle=0]{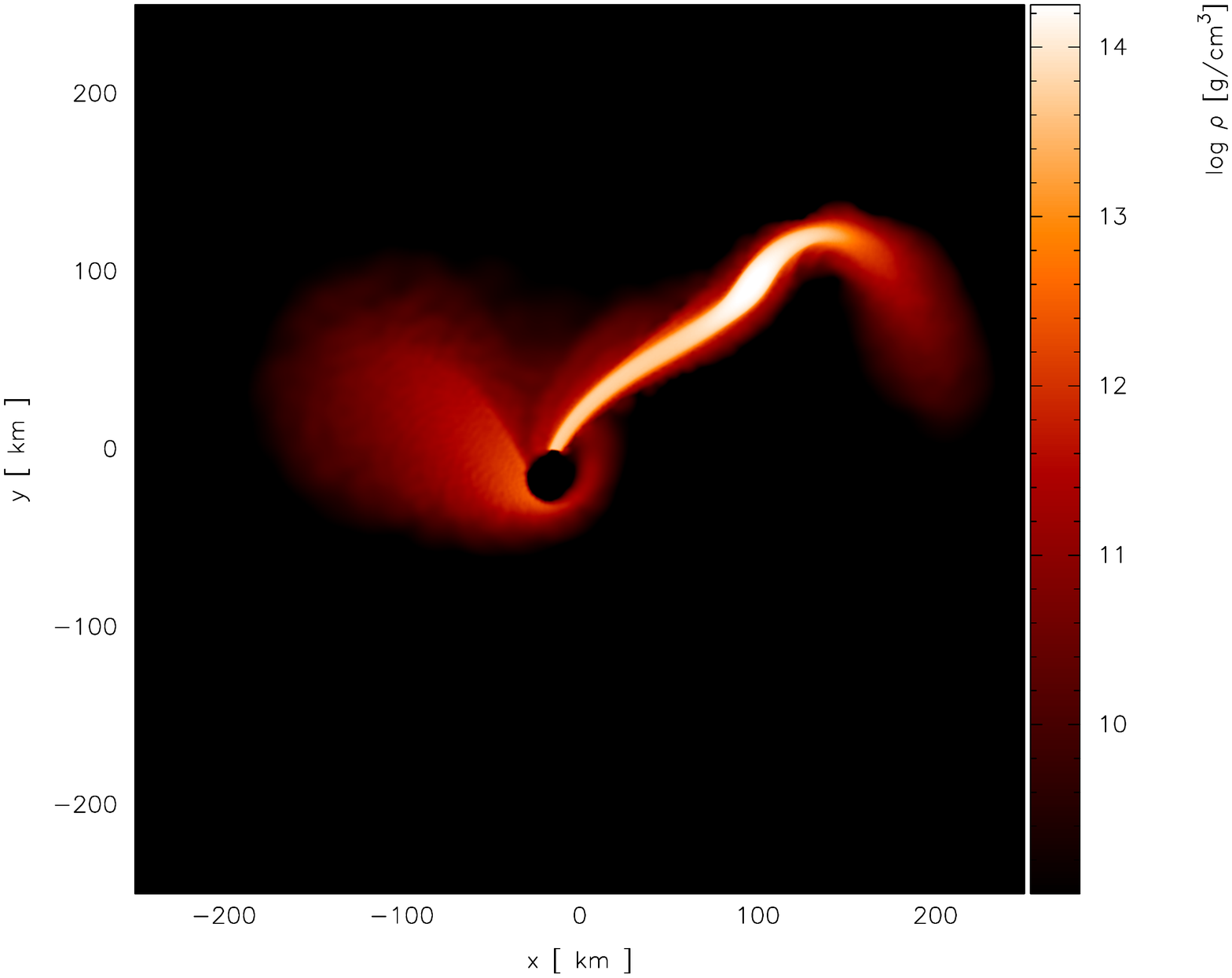}  \hspace*{-0.5cm}
    \includegraphics[width=10cm,angle=0]{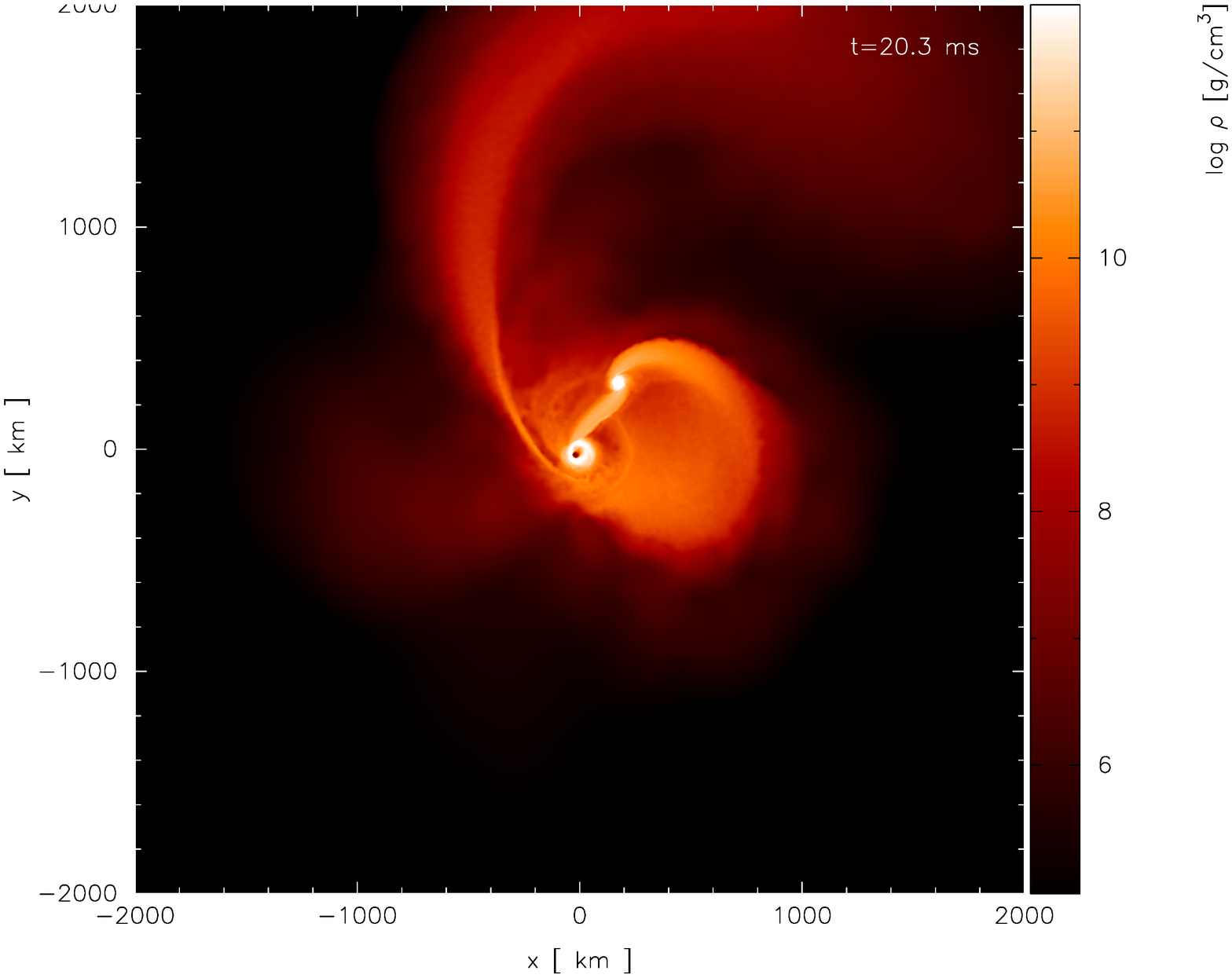} }   
   \vspace*{-0.5cm}
   \centerline{
   \includegraphics[width=10cm,angle=0]{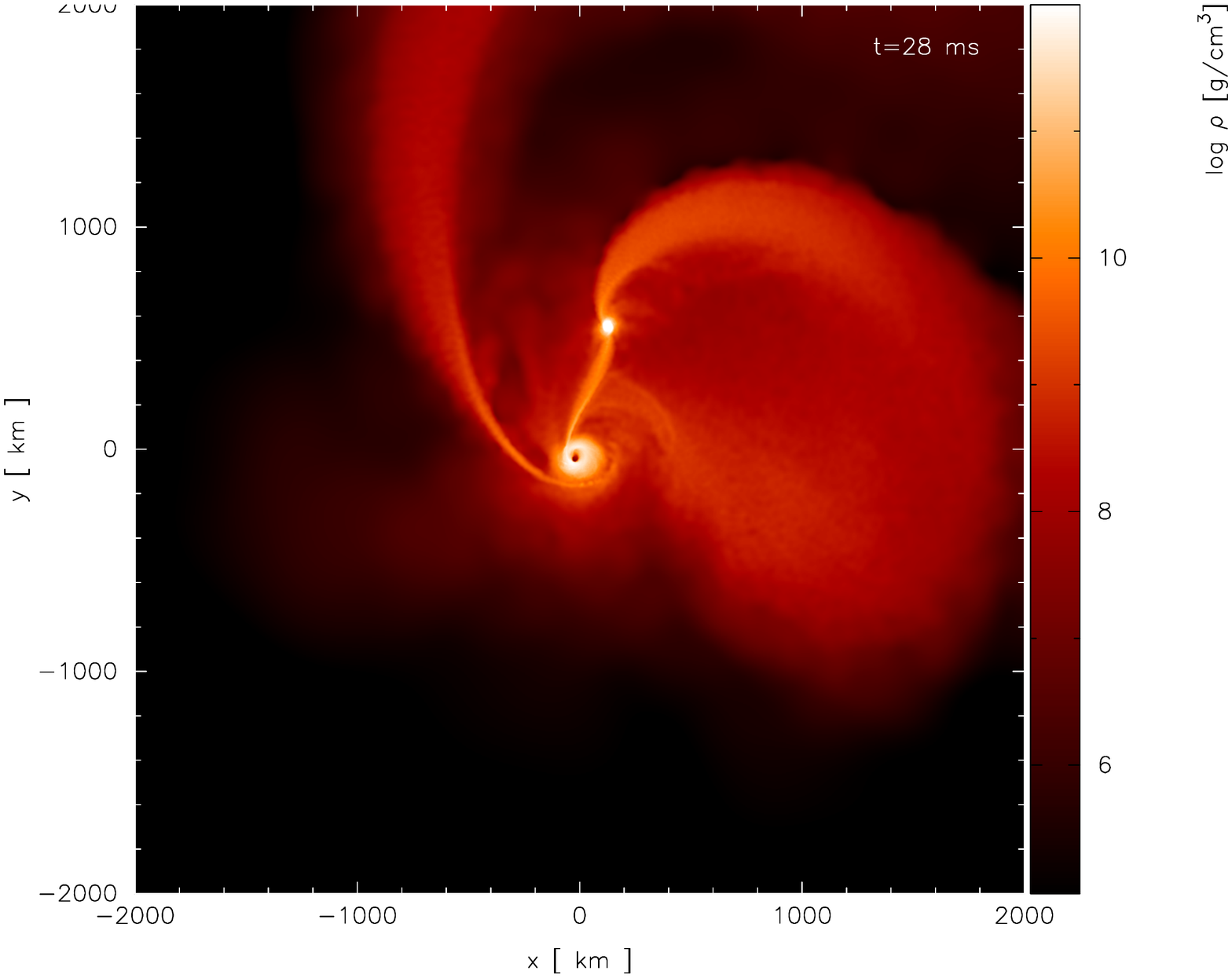} \hspace*{-0.5cm}
    \includegraphics[width=10cm,angle=0]{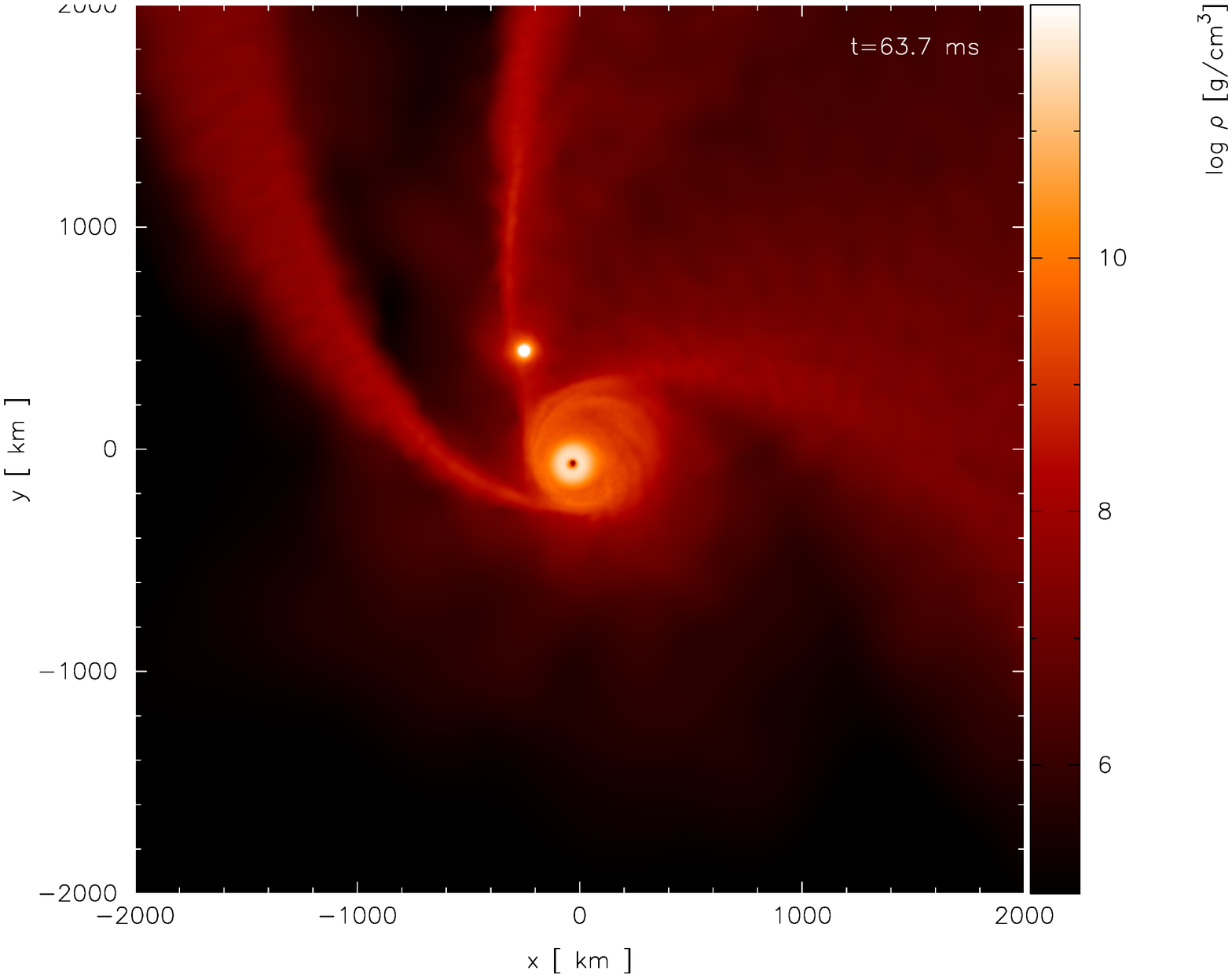} } 
   \vspace*{-0.5cm}  
    \centerline{
   \includegraphics[width=10cm,angle=0]{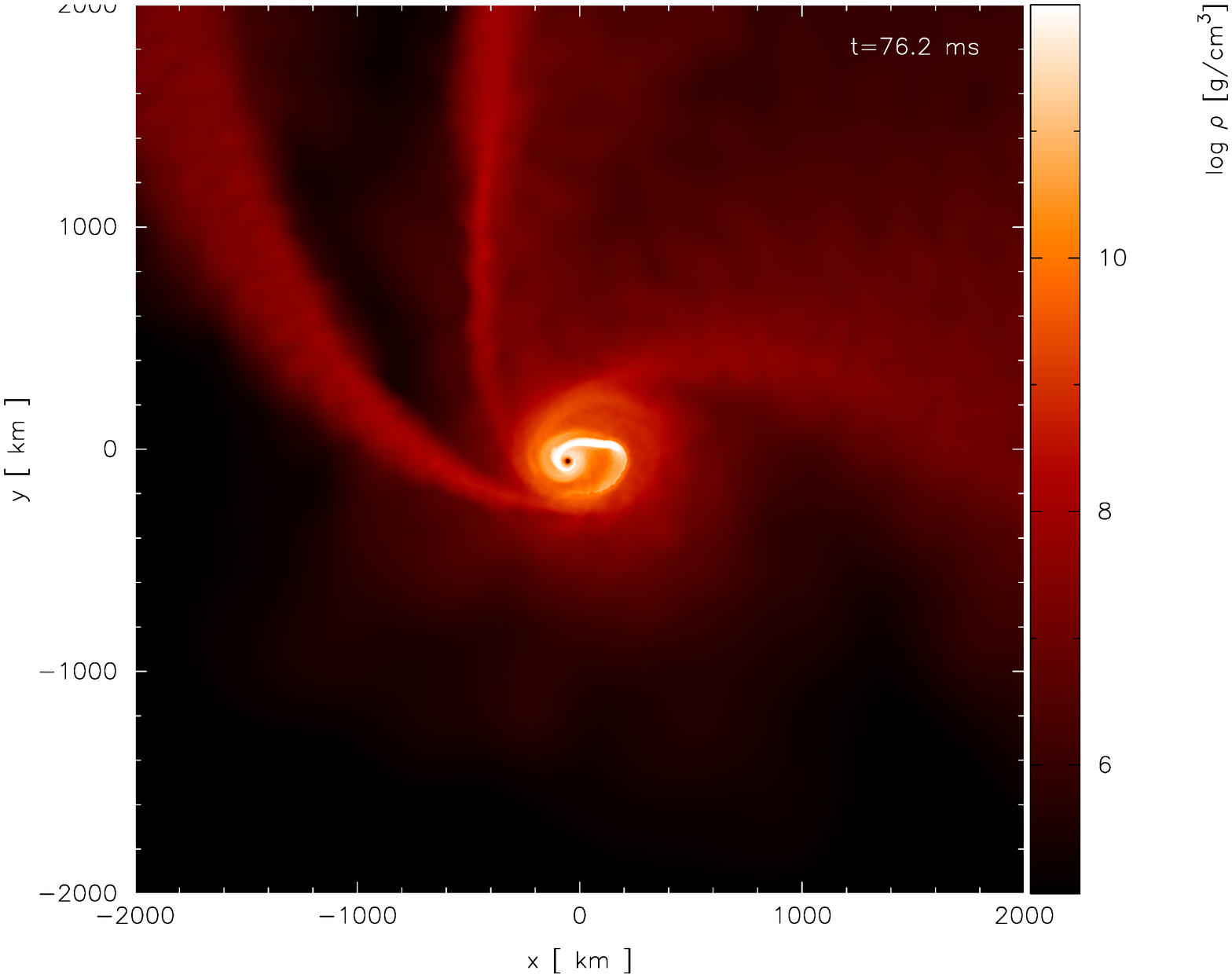}  \hspace*{-0.5cm}
    \includegraphics[width=10cm,angle=0]{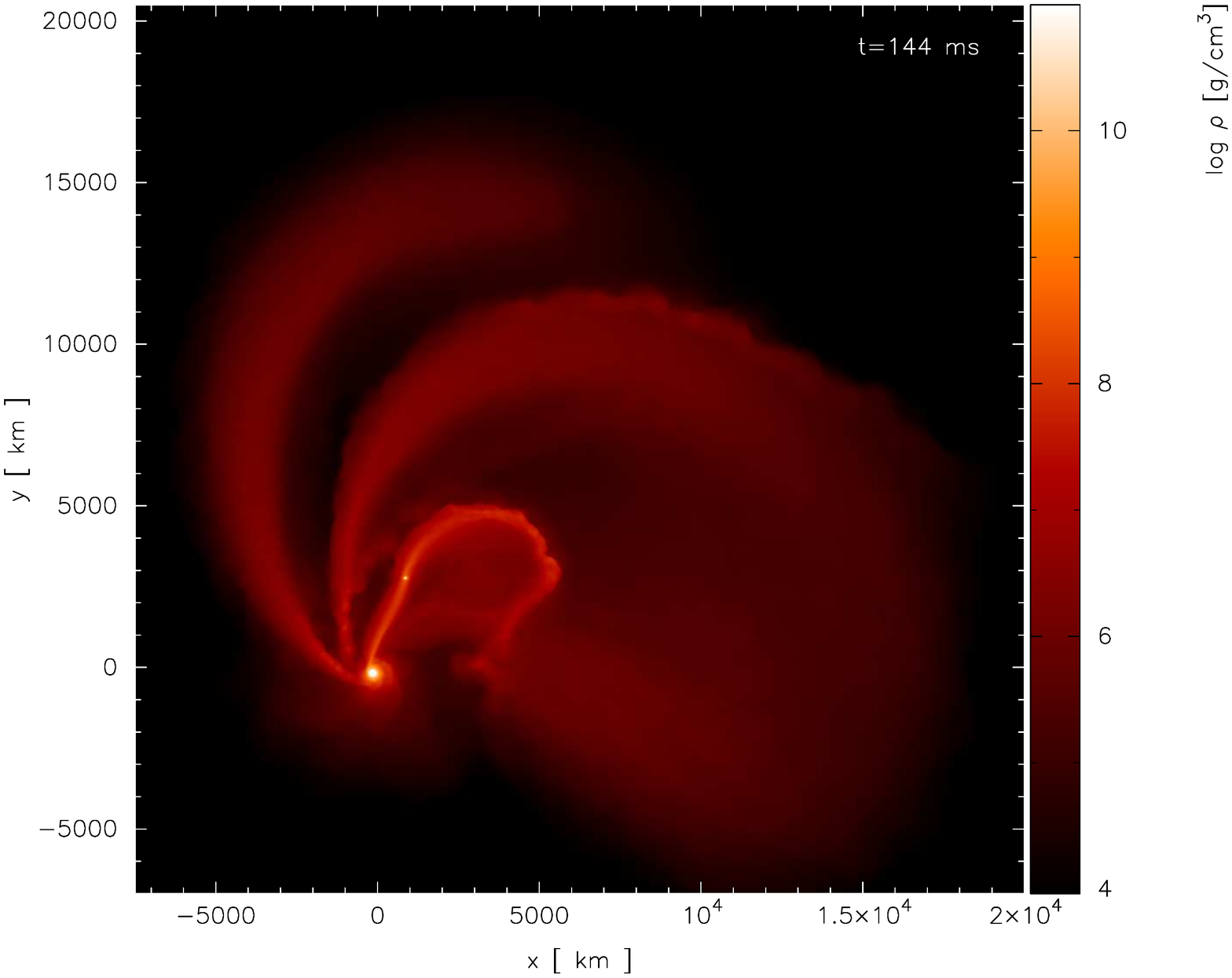} }   
   \caption{Density cut through orbital plane of run E. Panel one (numbering of 
    the panels is from left to right, from up to down)
   shows a snapshot just after the first, panel two after the second and panel five
   just after the third pericenter passage. Each pericenter passage produces a tidal
   tail. Note that the neutron star core survives even the third pericenter passage. At
   the end of the simulation it still has a mass of $\sim$ 0.1 \Msun and moves on 
   a close-to-parabolic orbit away from the black hole. Note that the scales are changing 
    between the different snapshots.}
   \label{fig:density_ns13_bh5}
\end{figure*}

\begin{figure*}
   \centerline{
   \includegraphics[width=10cm,angle=0]{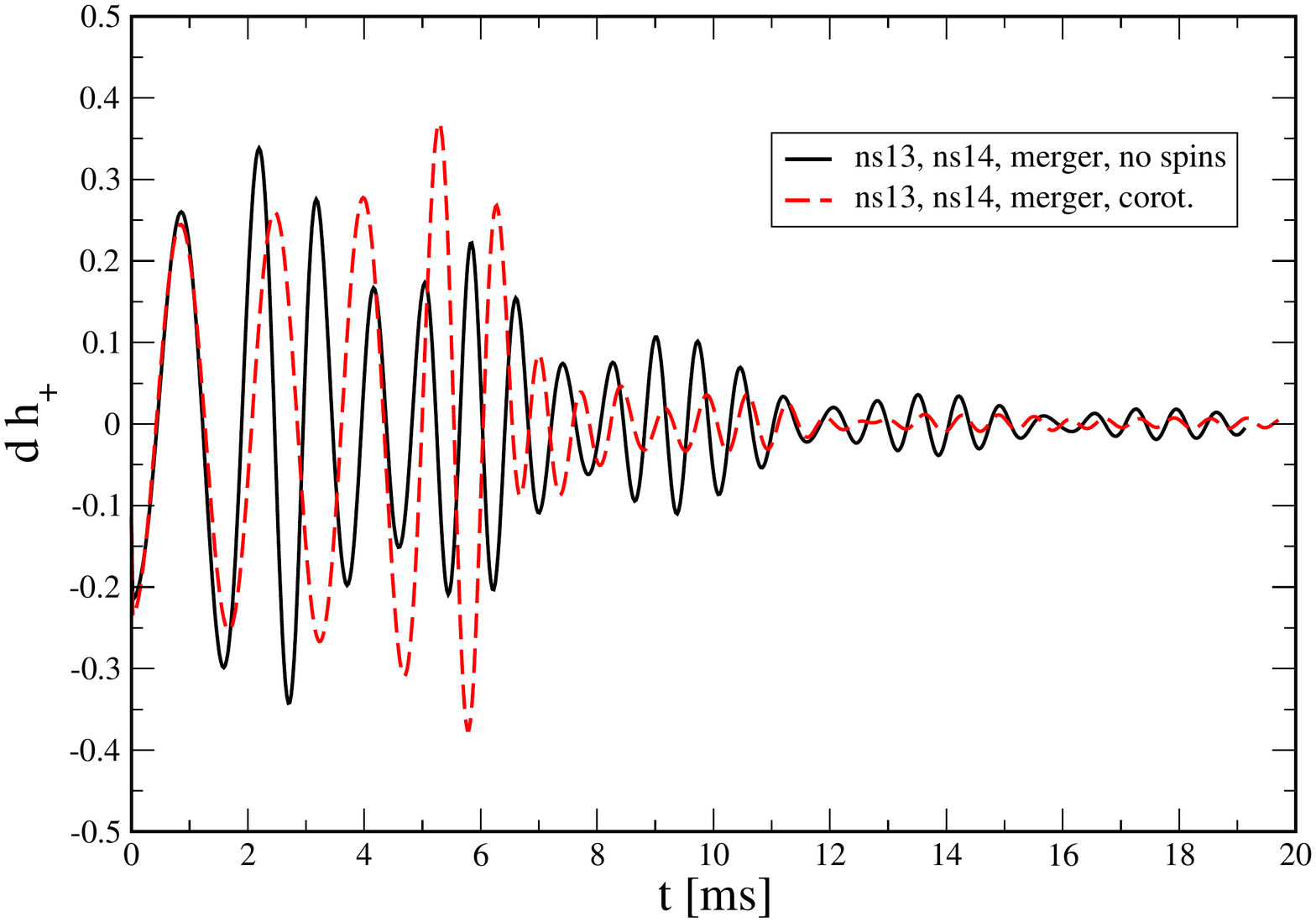} \hspace*{-1cm}
   \includegraphics[width=10cm,angle=0]{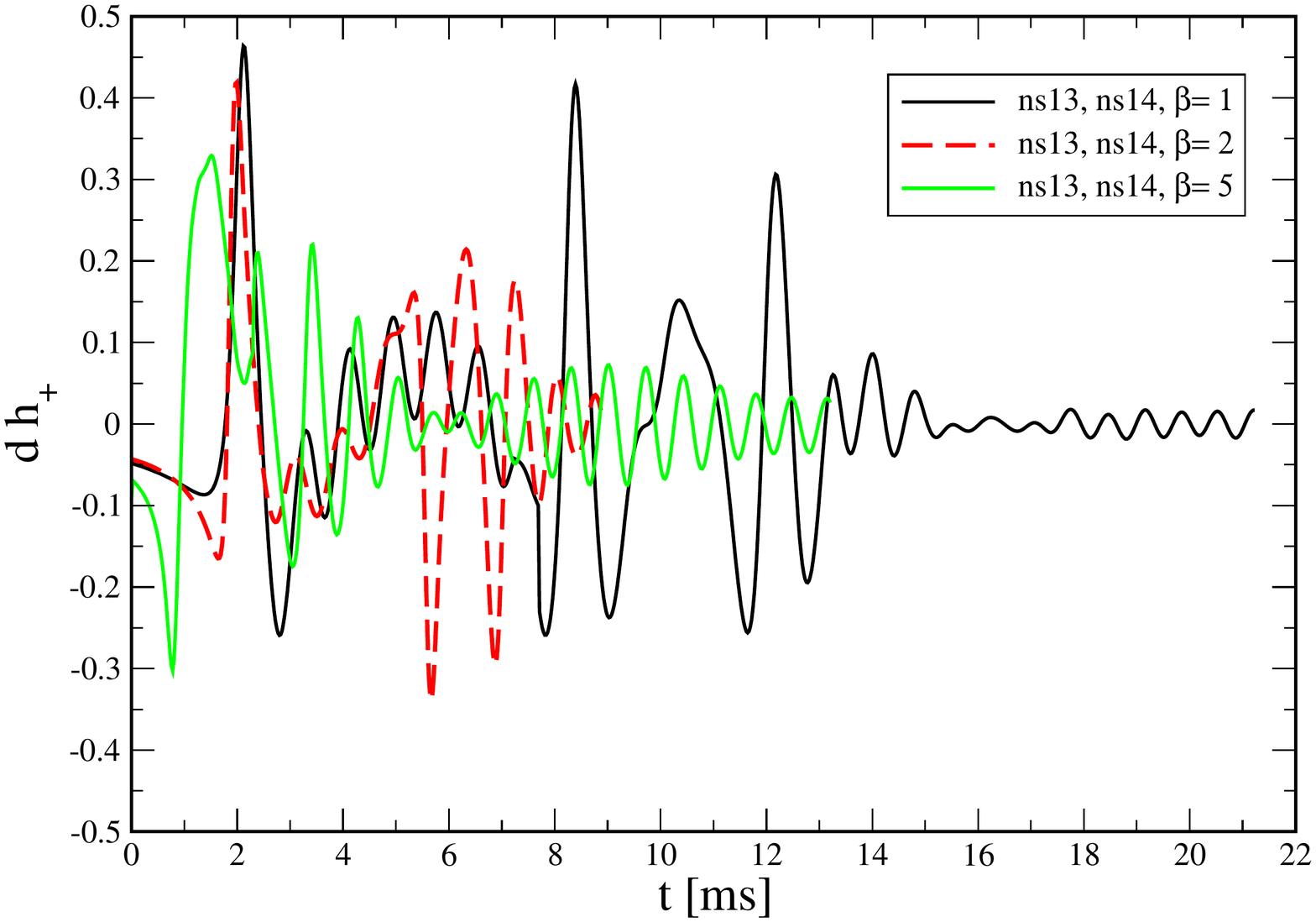}}
     \vspace*{-0.5cm}
   \centerline{
   \includegraphics[width=10cm,angle=0]{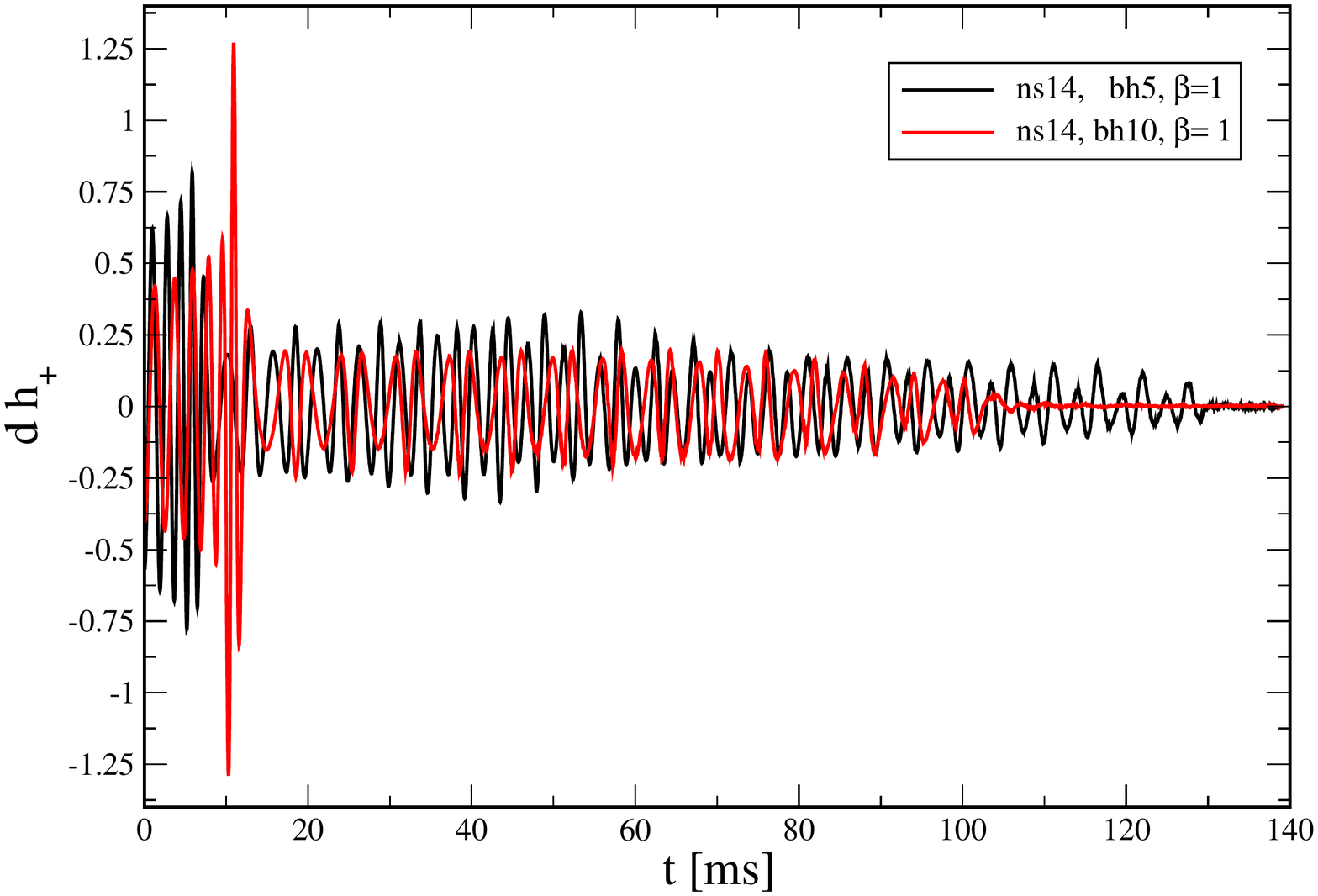} \hspace*{-1cm}
   \includegraphics[width=10cm,angle=0]{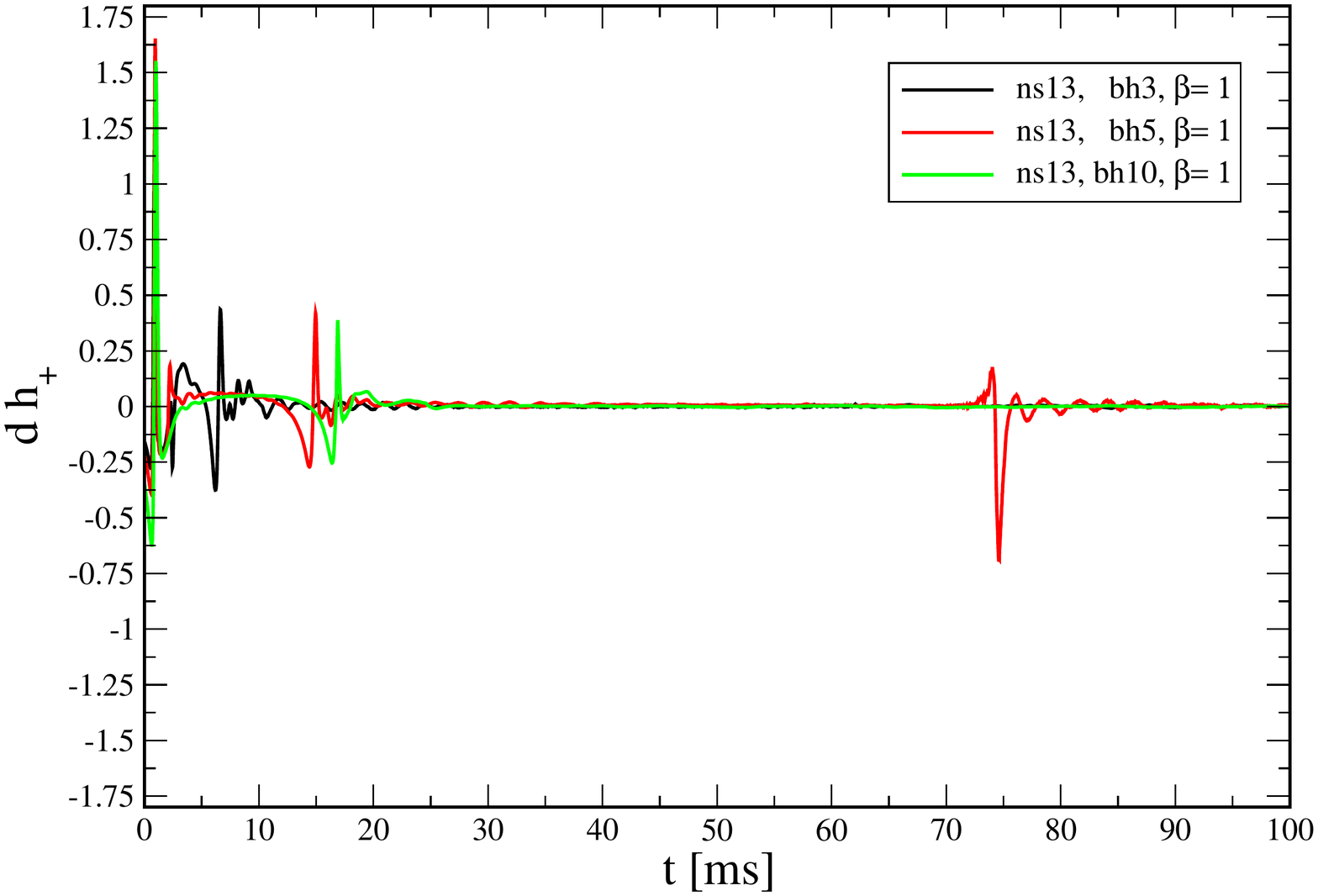}}    
   \caption{Gravitational wave amplitudes $h_+$ (times distance to source; code units = 1.5 km). 
            The upper two panels show neutron star neutron star encounters (mergers left, collisions right),
            the lower two panels show neutron star black hole encounters (mergers left, collisions right).}
   \label{fig:GW_amplitudes}
\end{figure*}

\begin{table*}
 \centering
 \begin{minipage}{140mm}
  \caption{Mass distribution: $t_{\rm ana}$ is the time at which we analyze the simulation data, $m_{\rm disk}$ 
           is the resulting disk mass, $m_{\rm fb}$ is the mass in fallback material, $m_{\rm esc}$ is the 
           dynamically ejected mass and $E_{\rm kin}$ the corresponding kinetic energy. Consistent with our 
           approach in Sect.~\ref{subsec:elmag} we capped the numerical velocities in the calculation of 
           $E_{\rm kin,esc}$ at 0.75 c.}
  \begin{tabular}{@{}crcccccl@{}}
  \hline
   Run   &  $t_{\rm ana}$ [ms] & $m_{\rm disk}$ [\msun] & $m_{\rm fb}$ [\msun] &  $m_{\rm esc}$ [\msun] & $E_{\rm kin}[10^{51} {\rm erg}]$ & $\langle v_{\rm esc}\rangle$ [c]& comment \\
\hline 
\\
        \underline{Collisions} \\
   A     &  21.2  & 0.27   & 0.10 & 0.060 &1.15 & 0.13&ns-ns\\
   B     &   9.0  & 0.40   & 0.10 & 0.009 &0.97 & 0.22&ns-ns\\
   C     &  13.2  & 0.32   & 0.03 & 0.030 &3.61 & 0.28&ns-ns\\
   D     & 127.5  & 0.24   & 0.11 & 0.142 &5.70 & 0.19&ns-bh\\
   E     & 143.6  & 0.14   & 0.04 & 0.172 &10.68& 0.24&ns-bh\\
   F     & 540.3  & 0.05   & 0.04 & 0.134 &8.73 & 0.24&ns-bh\\
\\
      \underline{Mergers} \\
   G     &  20.3  & 0.30      & 0.06 & 0.050 &1.15& 0.15&ns-ns, corot.\\
   H     &  19.1  & 0.25      & 0.04 & 0.014 &0.23& 0.12&ns-ns, no spins\\
   I     &  138.7 & 0.16/0.22 & 0.04 & 0.024 &0.61& 0.15& ``disk in disk'', inner/both disks\\
   J     &  139.3 & 0.21      & 0.03 & 0.049 &1.82& 0.18&ns-bh, no spins\\
\hline
\label{tab:mass}
\end{tabular}
\end{minipage}
\end{table*}

\subsection[]{Neutrino emission}
\label{subsec:nu}
\begin{table*}
 \centering
 \begin{minipage}{140mm}
  \caption{Neutrino emission. $L_{\nu}$ refers to the luminosity of all flavors, the average energies, $\langle E \rangle$, are measured in MeV
           at time $t_{\rm ana}$, where they are to a very good approximation constant.}
  \begin{tabular}{@{}crrrrrrl@{}}
  \hline
   Run   &  $t_{\rm ana}$ [ms] & $L_{\nu, {\rm peak}}$ & $L_{\nu}(t_{\rm ana})$  &  $\langle E_{\nu_e}\rangle$  &  $\langle E_{\bar{\nu}_e}\rangle$  &  $\langle E_{\nu_x}\rangle$  & comment \\
\hline 
\\
          \underline{Collisions} \\
   A     & 16.5 & $6.0 \times 10^{53}$  & $4.3 \times 10^{53}$ & 10.5  & 15.5  & 23.5 & ns-ns\\
   B     &  9.0 & $3.0 \times 10^{54}$  & $1.3 \times 10^{54}$ & 10.1  & 16.6  & 21.3 & ns-ns\\
   C     & 13.2 & $4.1 \times 10^{54}$  & $1.4 \times 10^{54}$ & 11.0  & 16.1  & 20.8 & ns-ns\\
   D     &127.5 & $3.1 \times 10^{53}$  & $8.1 \times 10^{52}$ &  7.1  & 10.1  & 21.4 & ns-bh\\
   E     &122.9 & $5.3 \times 10^{53}$  & $5.7 \times 10^{52}$ &  6.9  &  8.5  & 14.6 & ns-bh\\
   F     &540.4 & $1.0 \times 10^{53}$  & $5.6 \times 10^{51}$ &  9.0  & 10.6  &  8.9 & ns-bh\\
\\
 \underline{Mergers} \\
   G     & 17.5 & $1.8 \times 10^{53}$  & $1.3 \times 10^{53}$ & 9.3  & 15.5  & 27.2 & ns-ns, corot.\\
   H     & 16.0 & $1.3 \times 10^{53}$  & $9.9\times 10^{52}$  & 8.0  & 14.4  & 26.3 & ns-ns, no spins\\
   I     & 138.7  & $8.9 \times 10^{52}$& $5.7 \times 10^{52}$ & 6.5  & 10.0  & 13.8 & ns-bh, no spins\\
   J     & 138.7 & $1.0 \times 10^{53}$ & $6.5 \times 10^{52}$ & 6.3  & 11.0  & 14.9 & ns-bh, no spins\\
\hline
\label{tab:nu_props}
\end{tabular}
\end{minipage}
\end{table*}
\begin{figure}
   \includegraphics[width=9cm,angle=0]{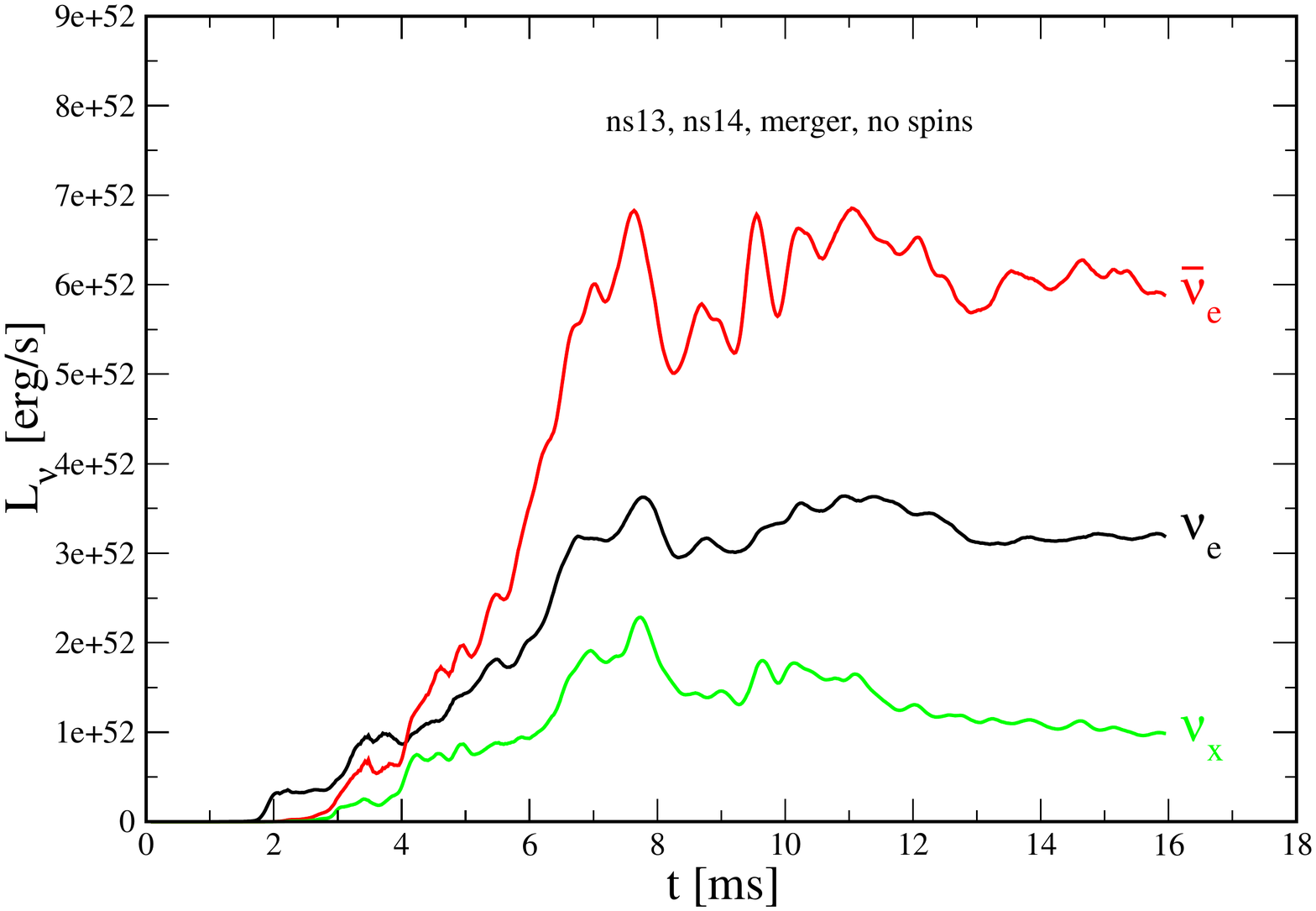}
   \includegraphics[width=9cm,angle=0]{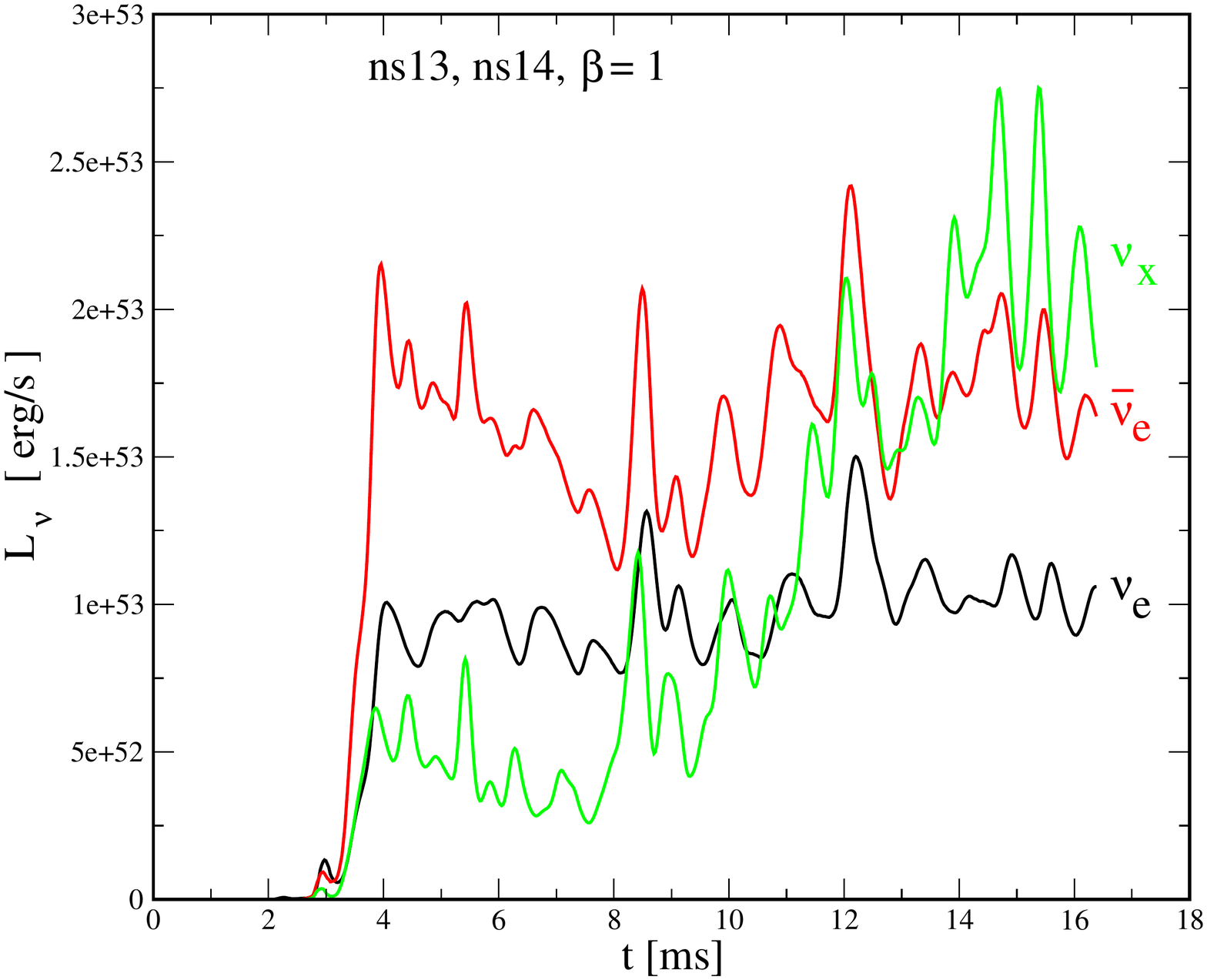}
   \includegraphics[width=9cm]{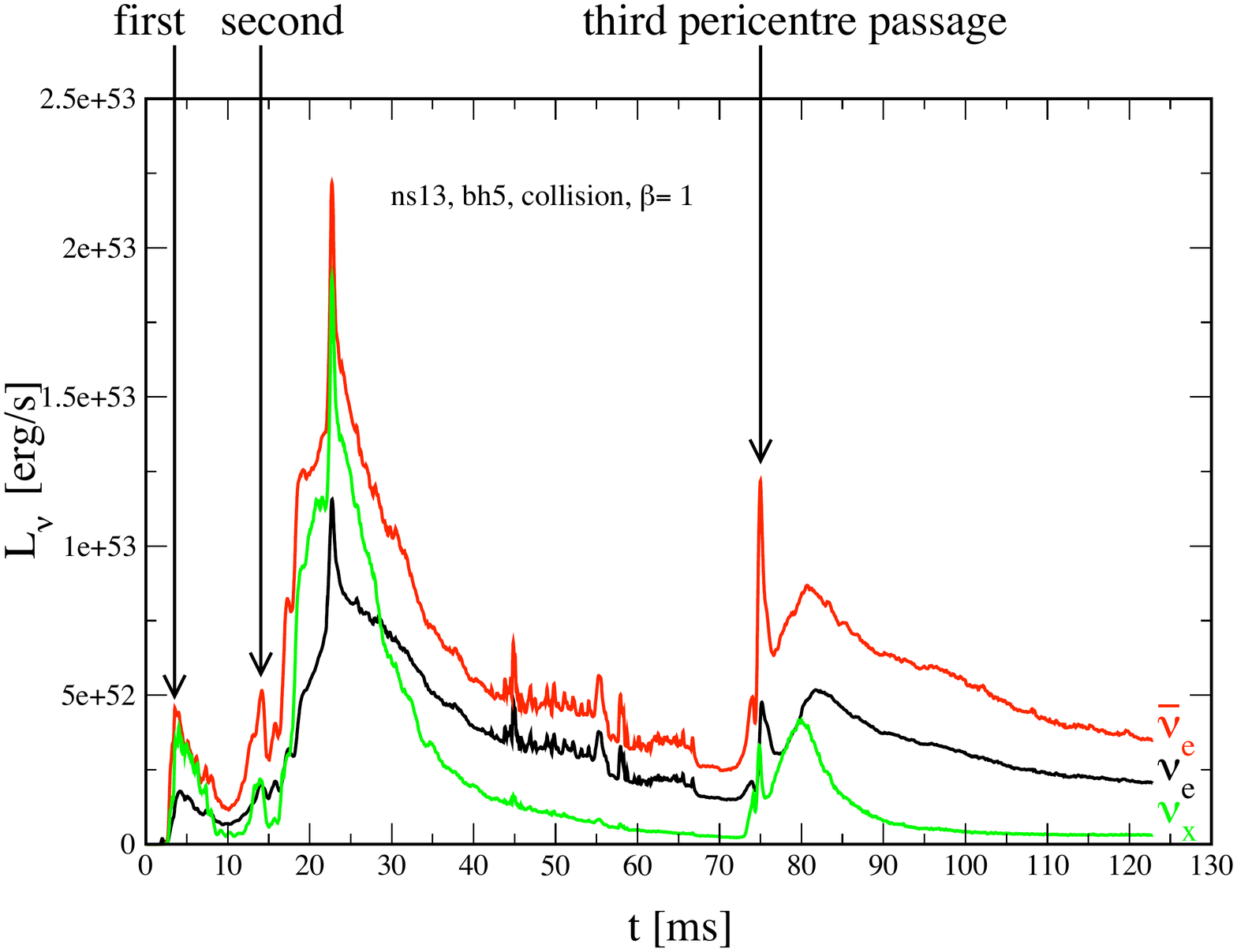}
   \caption{Examples of the emerging neutrino luminosities: the ``standard'' 
            merger case (run H, upper panel), the ns$^2$ collision with $\beta=1$ 
             (run A, middle), and the nsbh collision with $\beta=1$ and $m_{\rm bh}= 5$ \Msun 
             (run E, bottom).}
   \label{fig:neutrinos}
\end{figure}

We discuss the neutrino properties of one example of each encounter class 
(ns-ns: mergers and collisions; nsbh: mergers and collisions) in some detail, an overview 
over the properties of all cases is provided in Tab.~\ref{tab:nu_props}.
Again, we use the non-spinning neutron star merger case, run H, as a reference point
to gauge the other results. Here, the luminosities increase smoothly and peak about 
6 ms after contact (at $t\approx 7.7$ ms) with a total of $1.3 \times 10^{53}$ erg/s,
see upper panel of Fig.~\ref{fig:neutrinos}. 
The tidally locked case, run G, produces similar results, but due to the larger disk
mass, see Tab.~\ref{tab:mass}, slightly higher luminosities.
Since the debris is extremely neutron-rich, the neutrino luminosities are dominated by 
electron anti-neutrinos, followed in importance by electron neutrinos and 
heavy lepton neutrinos (collectively referred to as ``$\nu_X$''), consistent with
earlier findings \citep{ruffert97a,rosswog03a}. In a recent 2D neutrino-hydrodynamics 
calculation starting from the matter distribution that resulted from a 3D simulation
\citep{price06} \cite{dessart09} found that our leakage scheme
underestimates the heavy lepton neutrino emission, since it does not 
account for the nucleon-nucleon bremsstrahlung process. Therefore our heavy lepton neutrino 
emission results are robust lower limits on the true luminosities. In all cases apart from
run F we find the following hierarchy in the mean neutrino energies: $\langle E_{\nu_X}\rangle > 
\langle E_{\bar{\nu}_e}\rangle > \langle E_{\nu_e}\rangle$. The heavy lepton neutrinos are 
predominantly produced in hot, very dense regions and in the exceptional case, run F, 
the densities have already dropped substantially below nuclear matter density 
($\rho_{\rm max} < 10^{10}$ \gcc) at the end of the simulation when the mean neutrino energies 
(as given in Tab.~\ref{tab:nu_props}) are measured.\\
Even the most gentle ns$^2$ collision with $\beta=1$, run A, produces a neutrino luminosity that
is approximately three times larger than the standard double neutron star merger case (run H). 
The merger neutrino lightcurves are rather smooth, the collision cases, in contrast,
show a much larger variability with luminosity changes of up to a factor of two on
the dynamical time scale of the central object, $\approx$ 1 ms, see Fig.~\ref{fig:neutrinos}
middle panel. The more central collisions, run B and C, produce neutrino luminosities of about an order of 
magnitude larger than the reference case, see Tab.~\ref{tab:nu_props}.\\
As discussed earlier, we find long-lived, episodic mass transfer in the investigated nsbh cases.
During this phase the neutrino emission is moderate, but once the neutron star is finally disrupted
after dozens of orbital revolutions it reaches peak values of up to $10^{53}$ erg/s.
All investigated nsbh collisions robustly produce hot and massive
accretion disks with neutrino peak luminosities of at least 75\% of the standard merger case.
We display the neutrino luminosities for the case shown in Fig.~\ref{fig:density_ns13_bh5},
run E, in the last panel of Fig.~\ref{fig:neutrinos}. All pericenter passages are visible as a
peak in the neutrino luminosities. Passages two and three thereby substantially increase the disk mass,
the subsequent disk consumption each time produces an enhanced neutrino luminosity.
In the case of run E, the enhanced emission due to a pericenter passage is therefore double peaked 
with a first, short peak resulting from the neutron star remnant impact onto the disk (between panels 
four and five in Fig.~\ref{fig:density_ns13_bh5}), and a second, broader peak driven by the viscous 
consumption of the re-filled accretion disk. In summary, each pericenter passage produces both a
gravitational wave and at least one neutrino emission peak.

\subsection{Fallback and external feeding of accretion disks}
\label{subsec:fallback}
\begin{figure}
   \includegraphics[width=9cm,angle=0]{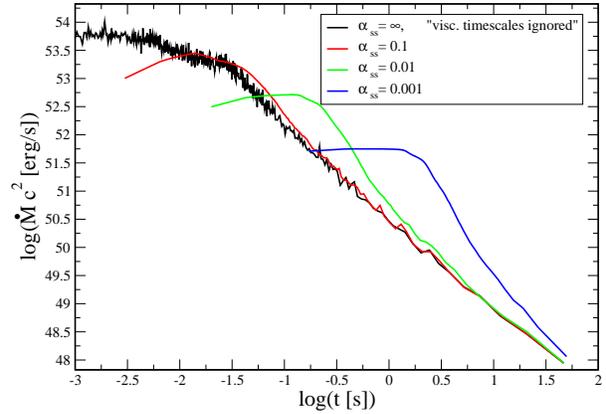}  
   \caption{The dependence of the matter energy accretion rate on the
            Shakura-Sunyaev viscosity parameter $\alpha_{\rm ss}$ is 
            illustrated for the case of a 
            merger between a ns and a 5 \Msun bh (run I). 
            Increasing the viscosity time scale (= reducing 
            the parameter $\alpha_{\rm ss}$) smoothly spreads out 
            the accretion to longer time scales.}
   \label{fig:fallback_luminosity_alpha}
\end{figure}

Swift observations have revealed that both short and long bursts can
show late time X-ray flaring activity \citep{burrows05,nousek06}.
Some SWIFT sGRBs are accompanied by a phase of
extended X-ray emission that lasts between $\sim$ 10 and 100 s 
\citep{norris08,perley09} and whose fluence 
can exceed that of the GRB itself. Such time scales are substantially
longer than the dynamical/viscous time scales, i.e. $\sim$ 1 ms/0.05 s,
that are expected for a merger remnant. In this context, it was pointed out 
that compact binary mergers possess in addition a much longer 
time scale due to nearly unbound material that will finally 
fall back to the central remnant \citep{rosswog07a,lee07,faber06b}. The
mass in the fallback of a merger however is substantially lower than
that of the accretion that is supposed to launch the GRB. Therefore it is
at least not obvious how fallback would produce extended emission with a larger
fluence than the sGRB.\\
Clearly, the accretion of the fallback material is a rather complicated phenomenon
and much of what will finally become visible as electromagnetic radiation 
will depend on the evolution of the inner remnant disk as it interacts with
the fallback material being delivered at super-Eddington rates. It has 
been argued \citep{rossi09} that --rather than being finally swallowed by
the likely emerging central black hole-- the fallback material could form 
an extended hot envelope around the remnant once neutrino cooling ceases to 
be an efficient cooling agent, but the matter is still opaque to photons. 
After about one week a soft X-ray signal near the Eddington luminosity should
become visible.\\
In addition, while the neutron star debris is initially receding from the central 
remnant it will be subject to heating from radioactivity 
\citep{li98,rosswog05a,metzger10a,roberts11} and it has been proposed that
this radioactive energy release might regulate the final fallback rate \citep{metzger10a}. 
At the same time, viscous effects drain the disk via accretion onto the central object 
and the disk expands due to outward angular momentum transport.  Winds are 
expected as a result of viscous heating and/or neutrino heating \citep{lee05b,dessart09}
which further complicates the situation. At late times, recombinations 
of nucleons into $\alpha$-particles are expected to unbind a substantial fraction 
of the late-time disk \citep{beloborodov08,lee09,metzger09b}.\\
Clearly, this interesting, but complicated evolution deserves further detailed studies,
but reliable predictions of the electromagnetic display require a reliable knowledge
of the properties in the emerging photosphere.
The treatment of these effects is beyond the scope of the current 
work and we restrict ourselves here to a simple semi-analytical model to 
estimate the time scales on which fallback material is delivered to central
object while ignoring for now additional complicating effects such emerging winds.\\ 
We assume that for 
most of its trajectory, the fallback material is reasonably well 
described as being ballistic. Under this assumption we can calculate 
quantities like eccentricity, circularization radii etc. as described 
in detail in \cite{rosswog07a} and we can in particular calculate the 
time, $t_{\rm fb}(R_{\rm circ})$, it takes each matter portion to return 
to its circularization radius, $R_{\rm circ}$. Under the simplifying assumption
that matter quickly settles into a disk,  
we assume that from this point of time, $t_{\rm fb,a}$, each fluid parcel (SPH particle) 
$a$ is accreted at a constant rate $\dot{m}_a= m_a/t_{\rm visc,a}$ on the viscous 
time scale of a thick accretion disk (with $H \sim R$), 
$t_{\rm visc,a}= \left(\alpha_{\rm ss} \Omega_K(R_{\rm circ,a})\right)^{-1}$,
where $\alpha_{\rm ss}$ is the Shakura-Sunyaev viscosity parameter 
\citep{shakura73}. The total mass accretion rate is then given as a sum 
over particle contributions
\be
\dot{M}(t)= \sum_a \dot{m}_a(t),   \\
\ee
where
\be
    \dot{m}_a(t)=\left\{\begin{array}{ll} \frac{m_a}{\tau_{\rm visc,a}}, & 
           t_{\rm fb,a} < t \le t_{\rm fb,a} +\tau_{\rm visc,a}  \\
           0, & {\rm else} \end{array}\right..
\ee
This simple model extends to original one \citep{rosswog07a} by accounting also
for the viscous dissipation time scale. Admittedly, this model is 
very simple and the complicated but important topic fallback accretion 
in the aftermath of a compact binary merger deserves more efforts
in future work.\\
We illustrate the effect of the viscosity parameter $\alpha_{\rm ss}$ in
Fig.~\ref{fig:fallback_luminosity_alpha} for the case of a merger 
between a ns and a 5 \Msun bh (run I). We show the resulting curves 
$\dot{M} c^2(t)$ for cases where either viscous time scales are 
ignored (i.e. $\alpha_{\rm ss}= \infty$) or varied in a plausible
range  ($\alpha_{\rm ss}= 0.1$, 0.01 and 0.001).\\
\begin{figure*}
   \centerline{
     \includegraphics[width=10cm,angle=0]{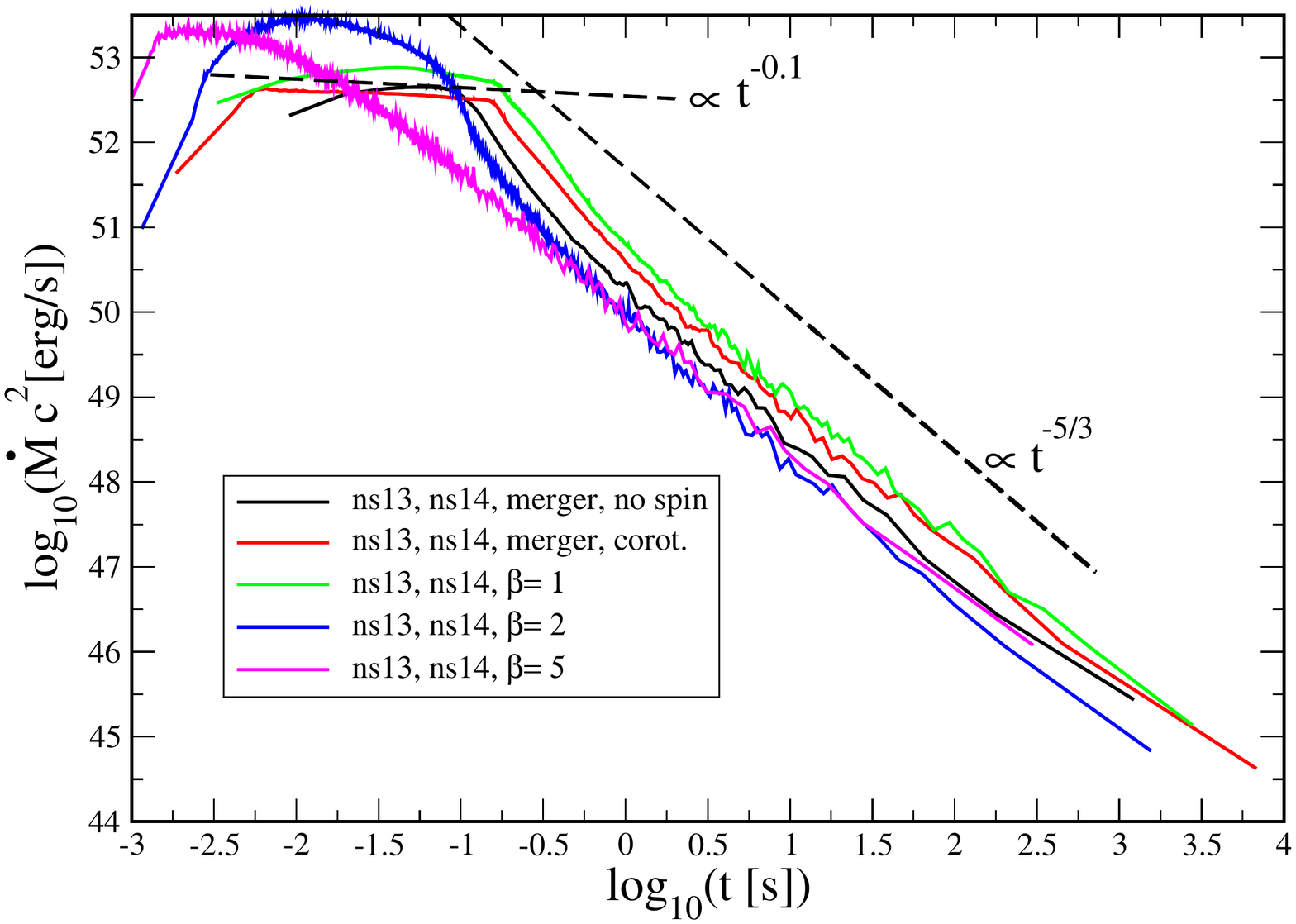} \hspace*{-1cm}
     \includegraphics[width=10cm,angle=0]{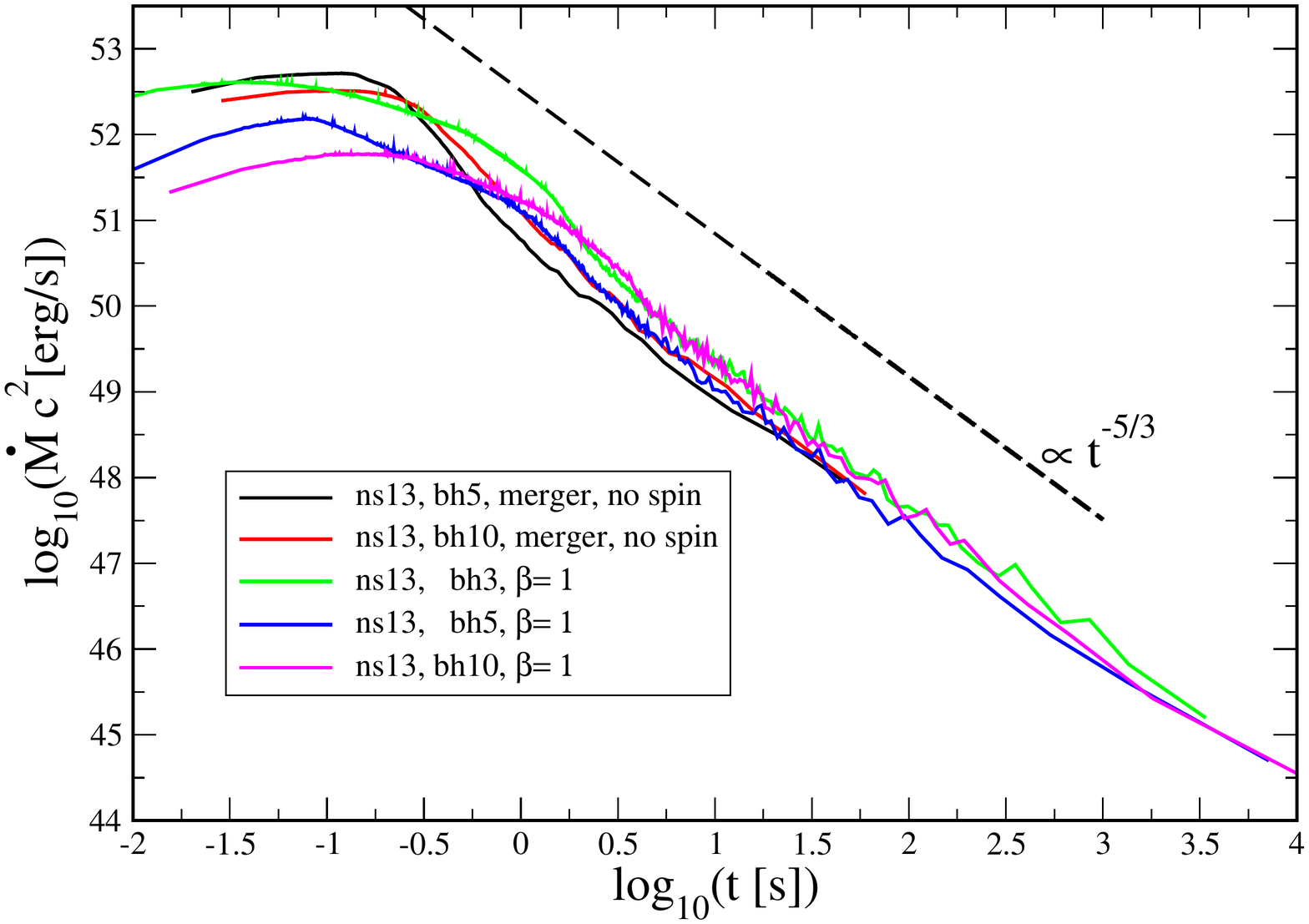}}
   \caption{$\dot{M} c^2$ from fallback material as predicted from the simple 
            model (ballistic fallback + disk accretion) described in the main text. 
            Left: neutron star - neutron star 
            cases, right: neutron star - black hole cases. In each panel both merger and collision
            cases are shown. For comparison we also show the power law for "standard" fallback 
            \citep{rees88,phinney89} that assumes a constant mass per energy, $dm/de=$ const,
                   and results in a time 
                   dependence $\propto t^{-5/3}$. For the figures a viscosity value of 
                      $\alpha_{\rm SS}= 0.01$ has been adopted, see text for details.}
   \label{fig:fallback_luminosity}
\end{figure*}
In Fig.~\ref{fig:fallback_luminosity} we show the values of $\dot{M} c^2$ from 
fallback material as predicted from our  model. For all shown curves
a value of $\alpha_{SS}= 0.01$ was adopted which is realistic, but maybe slightly 
on the low side. In our standard merger case,
run H, the fallback luminosity peaks around 0.1 s with $\sim 3 \times 10^{52}$ erg/s.
The somewhat academic corotating case, run G, shows a similar peak fallback rate
which is nearly constant for the first $\sim$ 0.15 s. The $\beta=1$ ns$^2$ collision case, run A, 
produces a roughly constant fallback rate until 0.15 s, before it enters the 5/3-powerlaw phase
that is expected for $dM/de\approx$ const \citep{rees88,phinney89}. 
During this phase the fallback luminosity is about a factor of four larger than in
the standard neutron star merger case. The other ns$^2$ collision cases peak earlier and 
their fallback rates are lower than for canonical ns$^2$ mergers. The early phases of the fallback
curves should generally be interpreted with a grain of salt since they
are sensitive to the exact location of the disk radius and the assumption of ballistic
motion is not necessarily well justified. At later times ($t > 0.2$ s) all cases are fairly 
well described by the $\dot{M} \propto t^{-5/3}$ powerlaw which is shown 
in the figure as a dashed line.\\
The nsbh cases are shown in the right panel. The mergers, run I and J, are qualitatively
similar to the ns$^2$ merger cases, after a ``plateau phase'' up to $\sim$ 0.3 s, the fallback 
luminosity drops off compatible with $\dot{M} \propto t^{-5/3}$ (the lack of late time 
data points for nsbh mergers is due to the substantially lower numerical resolution for 
these cases). The nsbh collisions show less clear initial plateau phases and only settle
into the 5/3-powerlaw phase after a few seconds.\\
The collision cases yield typically about a factor of two more fallback material
than the mergers, see Tab.~\ref{tab:mass}. This mass is located in tidal tails 
(one per pericenter passage) which can be considered as external mass reservoirs that keep 
replenishing the central accretion disks at their fallback rates. With an
available energy of $\sim 2 \times 10^{52} {\rm erg} \; (\epsilon/0.1) \; (m_{\rm fb}/0.1$ \msun)
the fallback material is in principle able to produce energetic post-merger activity and,
as evident from Fig.~\ref{fig:fallback_luminosity}, the fallback time scales are orders of
magnitude longer than the dynamical time scales. But as discussed above, the real situation 
is possibly too complicated to be adequately captured by this simple fallback + disk model. 
Therefore it remains an open question whether fallback is the trigger of the late activity that 
is observed in some sGRBs or not.

\subsection{Ejecta}
\label{subsec:ejecta}
\begin{figure*}
   \centering
   \centerline{\includegraphics[width=10cm,angle=0]{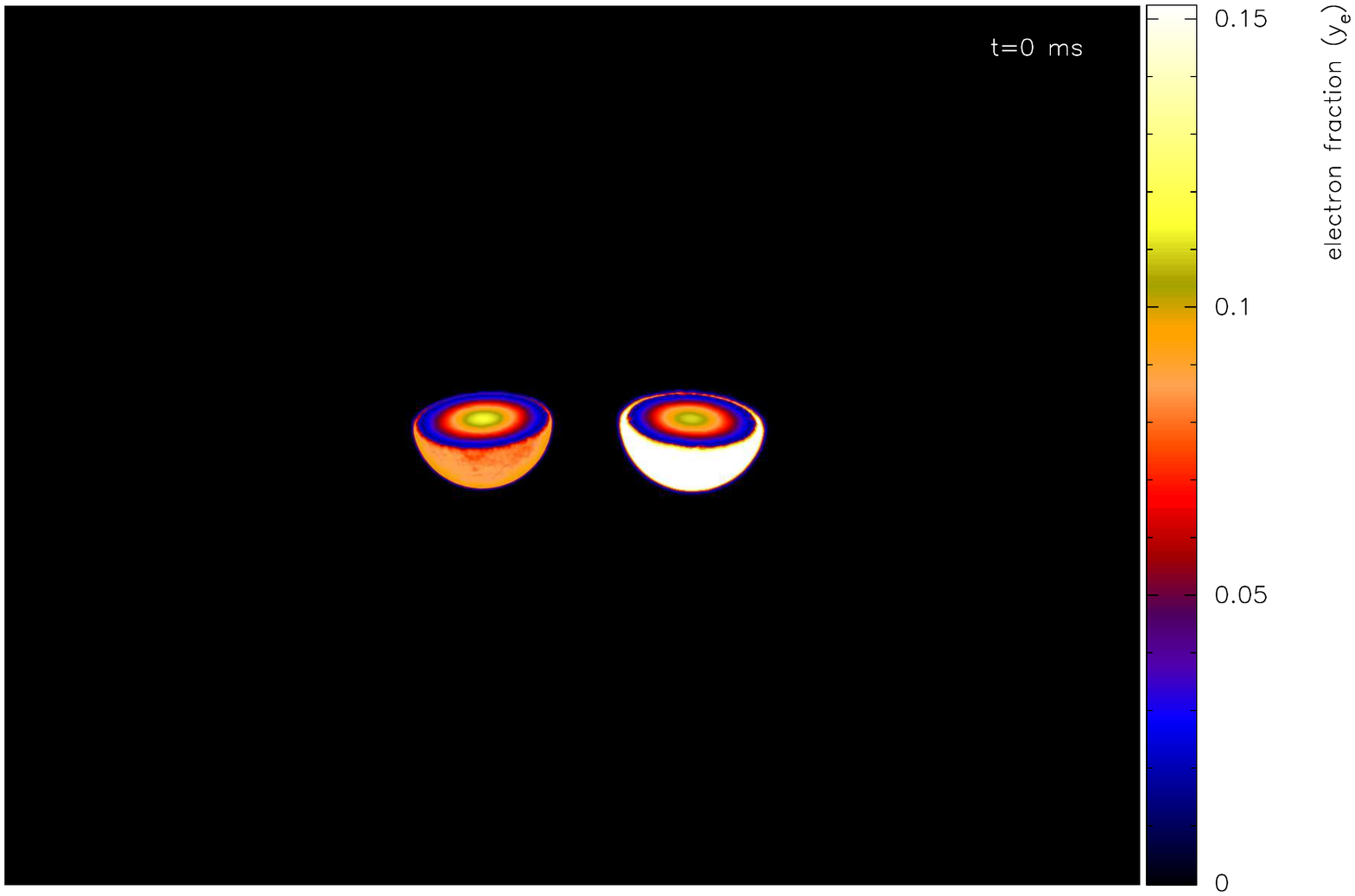}  \hspace*{-0.5cm}
               \includegraphics[width=10cm,angle=0]{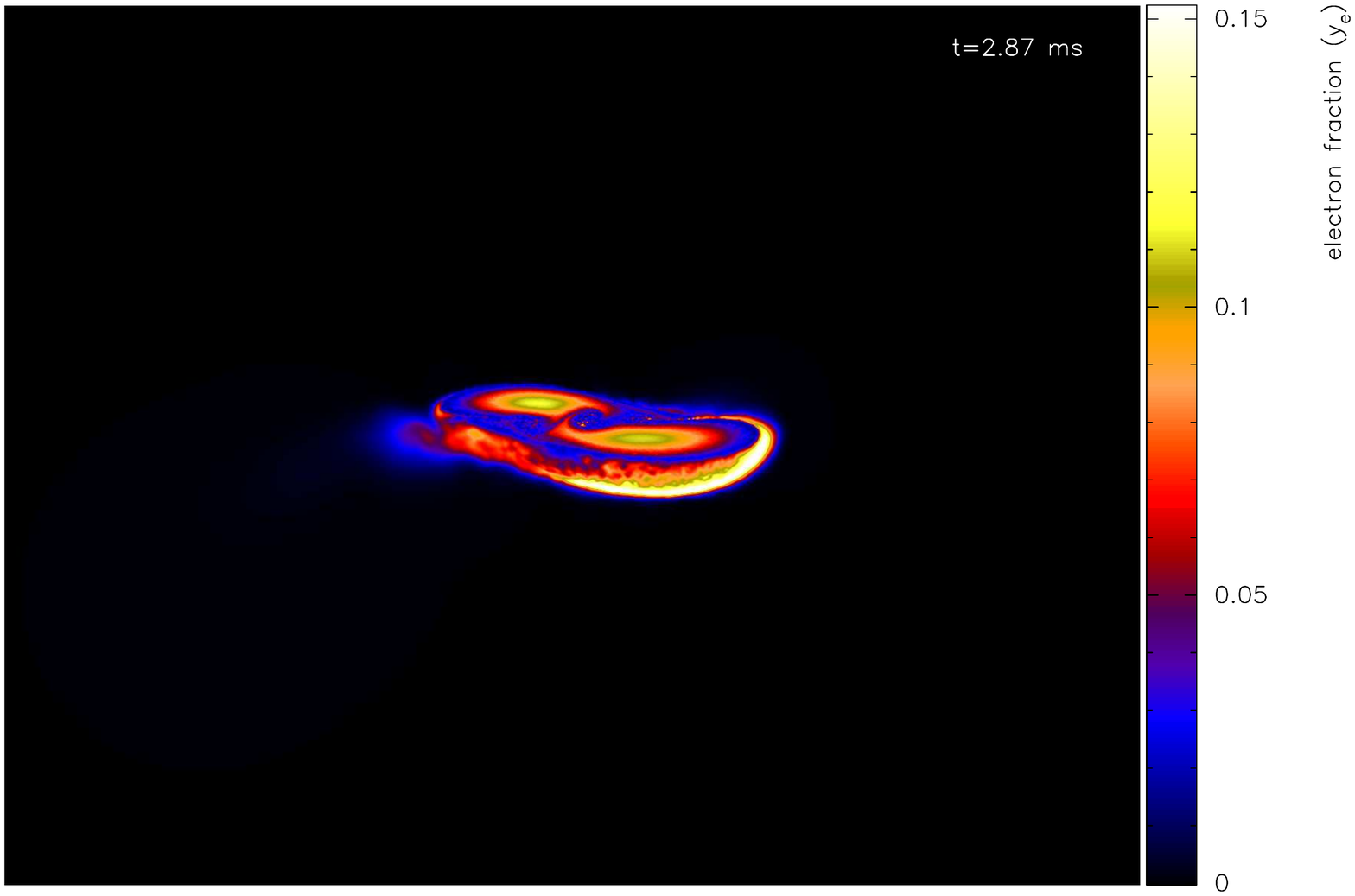}} 
   \vspace*{-0.5cm}
   \centerline{\includegraphics[width=10cm,angle=0]{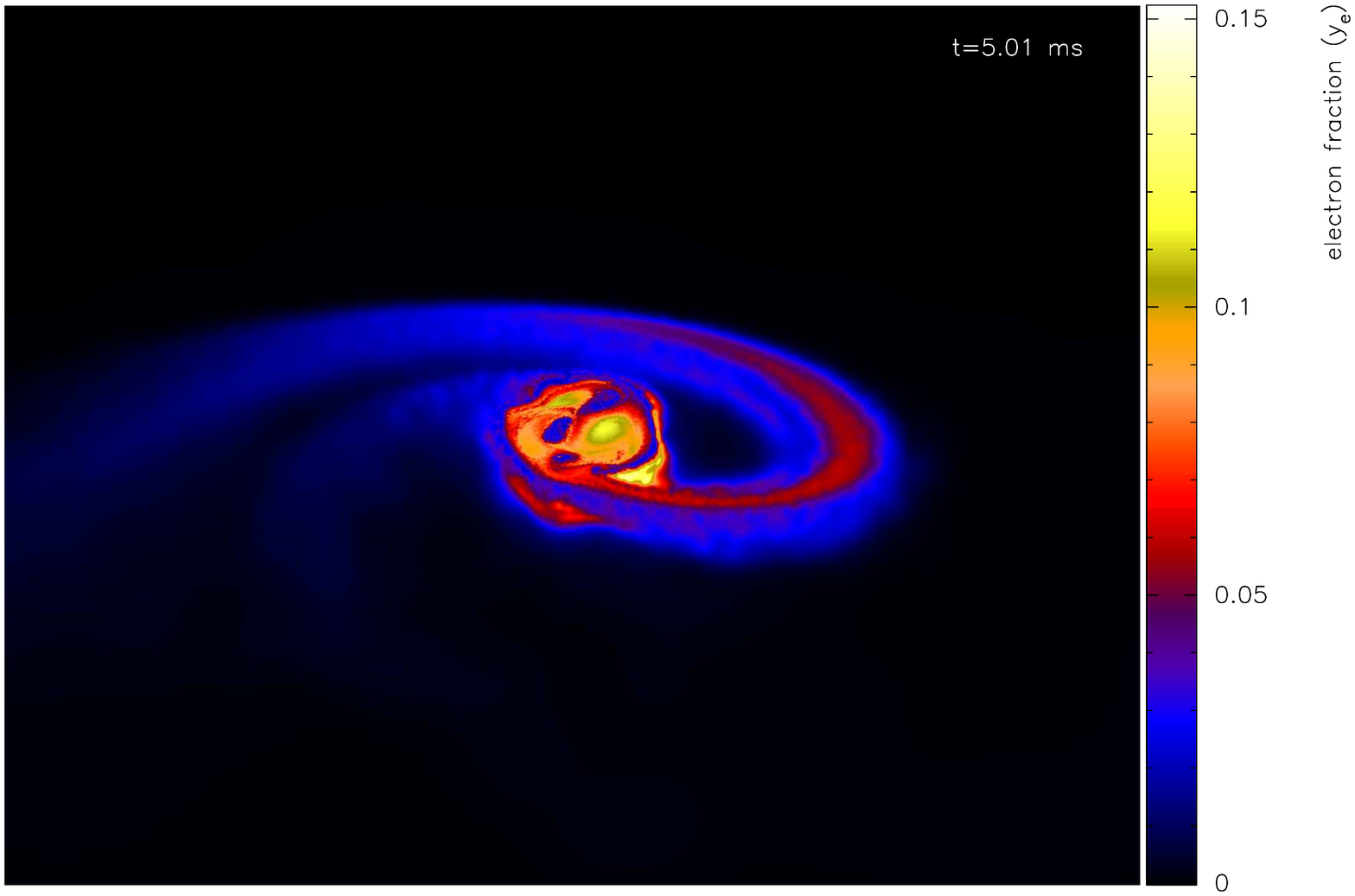} \hspace*{-0.5cm}
               \includegraphics[width=10cm,angle=0]{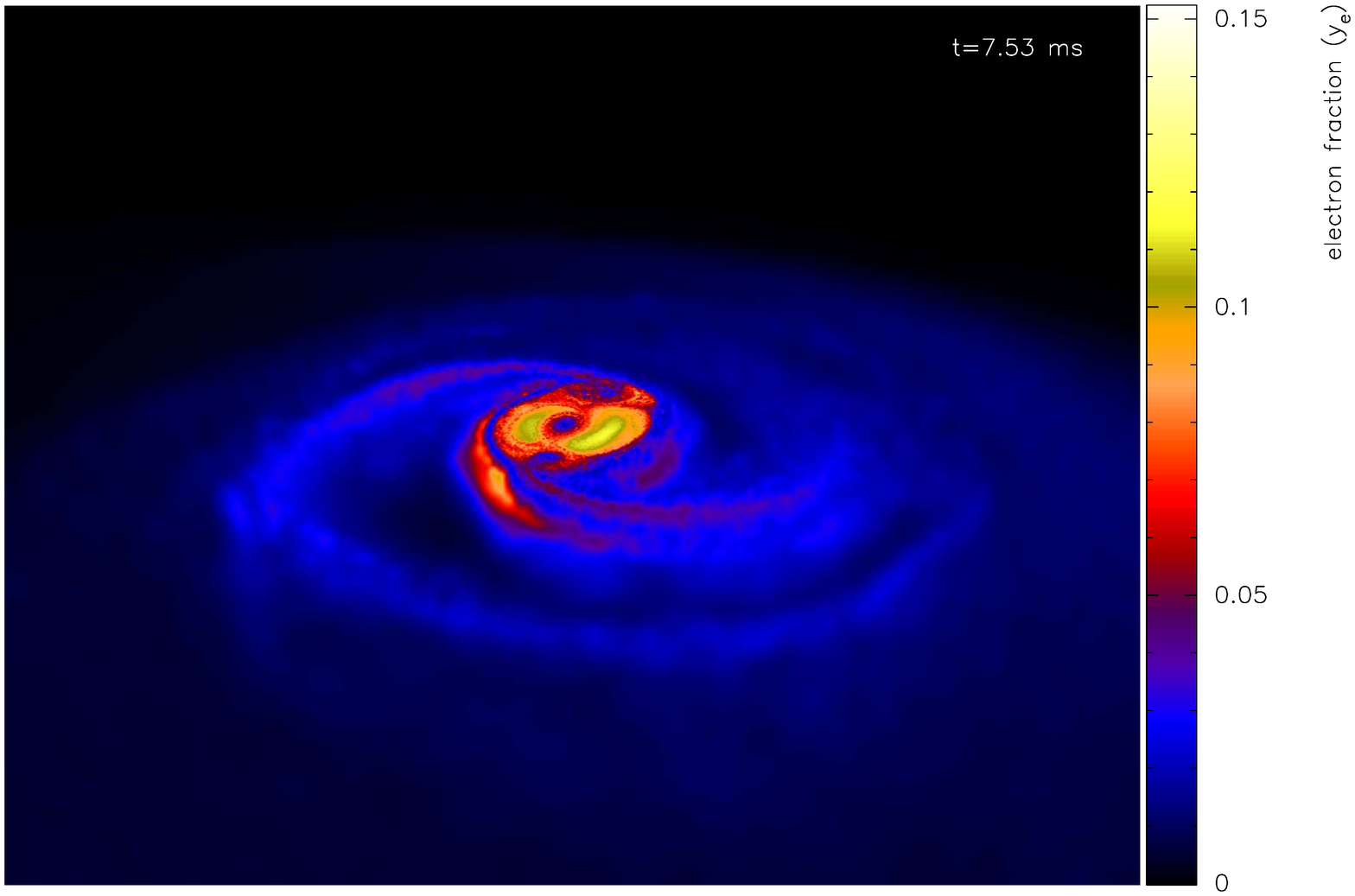}}
   \caption{3D rendering of the $Y_e$ distribution for the generic neutron star 
            merger case (1.3 \msun, 1.4 \msun, no spins; run H). Only matter below the orbital 
            plane is shown. The bulk of the dynamically ejected matter has \ye-values as low as $\sim 0.03$.}
   \label{fig:Ye_dist}
\end{figure*}
\begin{figure*}
   \centering
   \centerline{\includegraphics[width=10cm,angle=0]{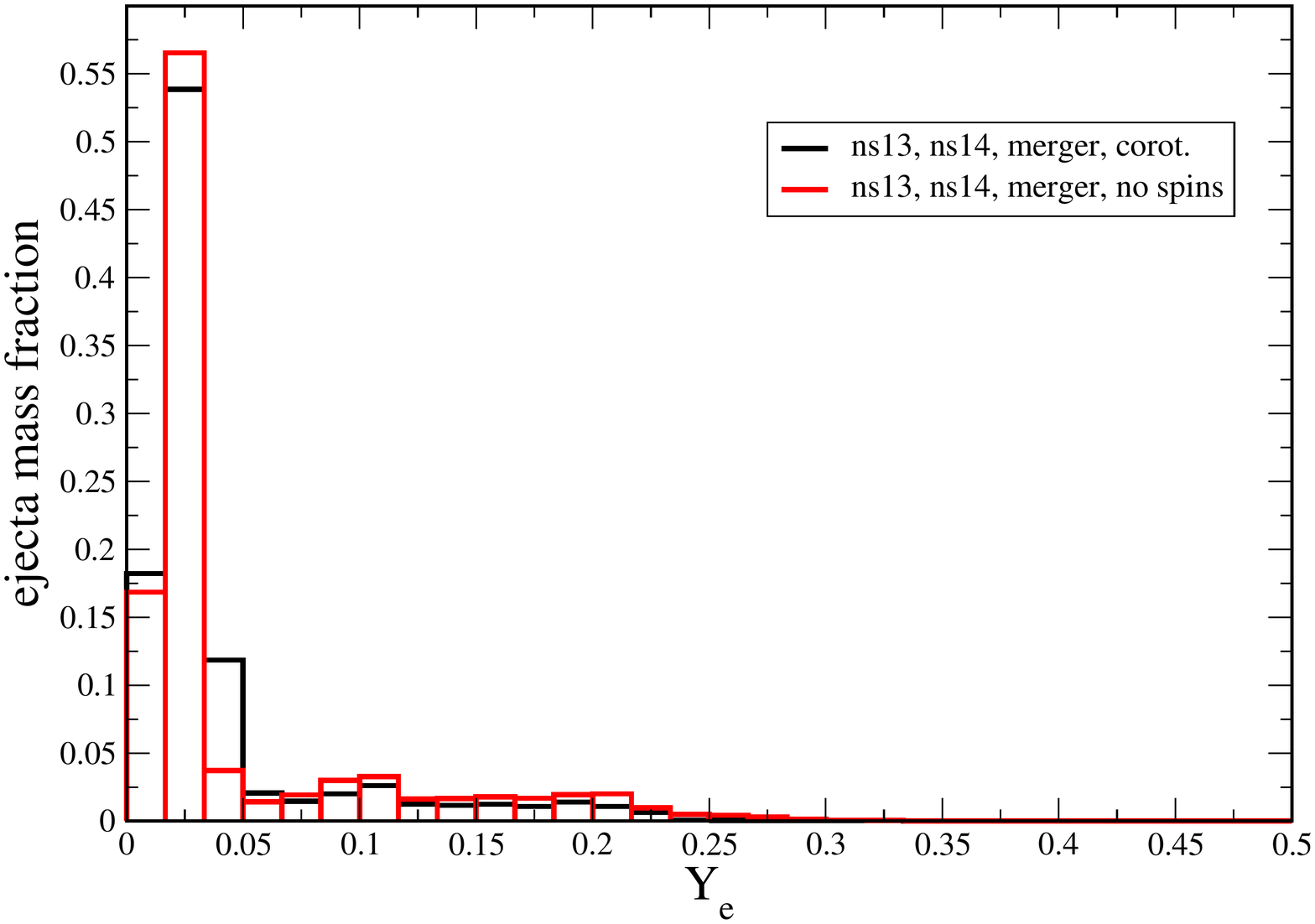} \hspace*{-0.7cm}
               \includegraphics[width=10cm,angle=0]{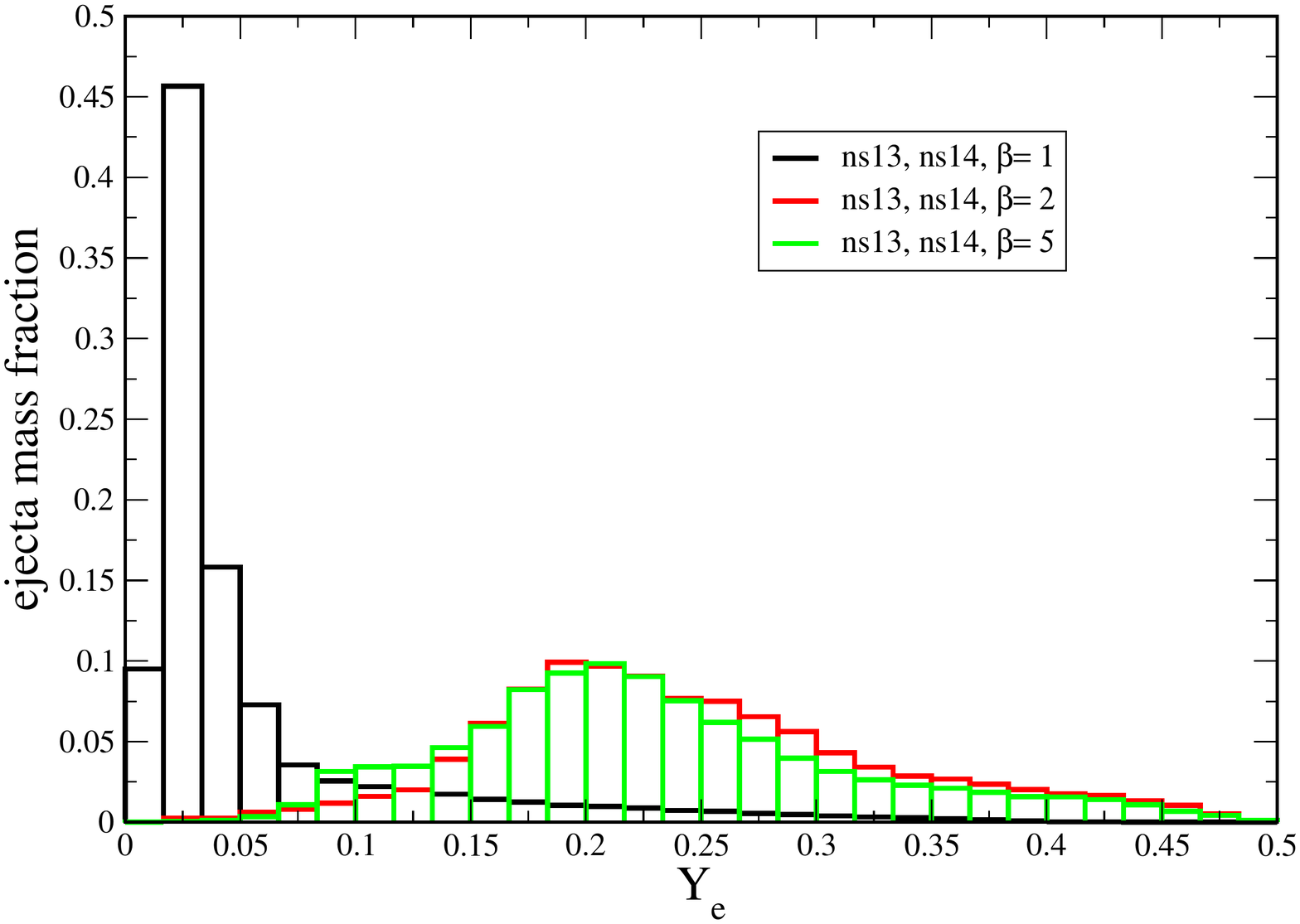}}
   	       \vspace*{-0.4cm}
   \centerline{\includegraphics[width=10cm,angle=0]{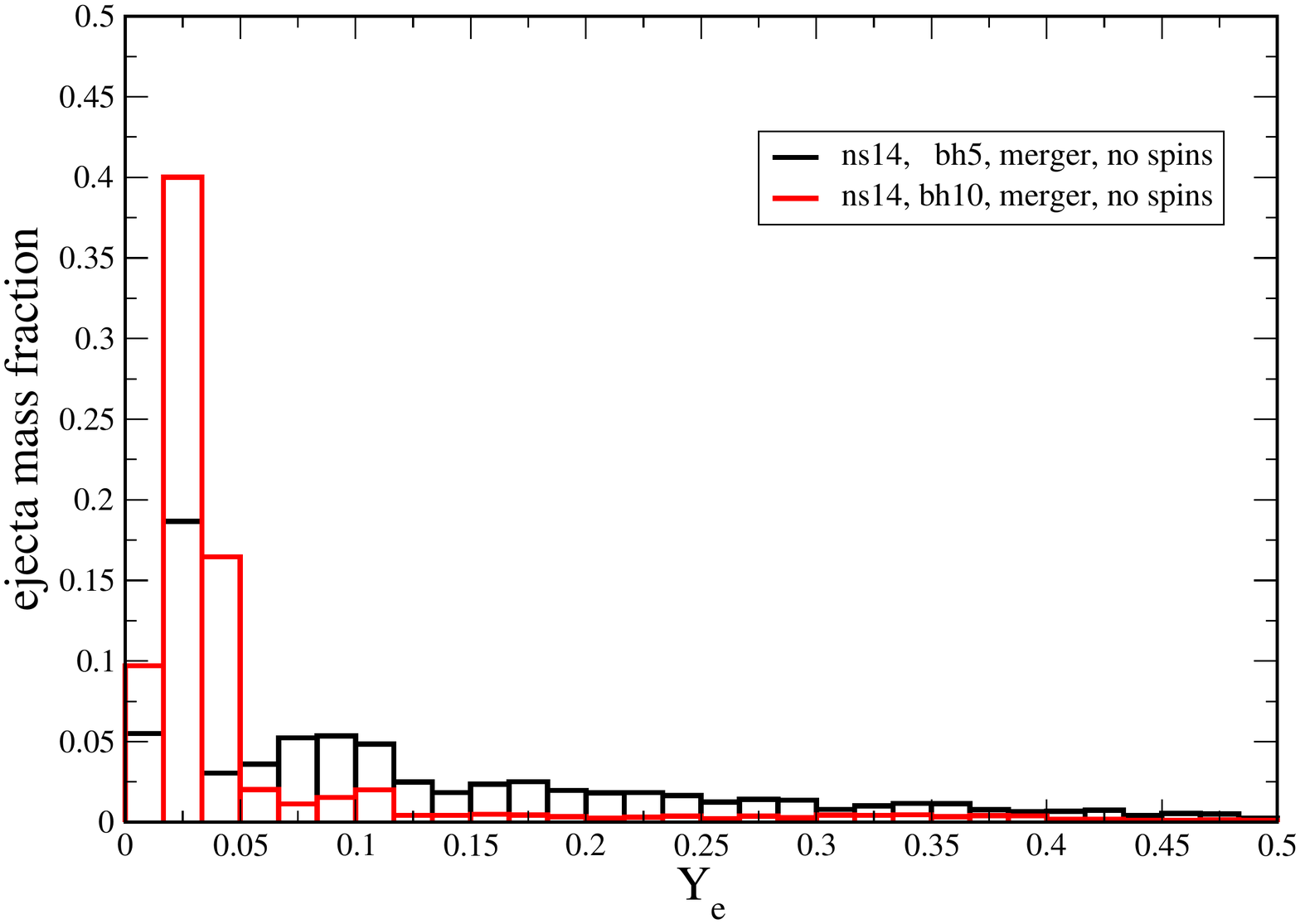}  \hspace*{-0.7cm}
               \includegraphics[width=10cm,angle=0]{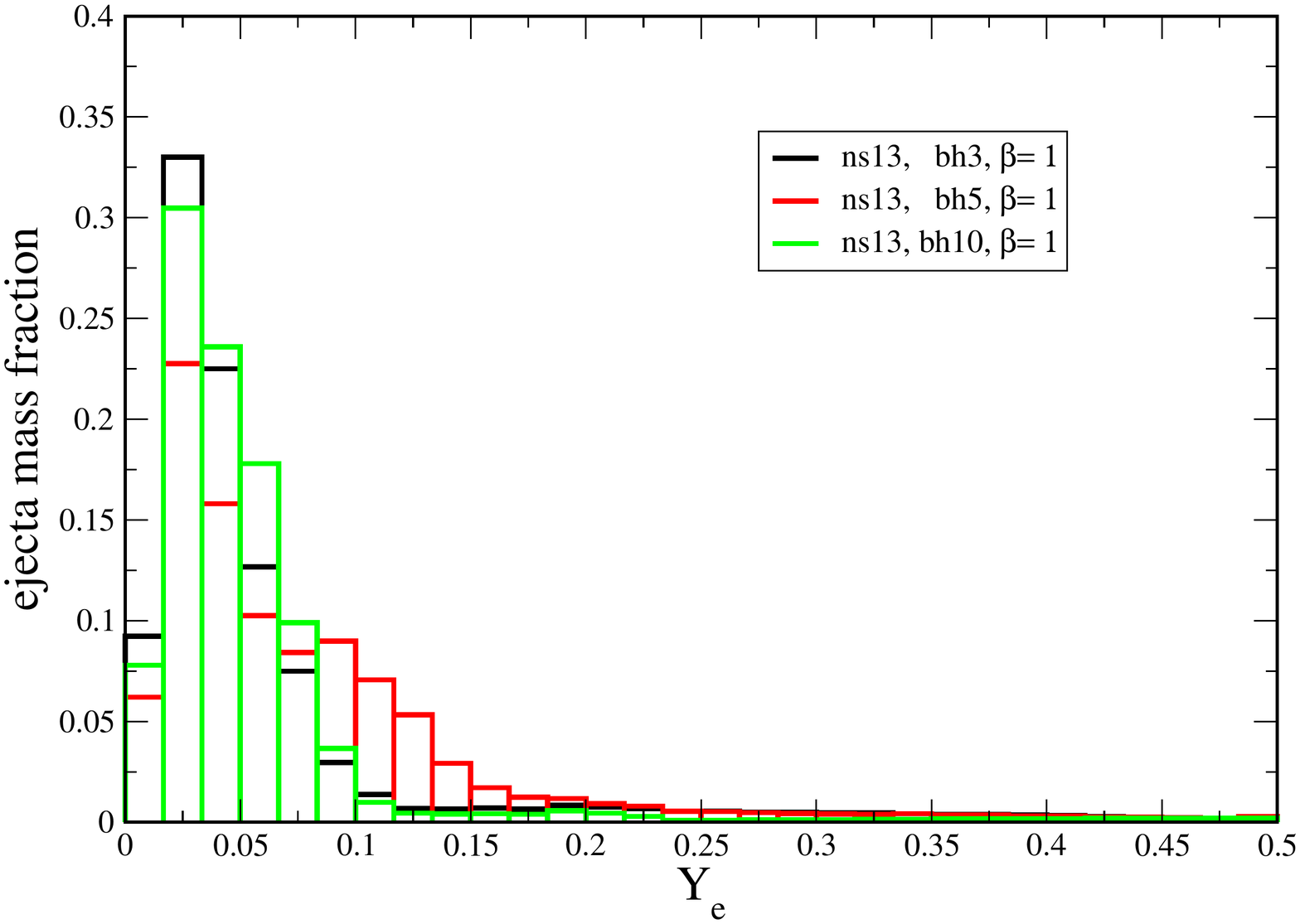}}
   \caption{$Y_e$ distribution in the dynamical ejecta binned by mass. The upper  
   row shows double neutron star encounters (mergers left, collisions right), the 
            lower row shows neutron star black hole encounters (mergers left, collisions right).}
   \label{fig:ejecta_Ye}
\end{figure*}

\begin{figure*}
   \centering
   \centerline{\includegraphics[width=10cm,angle=0]{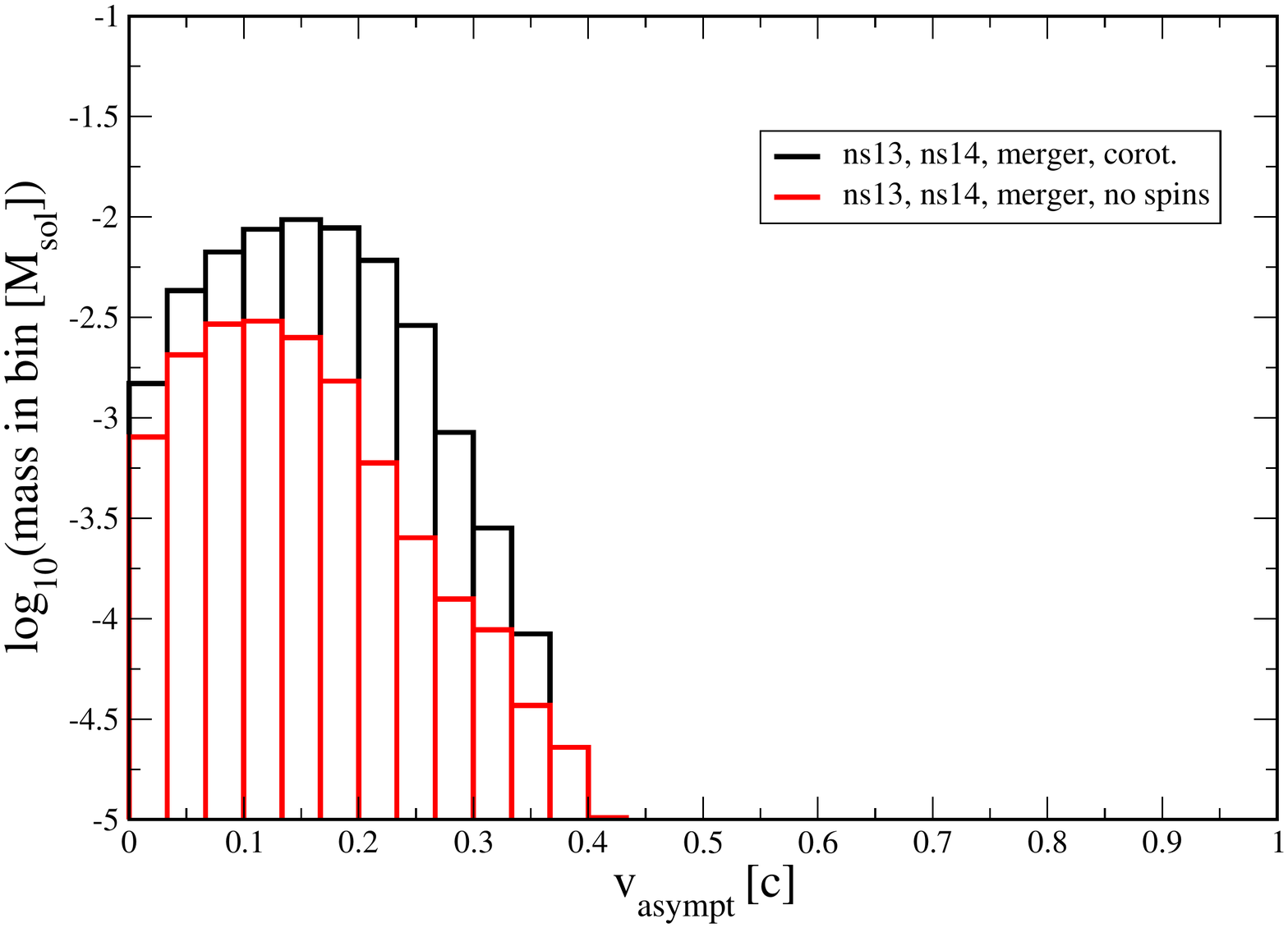} \hspace*{-0.7cm}
               \includegraphics[width=10cm,angle=0]{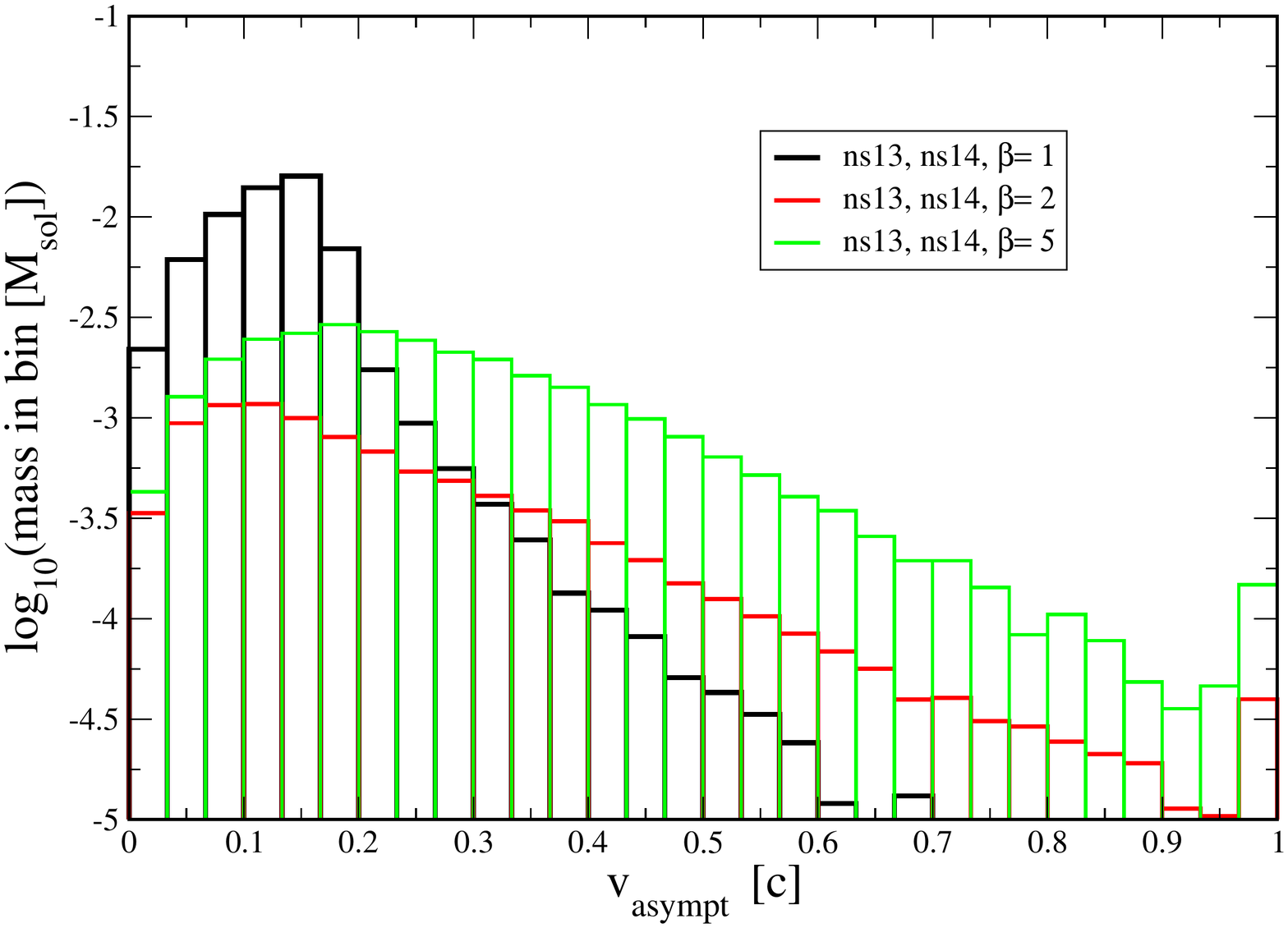}}
   	       \vspace*{-0.4cm}
   \centerline{\includegraphics[width=10cm,angle=0]{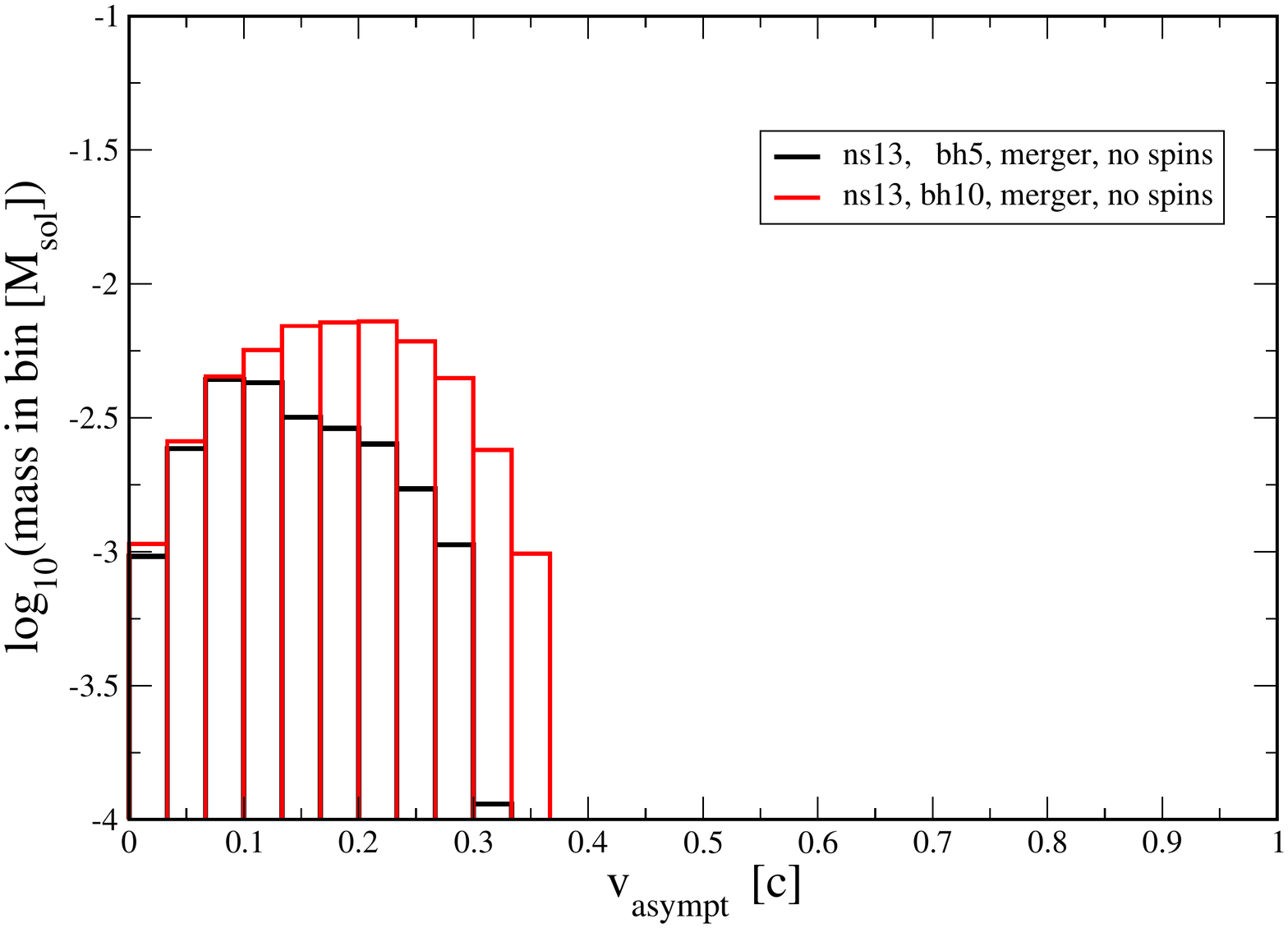}  \hspace*{-0.7cm}
               \includegraphics[width=10cm,angle=0]{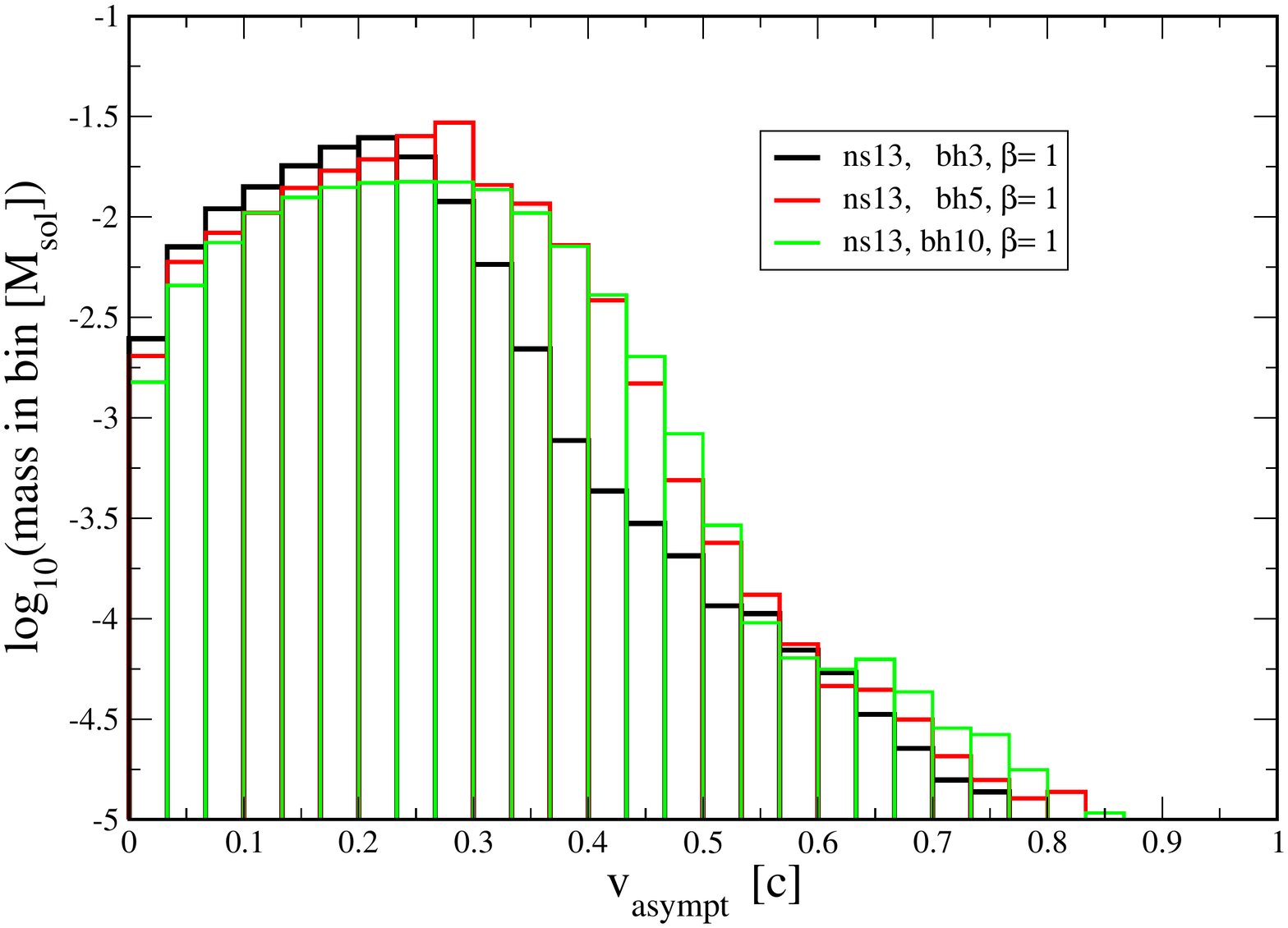}}
   \caption{Distribution of the asymptotic velocity within  the dynamical ejecta. 
            Upper row ns$^2$ encounters, mergers left, collisions right.
            Lower row, nsbh encounters, mergers left, collisions right.}
   \label{fig:ejecta_v_asym}
\end{figure*}
All investigated cases dynamically eject a substantial amount of neutron-rich material,
see Tab.~\ref{tab:mass}\footnote{An overview over the ejecta masses for more than 30 different 
cases can be found in \cite{rosswog12b}.}.
These dynamical ejecta are complemented further by neutrino-driven winds \citep{dessart09}
and  late-time evaporations of the accretion disks \citep{beloborodov08,lee09,metzger09b}.  
The amount of ejecta that we find in the presented simulations is very robust 
with respect to changing the numerical resolution, it is essentially unchanged even if we 
reduce the particle number by an order of magnitude. The amount of dynamical ejecta may change, 
however, with a different treatment of gravitational wave backreaction (reduction), general 
relativity (probably reduction) or a different nuclear equation of state (depending on EOS 
stiffness either reduction or enhancement). In earlier work \citep{rosswog00} we had seen 
the tendency of stiff nuclear equations of state to eject more material than softer ones. 
The recent discovery of a neutron star close to 2.0 \Msun  \citep{demorest10} suggests that 
the true neutron star EOS is indeed very stiff, therefore we consider the Shen et al. EOS 
as a good choice.\\
The amount of dynamically ejected material is summarized in Tab.~\ref{tab:mass}. The most likely case,
run H, ejects 1.4\% of a solar mass with a kinetic energy of $\sim 2 \times 10^{52}$ erg, the corotating 
neutron star merger case ejects three times as much mass. The collision cases eject substantially more, about 
four times as much for the most likely ns$^2$ collision, run A, and up to 10 times as much for nsbh collision, 
run E. The kinetic energies in the ejecta can reach up to $10^{52}$ erg (run E).\\
%
%
The ejected material is extremely neutron-rich and originally resides in the crust and the
outer core of the neutron star. We display a 3D rendering of the $Y_e$ distribution for our 
standard neutron star merger case in Fig.~\ref{fig:Ye_dist} (values are shown at a given artificial
optical depth, see \cite{price07d}). The fractional $Y_e$ distribution within the ejecta of the
different cases is displayed in Fig. \ref{fig:ejecta_Ye}.
Both neutron star merger cases, runs G and H, upper left panel, eject material with a pronounced 
peak near $Y_e\approx 0.03$ with (marginally resolved) higher $Y_e$-contributions from the  
neutron star crust, see Fig.~\ref{fig:initial_profiles}. 
For the most likely ns$^2$ collision, run A, the $Y_e$ distribution looks similar
(upper left panel).
The more extreme --and less likely-- more central collision cases ($\beta=2$ and 5) produce
strong shocks with very large temperatures ($T_{\rm peak} \approx 80$ MeV). In these cases 
positron captures substantially increase  the electron fraction. In both these cases the \Ye  
distribution has a broad peak near  $Y_e\sim 0.2$ (see upper right panel).\\
The nsbh cases show a similar tendency: where the ejecta are predominantly due to tidal torques
they still possess their original, very low \ye. If, in contrast, they suffered strong 
heating in shocks, the \Ye distribution is shifted to larger \Ye values. The nsbh cases experience
a different dynamical evolution. In the long episodic mass transfer phase, 
see Fig.~\ref{fig:GW_amplitudes}, they are continuously tidally heated. In the 
later stages of mass transfer a disk starts to emerge and there is a continuous 
hydrodynamic interaction between this growing disk and the ns remnant core.
Due to this heating history the \Ye distribution is different: it still shows a peak near $Y_e=0.03$
but also extends smoothly to values of $\sim 0.4$. The collision cases 
also show somewhat increased \Ye values ($\sim 0.1$) that are due to the various heating phases 
in the different pericenter passages.\\
%
%
The distribution of asymptotic velocities within the dynamic ejecta is shown in Fig.~\ref{fig:ejecta_v_asym}.
The distributions in  ns$^2$ merger cases (upper left panel) show  peak values around
$0.15~c$ and hardly any material faster than $\sim 0.4~c$\footnote{We chose the lower limit of the ordinate
depending on the numerical resolution. Since the nsbh merger cases have a lower resolution this value
differs from the nsns cases.}. The nsbh merger cases
show very similar velocity distributions.
The collision cases, in contrast, show a much broader distribution of velocities with small mass 
fractions reaching beyond $0.7~c$, see the right column of Fig.~\ref{fig:ejecta_v_asym}.  The bulk of ejecta
in the grazing ns$^2$ collision has velocities similar to the merger cases ($\sim 0.15~c$), but 
smaller amounts of matter can reach substantially higher velocities. The more central collisions show high velocity
tails reaching to close to the speed of light. Such close-to-$c$ velocities in the central collision cases 
($\beta= 2$ and 5) are likely an artifact of our essentially Newtonian simulations. The nsbh collisions 
show a very homogeneous behavior with the bulk of matter reaching about $0.3~c$ and peak velocities out 
to $0.8~c$. The impact of these properties on the electromagnetic signal is further discussed in 
Sec.~\ref{subsec:elmag}.

\subsection{Electromagnetic emission}
\label{subsec:elmag}
\subsubsection{Radio remnants from ejecta-ISM interactions}

At large radii the sub and mildly relativistic outflow from
mergers/collisions is decelerated by the external medium, driving a
fast shock into it. Shocks with similar velocities are seen in the
late phases of GRBs and in early phases of some supernovae, and are
known to produce a bright radio emission. This radio emission is
explained by the large fraction of the internal energy behind the
shock, $\sim$ 10\%, that is channeled into accelerated electrons and a
similar amount of energy that goes into  magnetic fields. In
\cite{nakar11a} and in Paper II \citep{piran12a} we describe the
calculation of the radio emission that follows compact binary
mergers. Here we use the same method to compare the radio remnants
of mergers and collisions.

\begin{figure*}
\centerline{
\includegraphics[width=9cm,angle=0]{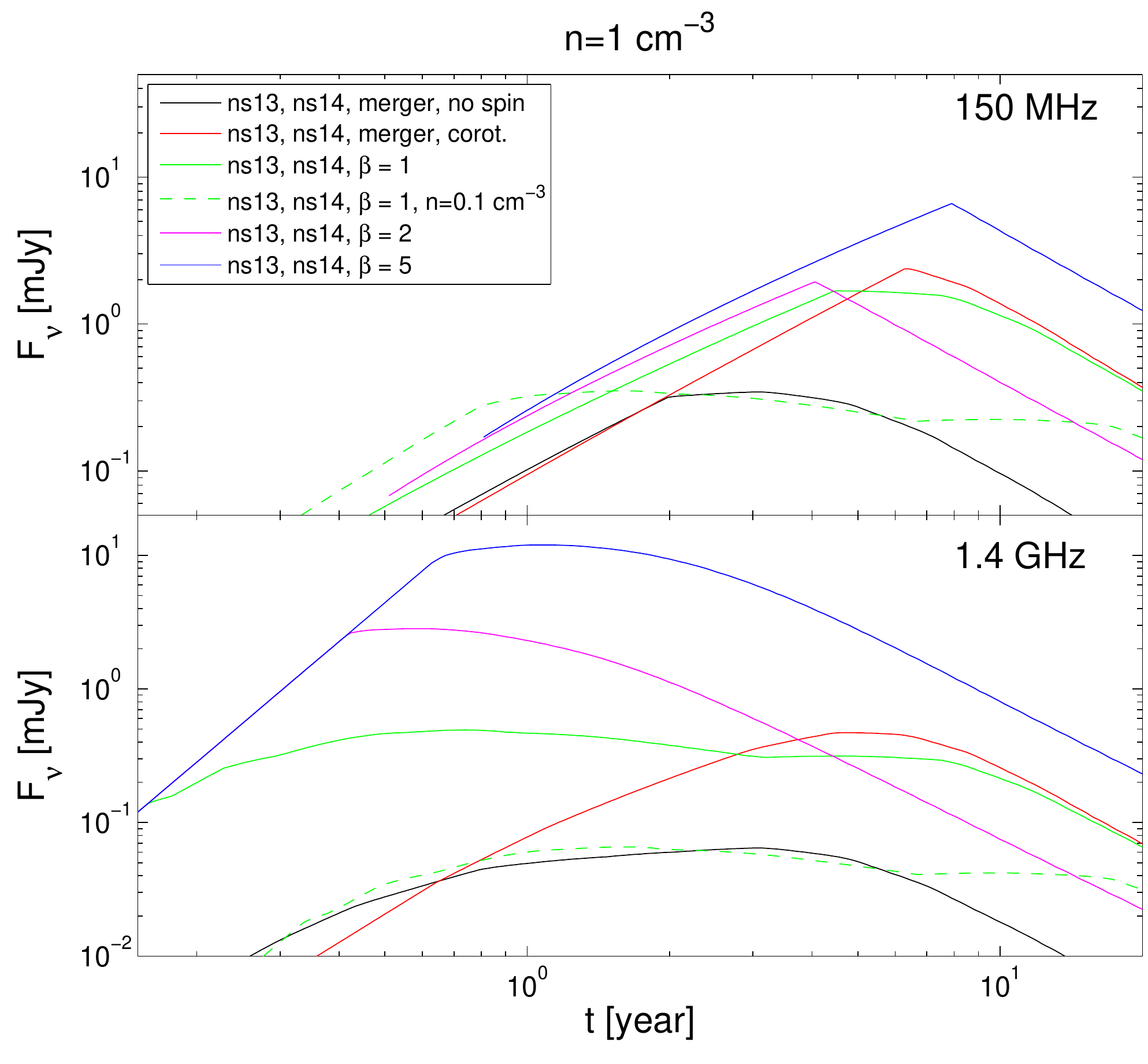}
\includegraphics[width=9cm,angle=0]{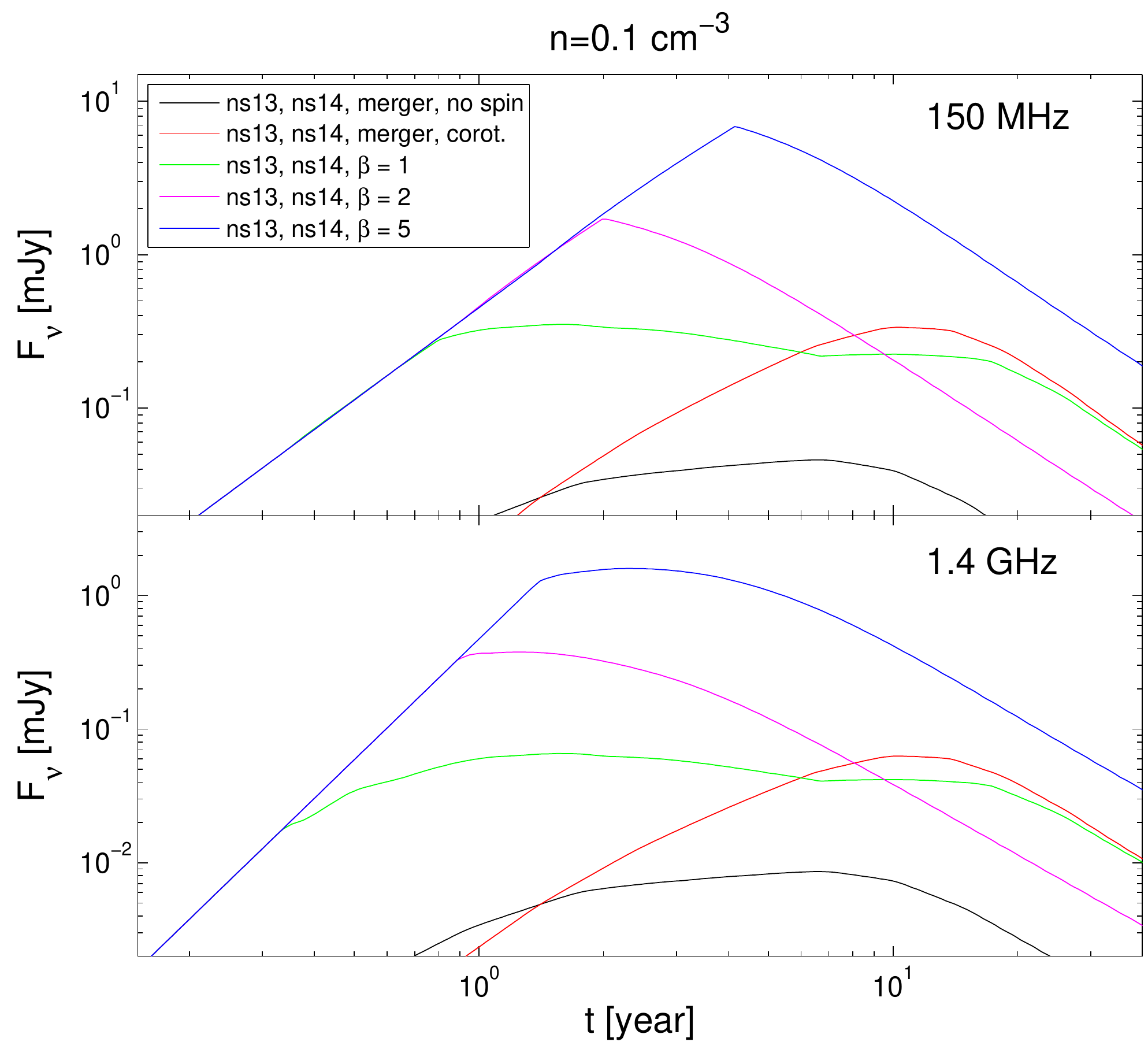}
}
  \caption{Radio lightcurves of ns$^2$ mergers and collisions at 150 MHz 
           and 1.4 GHz for two different densities of the external medium 
           (1 cm$^{-3}$, left and 0.1 cm$^{-3}$, right). The microphysics parameters are
           $\epsilon_e=\epsilon_B = 0.1$ and $p=2.5$, the flux normalization is for
events at a distance of $10^{27}$ cm, roughly the detection horizon for advanced
LIGO and VIRGO. }\label{fig:nsns_radio}
\end{figure*}

\begin{figure*}
\centerline{
\includegraphics[width=9cm,angle=0]{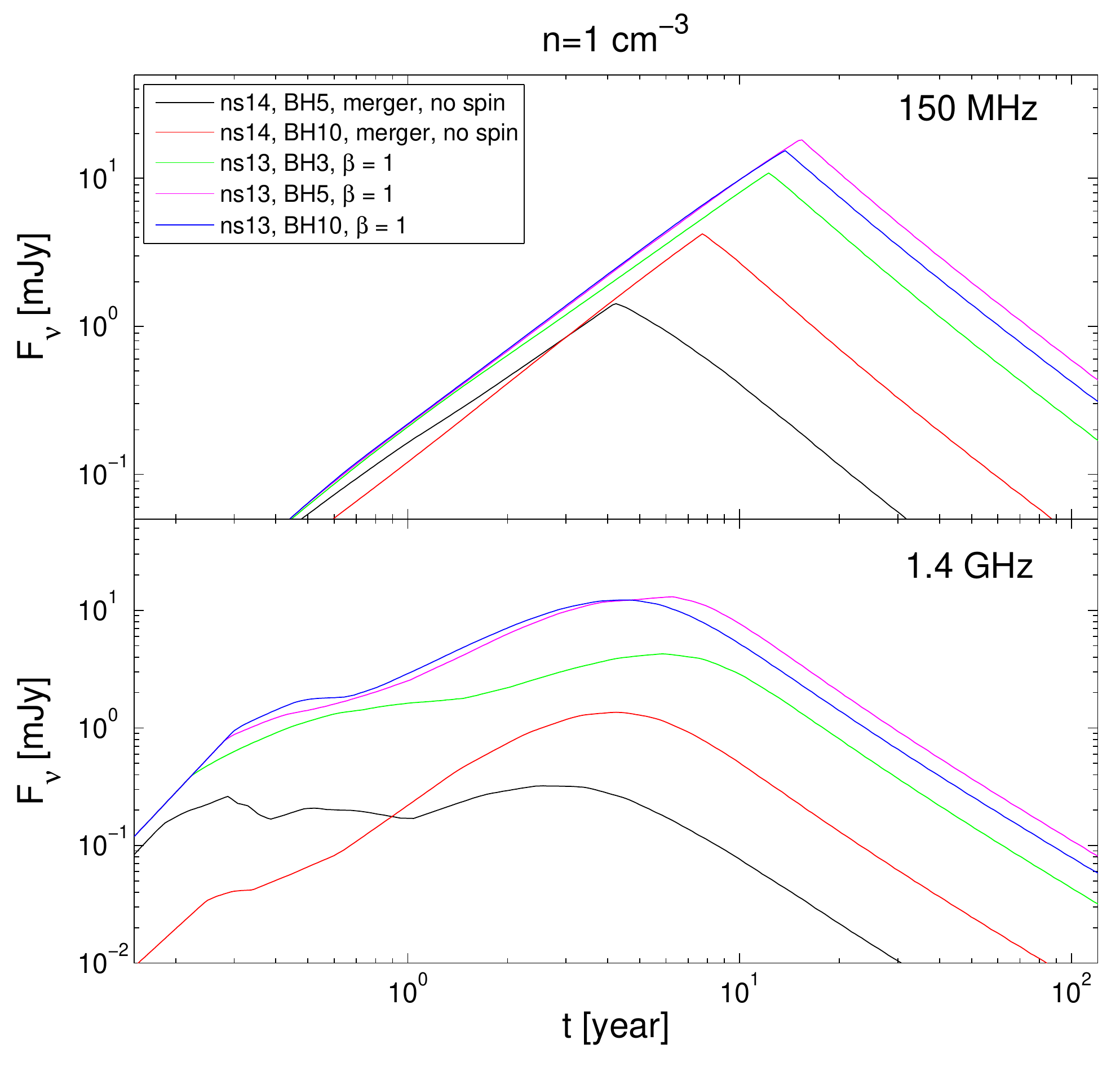}
\includegraphics[width=9cm,angle=0]{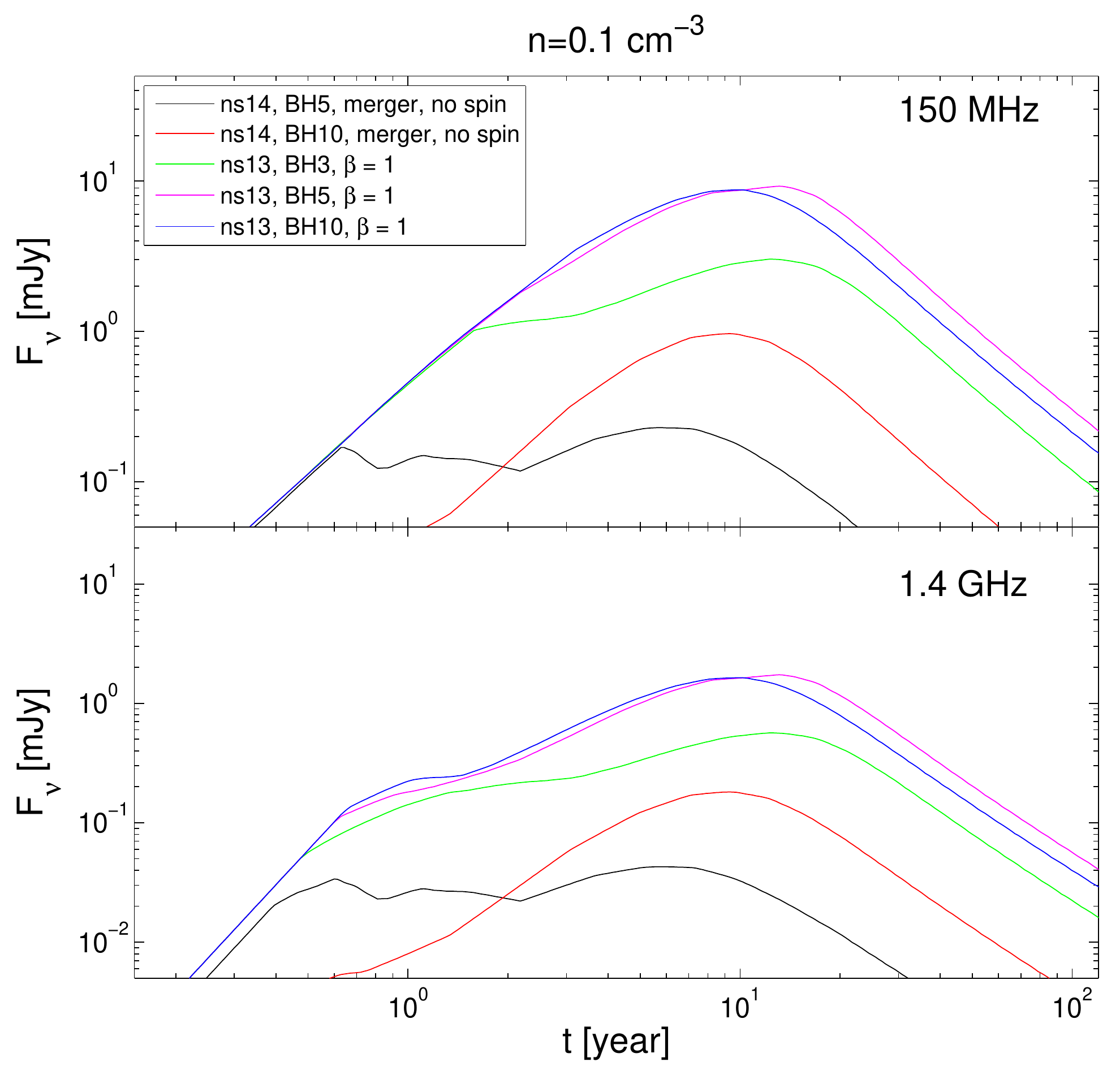}
}
  \caption{Same as Fig.~\ref{fig:nsns_radio}, but for the encounters 
           involving a black hole.}\label{fig:nsbh_radio}
\end{figure*}

In Figs.~\ref{fig:nsns_radio} and \ref{fig:nsbh_radio} we present the
radio signal that results from the dynamical ejecta being slowed down
by the external medium. The calculation is based on the velocity profiles 
presented in
Fig.~\ref{fig:ejecta_v_asym} and it
approximates the interaction as a spherical blast wave in an ambient
medium with a constant density, $n$. Behind the shock constant
fractions of the internal energy, $\epsilon_e=0.1$ and
$\epsilon_B=0.1$, are deposited in relativistic electrons and in
magnetic field, where the electrons are accelerated to a power-law
with an index $p=2.5$. The lightcurves are calculated following the
procedure described in Paper II \citep{piran12a}. Since the simulations are 
Newtonian we do not trust velocities that are close to the speed of light 
and we therefore conservatively restrict the simulation velocities to a 
maximum of 0.75 c when using them to calculate the radio emission. Note that 
the simulations presented here do  not include all possible sources of 
outflows from such events, and especially sources of mildly relativistic 
and relativistic ejecta (e.g., various outflow sources from  near the event 
horizon of the accreting black hole, such as Blandford-Znajek
\citep{blandford77}) are not accounted for. Thus, the true remnants may be
brighter than our prediction, especially on short time scales
(weeks-months) where the dissipation of the energy in the fastest moving
ejecta is most efficient.\\
The radio lightcurves of ns$^2$ and nsbh mergers/collisions, see Figs.
\ref{fig:nsns_radio} and \ref{fig:nsbh_radio}, are calculated at two 
frequencies, 1.4 GHz and 150 MHz, and for two values of external 
densities $n=1 {\rm ~ cm^{-3}}$ and $n=0.1 {\rm ~ cm^{-3}}$. The flux 
normalization is for events at a distance of $10^{27}$ cm, roughly at 
the detection horizon for ns$^2$ mergers by advanced LIGO and Virgo. 
The main difference between collisions and
mergers is that collisions produce brighter radio remnants that rise
faster. Among the collisions the peak flux increases with the
penetration factor $\beta$. In general the flux from collision
remnants rises fast, within less then a year\footnote{The rise time
is set in this case by the artificial cut-off (0.75 c) that we took
in the outflow velocity. A significant outflow with higher velocity
results in a shorter rise time.}, and remains bright (or even
continues to rise, but more slowly) for several years for ns$^2$ collisions
and for decades in the nsbh collision case. Merger remnants, in contrast, are
fainter and they rise over time scales ranging from a year to a
decade and remain bright for a decade or more. The reason for these
differences is that collisions are producing more energetic and
faster outflows. The larger energy and the faster ejecta increase
the remnants' brightness while the faster ejecta are the dominant
factor in the shorter rise time, see the detailed discussion in
Paper II \citep{piran12a} of the effects of the outflow's energy and
velocity on the radio signal. Interestingly, a lower external
density reduces the flux and slows its evolution, as long as the
observed frequency is above the synchrotron self-absorption at all
time. This is the case at 1.4 GHz and therefore, observing at this
frequency alone, one cannot distinguish between a ns$^2$ merger at
$n=1 {\rm ~cm^{-3}}$ and a collision (with $\beta=1$) that takes
place within a medium of density $n=0.1 {\rm ~cm^{-3}}$ (see Fig.
\ref{fig:nsns_radio}). The key for removing the degeneracy is to
observe the remnant also at 150 MHz, where the effect of the
external density on the self-absorption frequency leaves its mark on
the lightcurve and makes the two scenarios (merger in dense
environment vs. collision in sparse environment) distinguishable.

The major uncertainty concerning the detectability of collisions
(the detectability of mergers is discussed in Paper II \citep{piran12a})
is due to  the external density in globular cluster cores. While 
globular clusters usually reside in very low
density environments, the gas in their cores is expected to be
dominated by the mass loss from post main sequence cluster members.
The only globular cluster where gas is clearly detected is 47 Tuc,
where the density of ionized gas is of order $0.1 {\rm ~cm^{-3}}$
\citep{freire01}. This density corresponds to a total mass of 0.1
$M_\odot$ in the cluster core. \cite{vanLoon06} reports on a
tentative detection of 0.3 $M_\odot$ of neutral hydrogen in M15 and 
possible detections of similar amount of neutral hydrogen in two
other clusters (and upper limits of $\sim M_\odot$ in two other).
Therefore, it is realistic to expect a circum collision density of
$0.1 {\rm ~cm^{-3}}$, or even larger. nsbh cases are expected to dominate
the radio remnant population from collisions, since their volumetric rate
is higher \citep{lee10a} and they are about ten times
brighter. Since these remnants remain bright for decades the best
opportunity to identify a remnant is to catch it during its fast
rise phase, which takes about a year. Assuming the upper limit 
for nsbh collisions ($6 \times 10^{-6}$ per Milky way galaxy per 
year, see below) and taking a detection distance of 1 Gpc (as expected 
for a sub-mJy survey at 1.4 GHz) the number of remnants which are younger 
than 1 year, over the whole sky, is $<200$.

\subsubsection{Transients from radioactive decays, ``macronovae''}

As discussed in the previous section, the ejected material is extremely neutron-rich 
and rapidly expanding. Under such conditions rapid neutron capture is 
hard to avoid \citep{hoffman96,freiburghaus99b,roberts11}. The
r-process itself occurs on a very short time scale, but the freshly 
synthesized elements subsequently undergo nuclear fission, alpha- and beta-decay which occur
on much longer time scales.

\begin{figure*}
   \centerline{
   \includegraphics[width=10cm,angle=0]{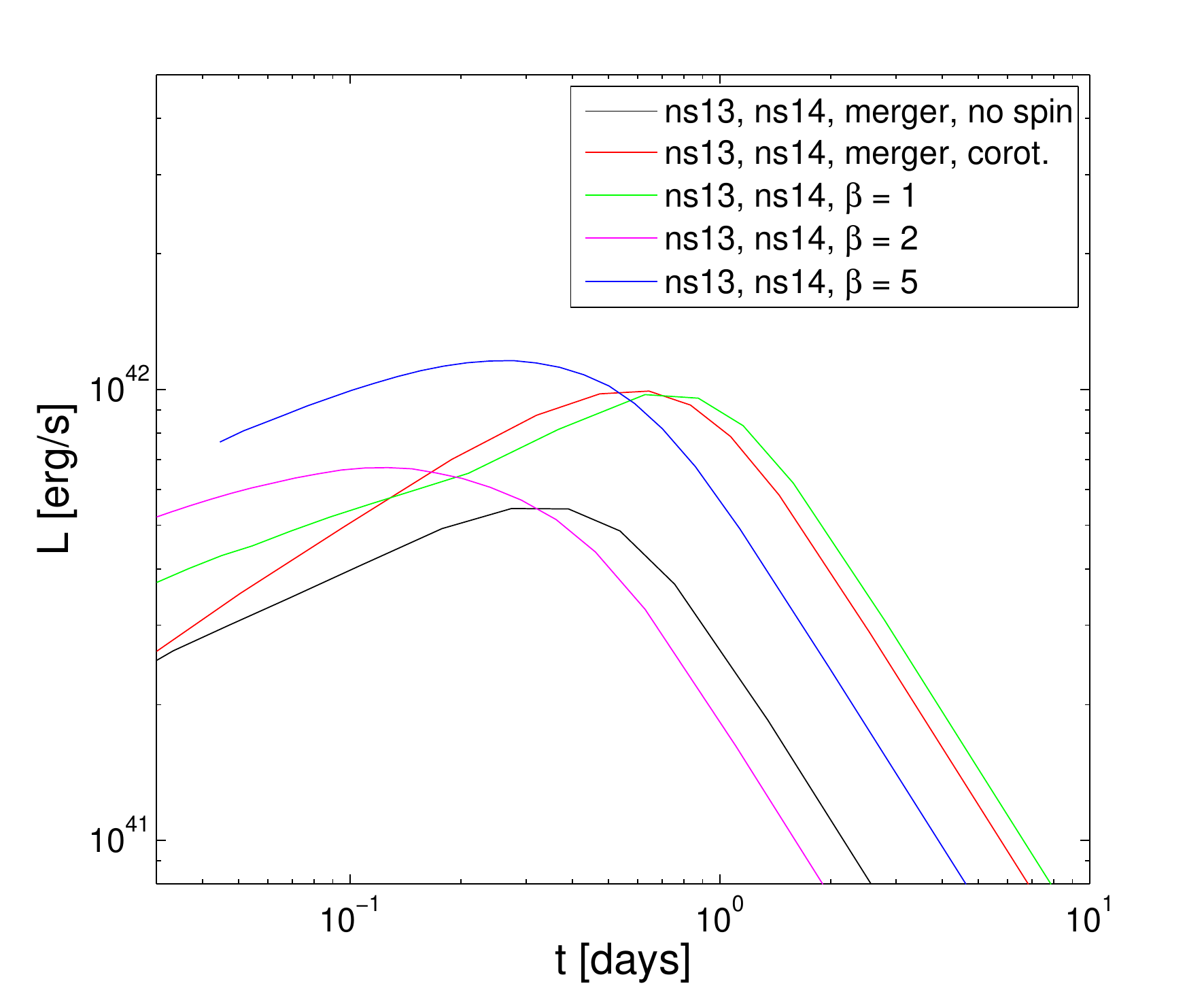}\hspace*{-0.7cm}
   \includegraphics[width=10cm,angle=0]{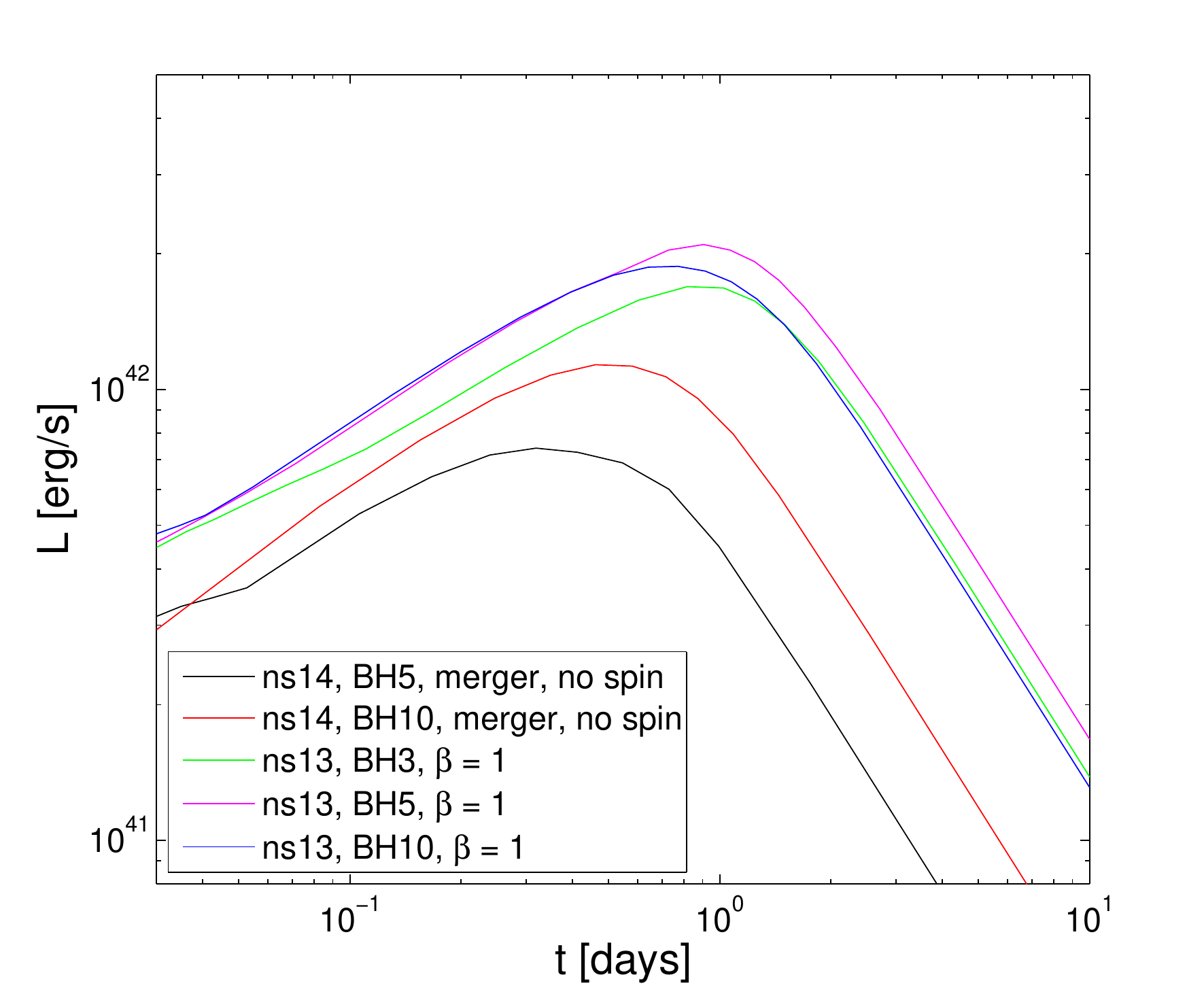}}
   \caption{Radioactive decays within the dynamical ejecta produce transient events
            (``macronovae'') in the optical-UV spectral range that share several properties  
            with supernovae, but evolve on substantially shorter time scales. The lightcurves
            for ns$^2$ encounters (mergers and collisions) are shown in the left panel,
            those of nsbh encounters in the right. See main text for details.}
   \label{fig:macronova_lightcurves}
\end{figure*}

The supernova-like emission powered by this radioctive decay of 
this expanding material was first suggested by \cite{li98}
and discussed later  by \cite{kulkarni05} and \cite{metzger10b}. Here we use
the ejecta velocity profile provided by our simulations to calculate
the observed light curve  following the formalism of 
 \cite{piran12a} - paper II.
\\
The resulting macronova lightcurves are shown in Fig.~\ref{fig:macronova_lightcurves}
for both ns$^2$ (left) and nsbh encounters (right). The canonical merger case (left, black)
peaks after 0.4 days with $\sim 5 \times 10^{41}$ erg/s. The collision cases show a
spread in peak times from 0.1 to 1 days and peak luminosities up to $10^{42}$ erg/s.
The macronova resulting from the merger of the 1.4 \Msun ns with a 5 \Msun bh (right, black)
is similar to the canonical ns$^2$ merger case, the 10 \Msun nsbh merger peaks at about
0.8 days with $10^{42}$ erg/s. The nsbh collision cases all peak beyond 1 day with $\sim
2\times 10^{42}$ erg/s.

\section{Summary and discussion}
\label{sec:summary}
Dynamical collisions between two neutron stars and a neutron star and a stellar mass black hole 
are naturally expected in stellar systems with large number densities such as globular 
clusters. A recent study \citep{lee10a} found that dynamical collisions between two neutron 
stars and a neutron star and a stellar mass black hole occurs more frequently than previously
estimated and that such collisions could contribute to the observed sGRB rate. In a large set
of simulations we have explored the multi-messenger signatures that are produced by both dynamical 
collisions and the more conventional gravitational wave driven compact binary mergers.
These simulations use Newtonian gravity, but benefit from the use of a nuclear
equation of state, a multi-flavour neutrino treatment and a numerical
resolution (up to $10^7$ SPH particles) that exceeds existing studies by far. 
Nevertheless, this study will need to be updated once relativistic simulations
with microphysics become available.\\
\\
{\em Dynamics}\\
Generally, collisions show a larger variety in all of their properties than binary mergers, mainly
due to the lack of strong constraints on their mass ratio and the impact parameter. Typical 
velocity dispersions in globular clusters are orders of magnitude smaller than the velocities
the compact objects reach due to their mutual gravitational attraction, therefore initially 
parabolic orbits are good approximations. A compact binary system is already strongly bound
at the onset of the dynamical merger phase, collisions, in contrast, possess a total orbital energy close 
to zero. To finally form a single remnant they have to get rid of energy and angular momentum by
gravitational wave emission and mass shedding episodes. Colliding 
systems therefore undergo several  close 
encounters before they can finally merge into a single central object surrounded by debris.
Each close encounter launches a tidal tail and the neutron star(s) is/are spun up to 
rotation frequencies close to breakup. The final remnant is then, like in the merger case, a
supermassive neutron star/black hole surrounded by a disk, but the disk is externally fed by 
tidal tails (one per close encounter). A good fraction of the mass is still bound to the 
central remnant and will fall back on times scales substantially exceeding those of the central 
engine.\\
\\
\begin{figure*}
   \centerline{
   \includegraphics[width=10cm,angle=0]{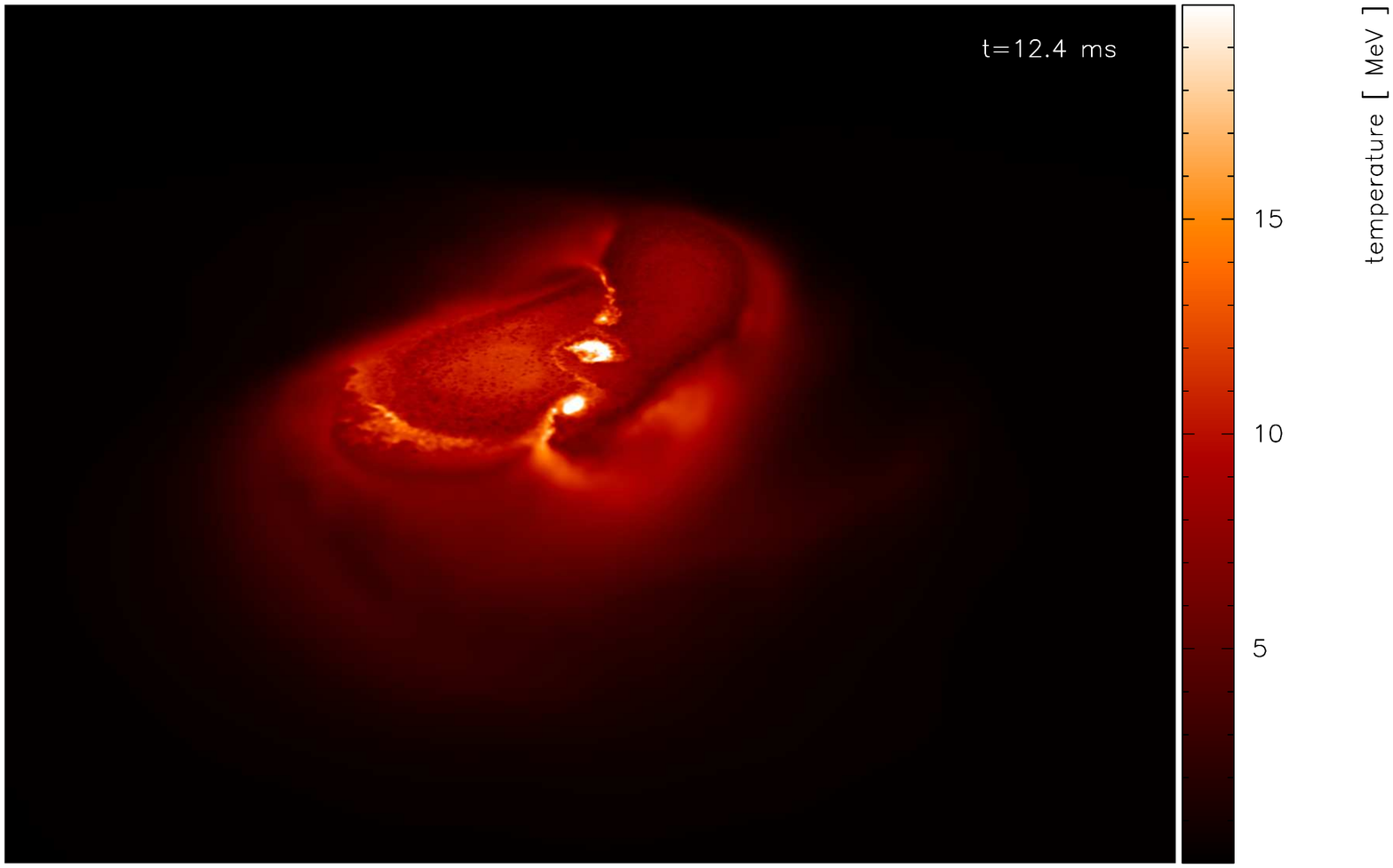} \hspace*{-0.5cm}
   \includegraphics[width=9.6cm,angle=0]{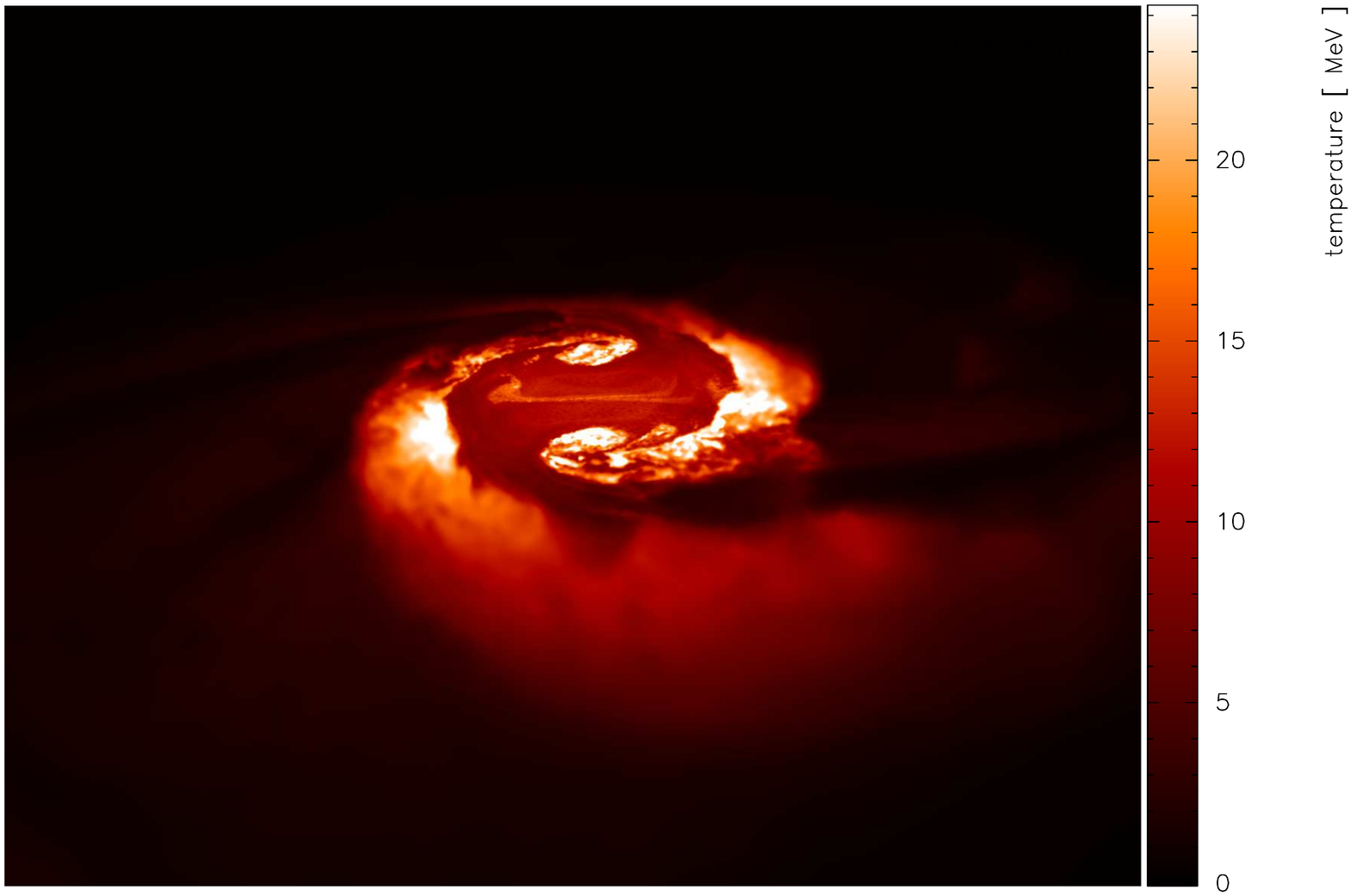}}
   \caption{Kelvin-Helmholtz unstable shear interfaces are a natural 
            consequence of ns$^2$ collision (illustrated here in 3D rendering of the temperature
            distribution; only matter below orbital plane is shown). Left: grazing inpact ($\beta= 1$, run A), 
           right: more central collision ($\beta= 2$, run B), colour bars restricted to 
           $T< 20/24$ MeV to enhance visibility.}
   \label{fig:shear}
\end{figure*}
{\em GRB engines}\\
We find that collisions are at least as promising as GRB engines as gravitational wave driven 
binary mergers. For example, ns$^2$ collisions naturally produce Kelvin-Helmholtz-unstable shear 
layers in each close encounter, see Fig.~\ref{fig:shear}, which have been found in previous
to amplify initial neutron star magnetic fields \citep{price06,anderson08b,obergaulinger10} and they
robustly form massive accretion disks between $\sim0.05$ to $\sim0.4$ \msun. 
Their neutrino luminosities are at least comparable to the standard neutron star merger case,
$\sim 10^{53}$ erg/s, in central ns$^2$ collisions this value can be exceeded by more than an order of 
magnitude, see Tab.~\ref{tab:nu_props}. 
The engine dynamics leaves a clear imprint in the neutrino luminosity. The oscillations of the freshly
formed central remnant of a ns$^2$ collision show up as ms oscillations in the neutrino luminosities, 
see Fig.~\ref{fig:neutrinos}, middle panel. In nsbh collisions the core of the neutron star can
survive several close encounters. Each of them ``re-fills'' the accretion disk and thus produces
apart from a GW burst also peaks in the  neutrino luminosities. We find examples of double peaks
where the first occurs  when the ns core impacts on the disk and the second results from the consumption 
of the ``refilled'' disk, see for example run E (ns of 1.3 \Msun and bh of 5 \msun), 
Fig.~\ref{fig:neutrinos}, last panel.\\
If neutrino annihilation should be the main driver for launching relativistic outflows, collisions 
would be at least as promising as binary mergers. One caveat, though, that applies to both 
mergers and collisions, is that neutrinos can also deposit part of their energy (either via absorption 
or annihilation) in the remnant matter and thus drive strong baryonic winds \citep{dessart09}. 
The latter work found that for at least as long as the central neutron star has not yet collapsed
into a black hole a strong baryonic wind is blown along the original rotation axis which may 
prevent relativistic outflow from forming. This threat may be specific for cases where a
central neutron star survives for many dynamical time scales. In such cases neutrinos deposit
energy in the outer neutron star layers and thus blow off baryons, see Fig. 10 in \cite{dessart09}. 
To which extent this effect also occurs in nsbh encounters is currently not known. This interesting 
topic certainly deserves more detailed work in the future.\\
\\
{\em Fallback and late-time activity}\\
A particularly attractive feature of the collision hypothesis is the possiblity to launch large
amounts of matter into eccentric ``fallback'' orbits. Collisions can deliver more than twice the
fallback mass of the standard neutron star merger case, see Tab.~\ref{tab:mass}.
Our fallback plus thick disk accretion model shows time scales in excess of 10$^2$ s, but it is too
simple to account for possible flares. Generally, $\dot{M}c^2$ has dropped below $10^{48}$ 
erg/s at times beyond $10^2$ s. An intriguing possibility, though, is the survival of the
neutron star core in nsbh collisions. In all investigated cases it survives at least the
first encounter, in one calculation even three of them and is finally ejected with $\sim 0.1$
\Msun in a close-to-parabolic orbit. The interaction with the debris from previous passages
may actually brake the core enough to fall back to the bh. This would trigger an impulsive accretion
event with $\sim 10^{52}$ erg at possibly very late times. Such scenarios may be responsible
for the observed late-time activity in some cases, but they are probably too rare (see below)
to account for extended emission in a substantial fraction of sGRBs.\\
\\
{\em Detectability of the neutrino signal}\\
Most of the energy of both mergers and collisions escapes as $\sim15$ MeV neutrinos. The 
neutrino signature is comparable in magnitude and duration to the signals of a 
typical core collapse supernova, though dominated by electron antineutrinos. Current  
facilities can only detect individual neutrino events from within the galaxy (or at most events 
within the local group). Even if more sensitive detectors are built in the distant future it 
will be difficult to distinguish the very rare merger or collision events among the much more 
frequent signals from core collapse SNe. So the neutrinos from compact object encounters may 
just make an individually undetectable $\sim 0.1\%$ contribution to the diffuse neutrino 
background.\\
\\
\noindent{\em Detectability of the gravitational wave signal}\\
In both mergers and collisions gravitational radiation carries also a significant fraction of the 
energy, second only to the neutrino signal.  The quasi-regular mergers' chirping signal can be 
detected up to distances of a few hundred Mpc, they are the prime targets for gravitational radiation 
detectors like advanced LIGO and Virgo. Since gravitational waves from eccentric binary systems
efficiently radiate angular momentum relative to energy, mergers of primordial binaries are
expected to occur at practically zero eccentricity. Binaries that have formed dynamically, say 
in nuclear or globular clusters, however, form at small orbital separations but with large eccentricities
and they may not have enough time to circularize up to merger. Thus, their close-to-encounter orbital dynamics 
and gravitational wave signal may differ substantially. Initially they produce a series of well-separated, 
repeated GW bursts that continues for minutes to days. This sequence of bursts gradually transforms into 
the powerful chirp inspiral signal of an eccentric binary system. \cite{kocsis06a} found the 
signal-to-noise ratio for GWs produced in single parabolic passages  to be 
significant only for rather deep encounters where the initial pericentre
separation is $r_{\rm p,0} < 6 M$, $M$ being the total mass ($G=c=1$).  
\cite{oleary09} estimate that nsbh mergers that result from tidal capture binaries  in galactic 
nuclei could be detectable by advanced LIGO type facilities at rates of
$\sim 1$ per year, i.e. they could contribute of order 1 \% to the overall detection rate.
Since a significant fraction of tidal capture binaries is expected to merge at non-negligible
eccentricities their properties are bracketed between those of the quasi-circular binary 
mergers and the $\beta\sim1$ collisions that were discussed in this paper. A first attempt to 
model such high eccentricity mergers has been undertaken by 
\cite{east12}, further modeling is left for future efforts.\\

\noindent{\em Rate constraints from ejected matter}\\
It had long been suggested that the cold decompression of neutron star material could produce a 
substantial contribution to the r-rprocess inventory of the Universe 
\citep{lattimer74,lattimer76,eichler89,freiburghaus99b,rosswog99}.
Core-collapse supernovae are often considered as the ``standard'' r-process production site, but recent 
studies suggest that they are seriously challended in producing the whole observed r-process 
pattern \citep{roberts10,fischer10,arcones11a} unless very specific SN events are invoked 
\citep{winteler12b}. 
These findings make an alternative/additional r-process production channel more welcome than ever.\\
Our reference case, a neutron star merger with masses of 1.4 \Msun (run H), ejects $0.014$ 
\Msun of extremely neutron-rich (\ye $\sim 0.03$) matter, mainly from the
outer core with small contaminations from the more proton-rich crust, see Fig.~\ref{fig:ejecta_Ye}.
Grazing ns$^2$ collisions (run A) eject material with very similar properties, while more
central collisions yield substantially higher temperatures/$e^+$-capture rates and therefore
larger $Y_e$ values (\ye $\approx 0.2$), though still very low in comparison to core-collapse
SNe.
If we assume an average Galactic r-process production rate of $10^{-6}$ \Msun per year \citep{qian00}
we can place robust upper limits on the admissible encounter rates under the assumption that the ejecta
are made entirely of r-process nuclei. Our recent study \citep{korobkin12a} finds indeed that the dynamic
ejecta of ns$^2$ and nsbh mergers are excellent candidates for the production sites of the 
robust heavy r-process patter with $A$ larger than $\sim 120$. 
\begin{figure}
   \hspace*{-0.9cm}\includegraphics[width=10cm,angle=0]{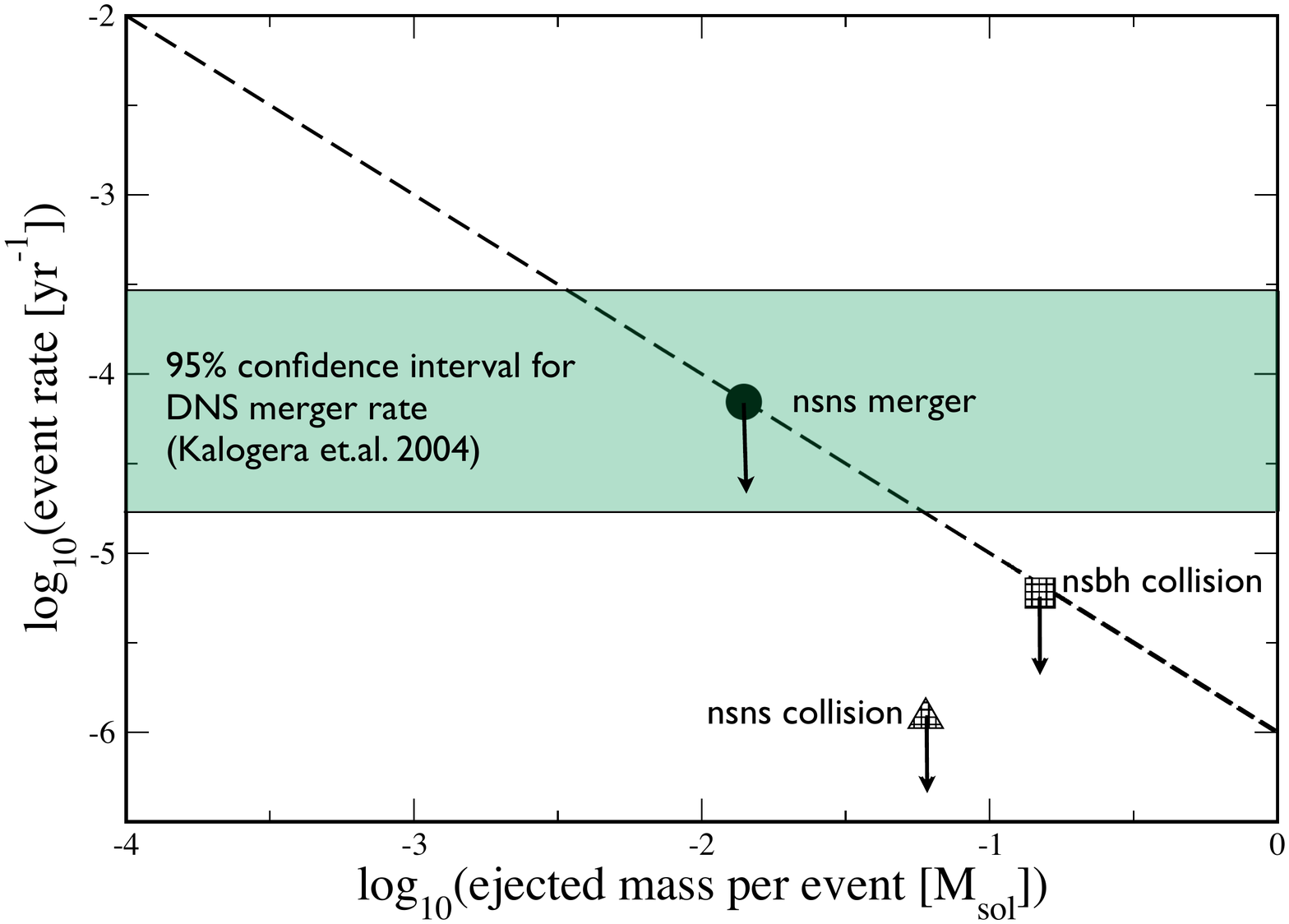}
   \caption{Nucleosynthetic constraints on the event rates. The double neutron star merger rate that
            is required in order to produce all the r-process material (using the average rate of $10^{-6}$ 
            \Msun yr$^{-1}$, dashed line, Qian 2000) is marked by the filled circle. This rate is promisingly
            consistent with the rate derived from observations by Kalogera et al. (2004) shown as green band. 
            If we make the (extreme) assumption that only ns$^2$ and nsbh collisions produce all the r-process 
            and their relative ratio is 1:5 (Lee et al. 2010) we can place upper limits on 
            their occurence rates. These limits are marked as triangle (ns$^2$) and square (nsbh).
            The rates realized in nature might be substantially below these values.}
   \label{fig:rate_constraints_nucleo}
\end{figure}
If neutron star mergers alone were responsible for all r-process material this would constrain
their galactic rate to $7 \times 10^{-5}$ \Msun yr$^{-1}$, right in the centre of the 95\% confidence
interval derived from the observed ns$^2$ binary distribution \citep{kalogera04a}, see 
Fig.~\ref{fig:rate_constraints_nucleo}. Within the given uncertainties 
they would be perfectly consistent with delivering a substantial contribution to the galactic r-process 
matter\footnote{It remains to be investigated, though, to which extent this is consistent with galactic chemical
evolution, see e.g. \cite{argast04}}. The collision cases with their larger ejection yields, however, are 
seriously constrained in their allowed occurrence rates. If we make the extreme assumption that ns$^2$ and nsbh
collisions produce all the r-process and use a nsbh rate, $R_{\rm coll,nsbh}$, that is five times larger than 
the ns$^2$ collision rate $R_{\rm coll,ns^2}$ \citep{lee10a}, we find upper limits of 
$R_{\rm coll,ns^2} < 1.2 \times 10^{-6}$ yr$^{-1}$ and $R_{\rm coll,nsbh} < 6 \times 10^{-6}$ yr$^{-1}$. Therefore 
the total (ns$^2$+nsbh) collision rate must be $R_{\rm coll} < 7.2 \times 10^{-6}$ yr$^{-1}$, or about 10\% of
the double neutron star merger rate. Given our extreme assumption that no other astrophysical event produces 
r-process material the true rate is likely substantially below this estimate.\\
\\
{\em Macronovae}\\
We estimate the properties of the electromagnetic transients due to radioactive decays for
spherically symmetric outflows. The standard neutron star merger case peaks after $\sim 0.4$ days
with a luminosity of  $5 \times 10^{41}$ erg/s. The other merger cases (ns$^2$ and nsbh) and
the ns$^2$ collisions peak with higher luminosities, both within a factor of about two in comparison
to the standard merger case. The neutron star black hole collisions form a distinct group: they all peak
beyond 1 day and are substantially brighter that the standard merger case ($\sim 2 \times 10^{42}$
erg/s).\\
\\
{\em Radio signal from ejecta-ISM interactions}\\
In analogy with supernova remnants,  the matter that is ejected at sub and mildly relativistic 
velocities produces a longer lasting radio flare \citep{nakar11a}. The flare 
is the most robust electromagnetic counterpart that is expected from compact 
mergers or collisions. It depends only on the total energy ejected in mildly or 
subrelativitic energies and on the density of the external matter. A flare from a 
canonical ns$^2$ merger would peak on a time scale of a year with a peak observed flux 
of $\sim 0.1$ mJy at 1.4 GHz from a source from a distance of $10^{27}$cm, roughly 
the detection horizon of  the gravitational radiation signal  by  the advanced LIGO/Virgo.  
The signals are longer and weaker if the external density is lower.  ns$^2$ collisions or bhns 
collisions eject much more material and with a higher velocity and hence the corresponding 
flares would by stronger and  longer. In addition the rise time of these flares will be faster 
(because of the higher velocities) and hence it will be easier to detect them. In general 
the rising phase  of the flares  depends  critically on the amount of fastest matter ejected. 
In these Newtonian calculations we had capped the maximal velocity at 0.75 c hence our results
on the early emission underestimate the true emission. This  part of the relativistic ejecta 
that we miss in the simulation may produce a brighter and a faster rising signal even if its 
energy is ten times smaller.

\section*{Acknowledgments}

It is a pleasure to thank Almudena Arcones, Brian Metzger, Gabriel Martinez-Pinedo,
Oleg Korobkin and B. Sathyaprakash for stimulating discussions. 
The work of SR was supported by DFG grant RO-3399, AOBJ-584282.
This research was supported in part by the National Science Foundation under Grant No. PHY05-25915. Parts of
this paper have been written at the University of Queensland, Brisbane, Australia. This 
visit was supported by the DFG by a grant to initiate and intensify bilateral collaboration.
S.R. thanks Enrico Ramirez-Ruiz/UC Santa Cruz and Andrew MacFadyen/NYU for 
their hospitality and for many stimulating discussions. 
It is a pleasure to acknowledge the use of the visualization software SPLASH developed by Daniel \cite{price07d}. 
The simulations of this paper were performed on the facilities of the 
H\"ochstleistungsrechenzentrum Nord (HLRN). T.P.'s research was supported by an Advanced ERC research grant. 
E.N. research was supported by ISF and IRG grants.\\
\\
Movies from our hydrodynamic simulations and ejecta trajectories can be 
downloaded from: http://compact-merger.astro.su.se/

\bibliography{astro_SKR}
\bibliographystyle{mn2e}
\bsp

\label{lastpage}

\end{document}